\documentclass[
  twocolumn, 
  twocolappendix, 
  resetfootnote]{aastex701}
\usepackage{amsmath}
\usepackage{mhchem}
\usepackage[range-phrase={--}, range-units=single]{siunitx}
\usepackage{makecell}
\usepackage{multirow}

\newcommand{\unitrange}[2]{\unit{#1}\text{--}\unit{#2}}

\newcommand{\ionname}[2]{#1\,#2}
\newcommand{\atomterm}[3]{$#1\, {}^{#2}\mathrm{#3}$}
\newcommand{\ionandterm}[3]{\ionname{#1}{#2}~#3}

\newcommand{\myrefsec}[1]{Section~\ref{#1}}
\newcommand{\myrefapp}[1]{Appendix~\ref{#1}}
\newcommand{\myreffig}[1]{Figure~\ref{#1}}
\newcommand{\myreftab}[1]{Table~\ref{#1}}
\newcommand{\myrefeq}[1]{Equation~\ref{#1}}

\shorttitle{Non-LTE modeling for He and Sr in NSM ejecta}
\shortauthors{Chiba et al.}
\received{2026 April 3}
\revised{2026 August 11}
\accepted{2026 September 4}
\graphicspath{{./}{figures/}}
\begin{document}

\title{Non-LTE Ionization Modeling for Helium and Strontium in Neutron Star Merger Ejecta}

\correspondingauthor{Koya Chiba}
\email{chiba.koya@astr.tohoku.ac.jp}

\author[0009-0009-6111-3985]{Koya Chiba}
\affiliation{Astronomical Institute, Tohoku University, Aoba, Sendai 980-8578, Japan}
\email{chiba.koya@astr.tohoku.ac.jp}

\author[0000-0001-8253-6850]{Masaomi Tanaka}
\affiliation{Astronomical Institute, Tohoku University, Aoba, Sendai 980-8578, Japan}
\affiliation{Division for the Establishment of Frontier Sciences, Organization for Advanced Studies, Tohoku University, Sendai 980-8577, Japan}
\email{masaomi.tanaka@astr.tohoku.ac.jp}

\author[0000-0002-4759-7794]{Shinya Wanajo}
\affiliation{Center of Gravitational Physics and Quantum Information,
Yukawa Institute for Theoretical Physics, Kyoto University, Kyoto, 606-8502, Japan}
\affiliation{Astronomical Institute, Tohoku University, Aoba, Sendai 980-8578, Japan}
\email{shinya.wanajo@yukawa.kyoto-u.ac.jp}

\author[0000-0001-6467-4969]{Sho Fujibayashi}
\affiliation{Frontier Research Institute for Interdisciplinary Sciences, Tohoku University, Sendai 980-8578, Japan}
\affiliation{Astronomical Institute, Tohoku University, Aoba, Sendai 980-8578, Japan}
\email{sho.fujibayashi@astr.tohoku.ac.jp}

\author[0000-0003-4443-6984]{Kyohei Kawaguchi}
\affiliation{Max Planck Institute for Gravitational Physics (Albert Einstein Institute), Am M\"{u}hlenberg 1, Potsdam-Golm, 14476, Germany}
\affiliation{Center of Gravitational Physics and Quantum Information,
Yukawa Institute for Theoretical Physics, Kyoto University, Kyoto, 606-8502, Japan}
\email{kyohei.kawaguchi@aei.mpg.de}

\author[0000-0002-2502-3730]{Kenta Hotokezaka}
\affiliation{Research Center for the Early Universe, Graduate School of Science,
The University of Tokyo, Bunkyo, Tokyo 113-0033, Japan}
\email{kentah@g.ecc.u-tokyo.ac.jp}

\begin{abstract}
The material ejected from a binary neutron star merger 
produces ``kilonova,'' a radioactively powered emission 
at ultraviolet, optical, and infrared wavelengths.
The early-phase spectra of the kilonova AT2017gfo, 
following the gravitational wave event GW170817, 
exhibit a strong absorption feature around $1\,\mathrm{\mu m}$.
Helium (He) and strontium (Sr) have been proposed 
as the candidate elements contributing to this feature. 
However, due to the lack of consistent modeling 
including these two elements simultaneously,
the exact contributions of each element to this feature remain unclear.
In this study, we develop 
non-local thermodynamic equilibrium ionization models 
for He and Sr that take into account ionization by high-energy electrons, 
and estimate the abundances of each element required to reproduce
the observed feature 
in the early-phase spectra of AT2017gfo. 
Our modeling indicates that about \qty{1}{\%} of He 
or \qtyrange{1}{10}{\%} of Sr in mass fraction are present 
in the ejecta moving at $v \sim 0.15 \, c$.
This Sr mass fraction nicely agrees with the mass fraction 
in the solar $r$-process pattern.
Based on comparison with nucleosynthesis calculations, our constraints suggest 
that $r$-process nucleosynthesis in GW170817 occurs 
at relatively low electron fraction ($Y_{\rm e} \lesssim 0.35$) 
and low entropy ($s \lesssim 30 \, k_\mathrm{B}/\mathrm{nucleon}$) conditions.
Generally, for $Y_{\rm e} \lesssim 0.15$, 
the observed feature can be reproduced by He at the mass fraction 
expected from the $\alpha$-decay of trans-Pb nuclei, 
and may therefore serve as an indirect signature for the production of elements 
beyond the third $r$-process peak in binary neutron star mergers.
\end{abstract}



%%%%%%%%%%%%%%%%%%%%%%%%%%%%%%%%%%%%%%%%
% Section: Introduction
%%%%%%%%%%%%%%%%%%%%%%%%%%%%%%%%%%%%%%%%

\section{Introduction} \label{sec:intro}
Binary neutron star (BNS) mergers provide 
an important astrophysical environment 
in which a wide range of physical processes 
can be simultaneously explored.
BNS mergers are widely recognized 
as central engines of short gamma-ray bursts  
\citep[GRBs, e.g.,][]{1989_Eichler+,1992_Narayan+}.
At the same time, thanks to the extremely dense, neutron-rich environments,
BNS mergers are compelling sites for 
rapid neutron-capture ($r$-process) nucleosynthesis
\citep[e.g.,][]{1982_Symbalisty&Schramm,1989_Eichler+}, potentially accounting for 
a significant fraction of the heavy elements in the Universe.

One of the most promising observational probes of BNS mergers is 
the radioactively powered emission from the merger remnants, 
known as ``kilonova'' \citep{1998_Li&Paczyski,2005_Kulkarni,2010_Metzger+}.
The radioactive decay of freshly synthesized $r$-process nuclei in BNS mergers
powers the electromagnetic emission,
resulting in the day-week timescale transient.

The discovery of the kilonova AT2017gfo \citep{2017_Abbott+_b} 
associated with the gravitational wave event GW170817 
\citep{2017_Abbott+_a} in 2017 
provided an unique opportunity 
to directly study the nucleosynthesis in the BNS merger.
In fact, recent studies have identified several atomic features 
in the optical and infrared spectra of AT2017gfo,
including absorption features attributed to strontium
\citep[Sr, $Z=38$,][]{2019_Watson+,2021_Domoto+,2022_Gillanders+}, 
yttrium \citep[Y, $Z=39$,][]{2023_Sneppen&Watson}, 
lanthanum \citep[La, $Z=57$,][]{2022_Domoto+}, cerium \citep[Ce, $Z=58$,][]{2022_Domoto+}, and gadolinium \citep[Gd, $Z=64$,][]{2025_Rahmouni+} 
in the early phase ($\lesssim \qty{5}{days}$),
as well as emission features 
attributed to tellurium \citep[Te, $Z=52$,][]{2023_Hotokezaka+} 
in the late phase ($\gtrsim \qty{5}{days}$).
These findings have enabled quantitative estimates of 
elemental abundances in the ejecta 
\citep[e.g.,][]{2022_Gillanders+,2024_Vieira+}, 
offering unique constraints on 
$r$-process nucleosynthesis in BNS mergers.
However, the estimated elemental abundances 
in GW170817 still carry large uncertainties.

A key limitation is the assumption of 
local thermodynamic equilibrium (LTE) in the spectral modeling.
Several previous studies have already demonstrated 
the importance of non-LTE modeling for the late-phase kilonova spectra, 
for which low density and low continuum radiation level lead to significant departures from LTE
\citep{2021_Hotokezaka+,2022_Pognan+_a,2022_Pognan+_b,2023_Pognan+,2025_Pognan+,2026_Jerkstrand+}.
In contrast, the modeling of early-phase kilonova spectra has largely 
relied on the LTE assumption 
\citep{2022_Domoto+,2022_Gillanders+,2023_Shingles+}.
However, significant departures from LTE are also expected for
the early-phase kilonova ejecta.
Since radioactive elements are distributed throughout the kilonova ejecta, 
non-thermal ionization by high-energy electrons in kilonovae is  
more important than in supernovae or stellar atmospheres.
Recently, \citet{2026_Brethauer+} demonstrated that
non-LTE treatment is important to reproduce early-phase 
color evolution of AT2017gfo. 
They also show the ionization states of heavy elements
can be deviated from LTE even in the early phase.
But the effects for the spectral feature are not fully understood.

An additional motivation for non-LTE modeling 
of early-phase kilonova spectra arises from
the potential presence of helium (He) in the kilonova ejecta.
It is well known that the He absorption lines observed in Type~Ib supernovae 
cannot be reproduced under the LTE assumption 
due to the high excitation energies \citep{1987_Harkness+}.
Instead, to reproduce He absorption features, non-thermal ionization by high-energy electrons produced 
by Compton scattering of $\gamma$-rays from \ce{^{56}Ni} decay
are required \citep[e.g.,][]{1991_Lucy,2012_Hachinger+}.

The spectra of AT2017gfo show a strong absorption feature around \qty{1}{\mu m} (hereafter, we call \qty{1}{\mu m} feature).
This feature has been attributed to \ionname{Sr}{II} \citep{2019_Watson+,2021_Domoto+,2022_Gillanders+,2023_Vieira+}.
But \ionname{He}{I} also shows a strong transition at 1.08 $\mu$m, which may reproduce the observed feature
\citep{2022_Perego+,2023_Tarumi+,2024_Sneppen+_c}.
Actually, a large amount of He can be synthesized 
under certain conditions in BNS mergers, and its abundance can be comparable to those of Sr \citep[e.g.,][]{2013_Fernandez&Metzger,2022_Perego+,2020_Fujibayashi+_b,2023_Fujibayashi+_a, 2025_Bernuzzi+}. 
Therefore, to study the origin of the observed \qty{1}{\mu m} feature, 
the contributions of both He and Sr should be considered simultaneously.

Interestingly, \citet{2026_Sneppen+_a} pointed out that 
neutrino-driven winds from a long-lived remnant NS can enhance 
the He abundance in the ejecta, leading to an overproduction of the He absorption feature.
This work suggests that the He abundance can also serve as an indicator of the nuclear equation of state through the lifetime of the remnant NS.

In this work, we develop self-consistent non-LTE ionization models including both He and Sr to obtain 
the constraint on the abundance of these elements in GW170817.
The works by \citet{2024_Sneppen+_c,2026_Sneppen+_a}  considered only He, and thus, the contribution of Sr was not investigated. The work by \citet{2023_Tarumi+} considered both He and Sr, but in a separated manner. Also, they tested only a representative mass fraction 
for each element, and did not explore possible ranges of 
their mass fractions.
In this work, we systematically explore the abundances required for both elements 
to reproduce the $\qty{1}{\mu m}$ feature in AT2017gfo, 
considering their ionization states self-consistently (see \citealt{2026_Arya+} 
for recent work with a similar approach).

This paper is organized as follows.
In \myrefsec{sec:obs_feature}, 
we outline the properties of the $\qty{1}{\mu m}$ feature
in the early-phase spectra of AT2017gfo.
In \myrefsec{sec:methods}, 
we introduce a non-LTE framework to evaluate the ionization degree
and level populations of He and Sr.
In \myrefsec{sec:results}, we demonstrate the impact of non-LTE effects 
for each element and present the constraints on 
their abundances in GW170817.  
In \myrefsec{sec:discussion}, 
we discuss the nucleosynthesis conditions and the mass-ejection mechanisms 
in GW170817 based on the derived constraints.
Finally, we give conclusions of this work in \myrefsec{sec:conclusions}.

%%%%%%%%%%%%%%%%%%%%%%%%%%%%%%%%%%%%%%%%
% Section
%%%%%%%%%%%%%%%%%%%%%%%%%%%%%%%%%%%%%%%%

\begin{figure*}
  \centering
  \begin{tabular}{cc}
  \includegraphics[width=0.5\linewidth]{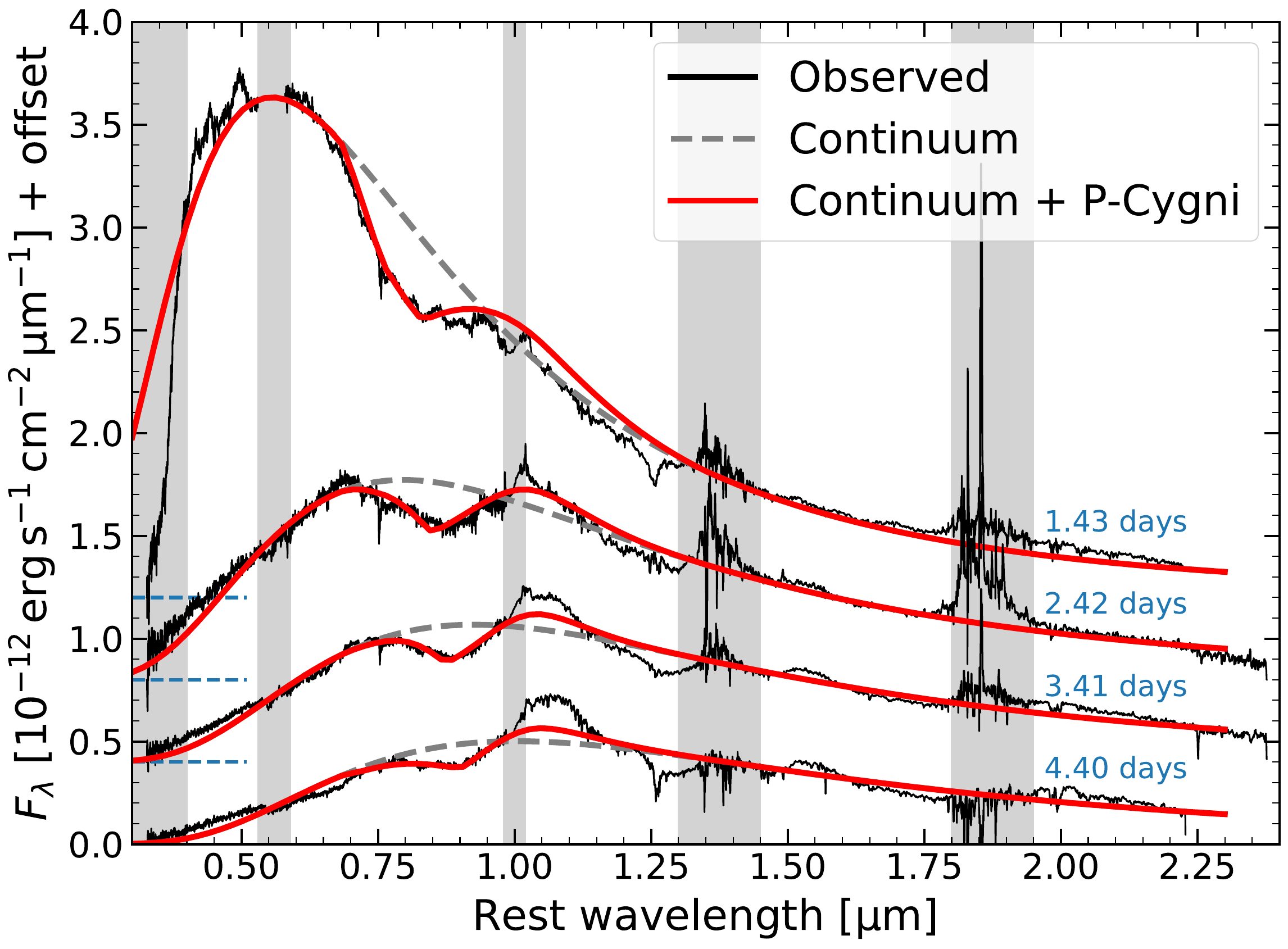} &
%  \medskip
  \includegraphics[width=0.4\linewidth]{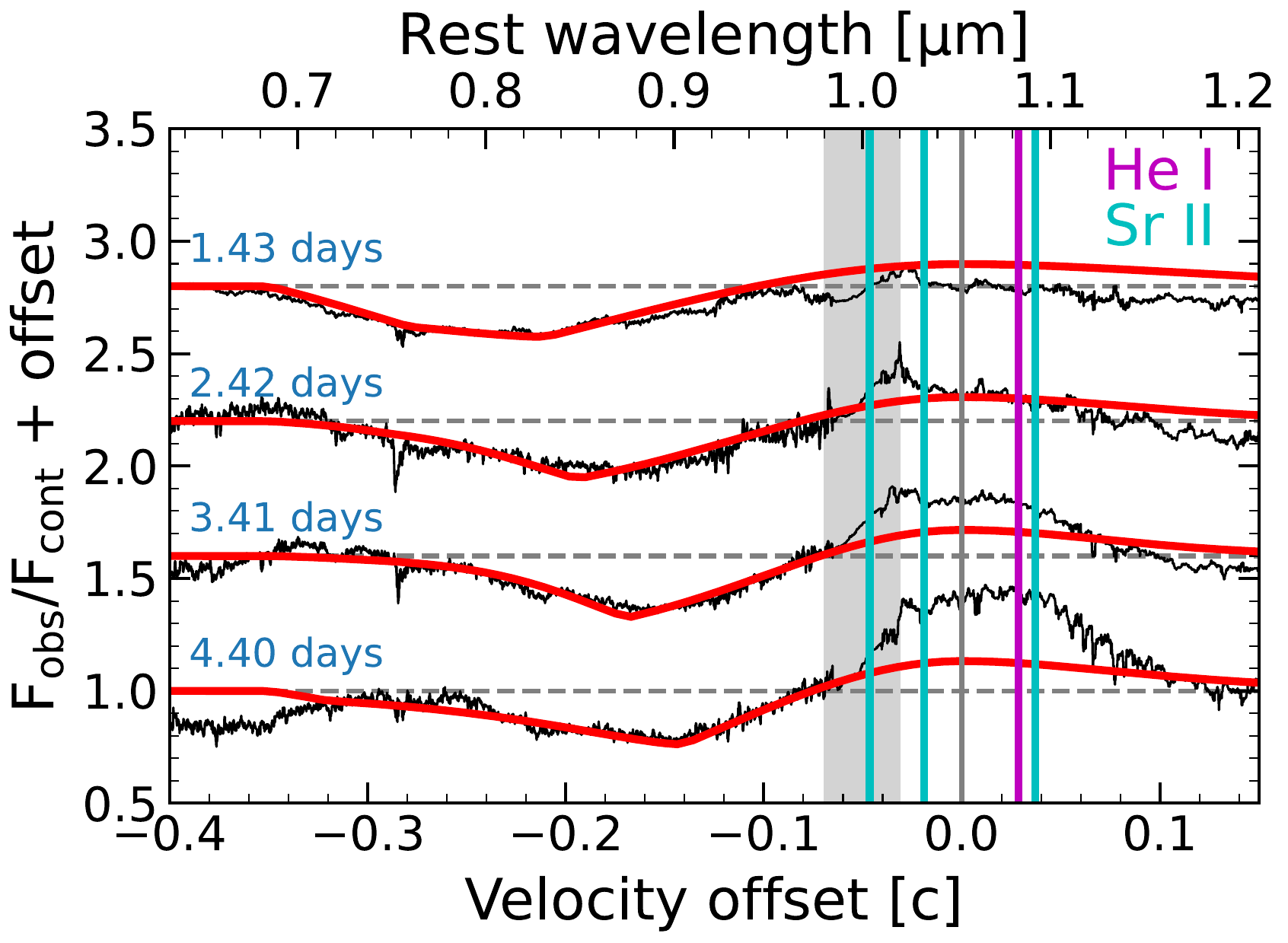}
    \end{tabular}
  \caption{The observed spectra of AT2017gfo
    over the full wavelength range (left) and 
    around the $\qty{1}{\mu m}$ feature (right).
    The red solid lines show the best-fit models 
    and the gray dashed lines show the continuum component.
    The wavelength ranges represented by gray shaded regions
    are not included in the fit due to the telluric contamination
    ($\sim \qty{1.4}{\mu m}$ and $\sim \qty{1.9}{\mu m}$),
    the large noise levels at the edge of the detectors 
    ($\sim \qty{0.6}{\mu m}$ and $\sim \qty{1}{\mu m}$), 
    and the line blanketing of many $r$-process elements ($\lesssim \qty{0.4}{\mu m}$).
    The horizontal blue dashed lines in the left panel represent
    the offset levels for the flux in the first three epochs.
    The spectra in the right panel are normalized by the model continuum
    and shown as a function of velocity offset relative to
    the arithmetic-mean line wavelength
    ($\lambda_\mathrm{line} \approx \qty{1.05}{\mu m}$).
    Magenta and cyan lines represent the rest wavelengths of 
    the \ionname{He}{I} and \ionname{Sr}{II} lines, respectively.
    The observed spectra are taken from 
    \citet{2017_Pian+} and \citet{2017_Smartt+}. \label{fig:obs_feature}}
\end{figure*}

\section{The 1 \lowercase{$\mu$m} Feature in AT2017\lowercase{gfo}}
\label{sec:obs_feature}

\begin{figure*}
\begin{center}
  \includegraphics[width=0.75\linewidth]{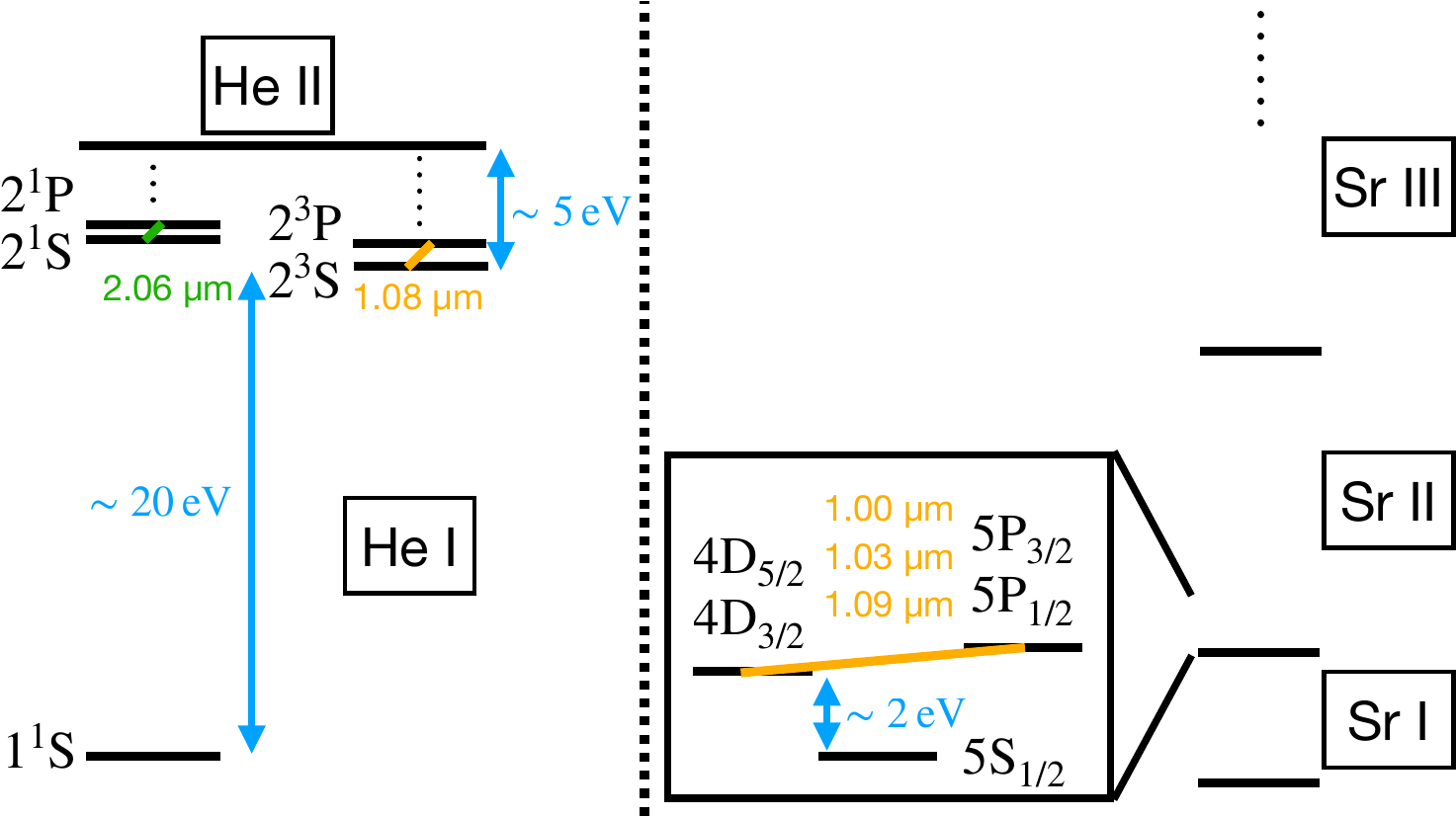}
  \caption{Schematic atomic structure of He (left) and Sr (right).
  Orange lines represent the transitions producing 
  the $\qty{1}{\mu m}$ feature in the spectra of AT2017gfo.
  A green line in the left shows the singlet $\qty{2}{\mu m}$ transition
  of \ionname{He}{I}.
  The blue arrows show approximated energy differences for the corresponding levels.
  \label{fig:atom_structure}}
\end{center}
\end{figure*}

We first investigate the required conditions to reproduce the \qty{1}{\mu m} feature
observed in the spectra of AT2017gfo.
In an expanding atmosphere such as the ejecta of BNS mergers, 
the absorption feature originating from the bound-bound transitions 
of atomic species can be evaluated 
using the Sobolev optical depth, $\tau_\mathrm{sob}$ \citep{1960_Sobolev}:
\begin{equation}
  \label{eq:def_tau_sob}
  \begin{aligned}
    \tau_\mathrm{sob} 
    &= \frac{\pi e^2}{m_e c} \lambda f_{lu} t n_l 
    \left(1-\frac{g_l}{g_u}\frac{n_u}{n_l}\right) \\
    &\approx 0.23\, \lambda_\mathrm{\mu m} f_{lu} t_\mathrm{d} n_l,
  \end{aligned}
\end{equation}
where $\lambda$ is the line wavelength 
($\lambda_\mathrm{\mu m}$ as in units of micro-meter), 
$f_{lu}$ is the oscillator strength of the transition
from lower to upper states (labeled as $l$ and $u$, respectively),
$t$ is the time after the merger
($t_\mathrm{d}$ as in units of day),
$n_l$ and $n_u$ are the number densities of the lower  
and upper levels corresponding to the transition,
and $g_l$ and $g_u$ are the statistical weights of each level.
The correction term due to stimulated emission
is expected to be negligible\footnote{
For the lines considered in this study
(see \myreffig{fig:atom_structure}), the transitions are 
governed by radiative bound-bound processes with diluted blackbody
radiation which we assume as a continuum emission. 
As a result, even in the non-LTE case, 
the population ratios, $n_u/n_l$, 
can be approximated as their LTE values,
$n_u/n_l \approx (g_u/g_l) e^{-\Delta E/kT}$,
where $\Delta E$ is the energy difference of the two levels
and $T$ is the radiation temperature.
Adopting $\Delta E \sim \qty{1}{eV}$ for the \qty{1}{\mu m} lines
and $T \sim \qty{4e3}{K}$, 
we obtain $(g_l/g_u)(n_u/n_l)\sim0.05$. 
Thus, the correction term for the stimulated emission is negligible.}.

under the conditions considered in this work.
Since $\lambda$ and $f_{lu}$ are known from
the atomic property and 
$t$ is known from the observation,
we can place constraints on $n_l$ by estimating $\tau_\mathrm{sob}$ from the observed spectra. 

To infer the physical conditions of the ejecta, 
such as the temperature, velocity, and $\tau_\mathrm{sob}$, 
we perform spectral fitting with a simple model composed of 
a relativistic blackbody and a P-Cygni profile.
We use the simple P-Cygni profile modeling \citep{1990_Jeffery&Branch} 
on the spherically symmetric atmosphere.
For the radial distribution of Sobolev optical depth, we assume
$\tau_\mathrm{sob}(v) = \tau_\mathrm{ph}e^{-(v-v_\mathrm{ph})/v_\mathrm{e}}$, where $v_\mathrm{ph}$ is the photospheric velocity,
$v_\mathrm{e}$ is the e-folding velocity of the optical depth,
and $\tau_\mathrm{ph}$ is the Sobolev optical depth at the photosphere.
The line wavelength is given by the arithmetic mean of 
the corresponding line wavelengths of \ionname{He}{I} and \ionname{Sr}{II}
($\lambda_\mathrm{line} \approx \qty{1.05}{\mu m}$).
Note that we employ a non-relativistic treatment of the P-Cygni profile. 
While relativistic effects could modify the line shape of the P-Cygni profile 
\citep{1990_Hutsemekers&Surdej,1993_Jeffery}, 
their impact is likely to be small for the kilonova ejecta \citep{2023_Sneppen+_b}. 

For continuum, we assume the existence of 
a sharply defined photosphere at $v_\mathrm{ph}$.
At the photosphere, blackbody radiation with a temperature $T_0$ in the ejecta-comoving frame is assumed.
The continuum spectrum is modeled by the following formula
\citep{2023_Sneppen,2025_Sadeh}:
\begin{equation}
  F_\lambda = 2\pi \left(\frac{r_\mathrm{ph}}{D}\right)^2 
  \int_{\beta_\mathrm{ph}}^{1} B_\lambda(T(\mu)) 
  \frac{1}{(1-\beta_\mathrm{ph}\mu)^2} \mu \mathrm{d}\mu,
\end{equation}
where $r_\mathrm{ph} = v_\mathrm{ph} t$ is the photospheric radius at each epoch,
$D=40.7\, \mathrm{Mpc}$ is the distance to the host galaxy
of GW170817 \citep{2018_Cantiello+},
$\beta_\mathrm{ph}=v_\mathrm{ph}/c$ is the photospheric velocity 
in units of the speed of light,
and $B_\lambda(T)$ is the Planck function. 
The temperature in the observer's frame is corrected 
due to the relativistic beaming effect \citep{1979_Rybicki&Lightman}:
\begin{equation}
  T(\mu) = \frac{T_0}{\gamma(1-\beta_\mathrm{ph}\mu)},
\end{equation}
where $\gamma$ is the Lorentz factor 
and $\mu$ is the direction cosine with respect to the line of sight.

\begin{deluxetable}{ccccc}[t]
\tablecaption{Best-fit parameters from the spectral fitting \label{tab:fit_params}}
\tablehead{
    \colhead{Epoch [days] \tablenotemark{a}} 
    & \colhead{$T_0$ [K] \tablenotemark{b}} 
    & \colhead{$v_\mathrm{ph}$ [$c$] \tablenotemark{c}} 
    & \colhead{$\tau_\mathrm{ph}$ \tablenotemark{d}}
    & \colhead{$v_\mathrm{e}$ [$c$] \tablenotemark{e}}
}
%\colnumbers
\startdata
1.43 & \num{4.5e3} & 0.21 & 0.58 & 0.12 \\
2.42 & \num{3.2e3} & 0.19 & 1.1 & 0.043 \\
3.41 & \num{2.8e3} & 0.17 & 1.3 & 0.036 \\
4.40 & \num{2.6e3} & 0.14 & 0.70 & 0.072 \\
\enddata
\tablecomments{
  \tablenotetext{a}{Time since the merger}
  \tablenotetext{b}{Photospheric temperature in the ejecta-comoving frame}
  \tablenotetext{c}{Photospheric velocity}
  \tablenotetext{d}{Reference optical depth for the P-Cygni profile}
  \tablenotetext{e}{e-folding velocity for the Sobolev optical depth}
}
\end{deluxetable}

We fit our model to the VLT/X-shooter spectra of AT2017gfo 
at the first four epochs
\citep{2017_Pian+,2017_Smartt+}.
Comparison of our models with the observed spectra of AT2017gfo 
is shown in \myreffig{fig:obs_feature}.
The result of our spectral fitting is summarized in \myreftab{tab:fit_params}.
As also mentioned in \citet{2023_Tarumi+}, $\tau_{\rm ph} \approx 1$ is required 
at all epochs between \qty{1.43}{days} to \qty{4.40}{days} to reproduce the \qty{1}{\mu m} feature.
To include systematic uncertainties of 
our simple spectral model\footnote{
In fact, the observed emission component of the \qty{1}{\mu m} feature 
at 3.41 and 4.40 days is stronger than our best-fit model. 
This reflects the difficulty in estimating the continuum level 
under the presence of the very broad feature. 
Also, there might be an additional contribution from emission lines 
(e.g., [Te IV] line as suggested by \citealt{2026_Mulholland+}).},
we adopt $0.5 \le \tau_\mathrm{ph} \le 2$ 
as a required condition to reproduce 
the \qty{1}{\mu m} feature in AT2017gfo.

Hereafter, we mainly focus on the plasma conditions just above the photosphere 
and refer to $\tau_\mathrm{ph}$ as $\tau$.
In the spectral fitting, we assume that 
the depth of the absorption feature is mainly determined 
by the plasma conditions just above the photosphere at each epoch.
Under this assumption, 
the elemental abundances inferred from the spectra at each epoch
represent the local quantities at the photospheric layer, 
rather than the averaged abundances outside of the photosphere. 
Therefore, our constraints are not significantly affected by 
ejecta stratification.

However, the exact line profile of the P-Cygni profile 
can be affected by stratified ejecta 
(i.e., the non-uniform distribution of the elemental abundances).
Such effect should be investigated by more detailed radiative transfer simulations, 
which is left for future work.

%%%%%%%%%%%%%%%%%%%%%%%%%%%%%%%%%%%%%%%%
% Section 
%%%%%%%%%%%%%%%%%%%%%%%%%%%%%%%%%%%%%%%%

\section{Methods} \label{sec:methods}

\subsection{Ionization by High-Energy Electrons}
\label{subsec:ionization}
High-energy electrons are created following radioactive decays of heavy elements 
in BNS merger ejecta. 
Such high-energy electrons have energies of \unitrange{\keV}{\MeV},
while a typical radiation temperature of kilonova is 
\qtyrange{0.1}{1}{\eV} (\qtyrange{e3}{e4}{\K}).
In these conditions, deviations from LTE in the ionization state 
are induced by non-thermal ionization due to these high-energy electrons.

In the first few days after the merger, 
the ejecta is weakly ionized with a sufficiently high density.
Non-thermal electrons interact with surrounding ions and thermal electrons 
with very short mean free paths.
As a result, the energy of non-thermal electrons released by 
radioactive decay is instantaneously and locally deposited 
through the excitation and ionization of ions, 
as well as the heating of thermal electrons 
\citep[e.g.,][]{2016_Barnes+,2021_Hotokezaka+}.
It is known that typically a few percent of the total energy 
of non-thermal electrons is spent for ionization
\citep[e.g.,][]{2022_Pognan+_b}.
This fraction can be quantitatively evaluated 
by solving the energy degradation of electrons
\citep{1954_Spencer&Fano,1992_Kozma&Fransson,2026_VandenBerg&Hotokezaka}.
The effective energy required 
to ionize a certain ion $i$ by non-thermal electrons is 
often expressed by an effective ionization potential $w_i$.
Then, ionization rate by non-thermal electrons is represented as follows 
\citep{2022_Pognan+_a,2023_Tarumi+}:
\begin{equation}
  \label{eq:def-nonth_ion_rate}
  \Gamma_i = \frac{\dot{q}}{w_i},
\end{equation}
where $\dot{q}$ is the radioactive heating rate per ion.
Here $\dot{q}$ can be evaluated as follows:
\begin{equation}
  \label{eq:def-heat_rate}
  \dot{q} = \mu_\mathrm{ion} m_\mathrm{u} (1-X_\mathrm{He}) f_\mathrm{th} \dot{Q},
\end{equation}
where $\mu_\mathrm{ion}$ is a mean atomic mass per ion in the ejecta,
$m_\mathrm{u}$ is the atomic mass unit,
$X_\mathrm{He}$ is a mass fraction of He,
$\dot{Q} \approx 10^{10}\,t_\mathrm{d}^{-1.3}\,\unit{erg.s^{-1}.g^{-1}}$
is the typical radioactive power through the $\beta$-decay via $\gamma$-rays and electrons, excluding the energy carried by neutrinos
\citep{2010_Metzger+,2014_Wanajo+,2015_Lippuner&Roberts,2020_Hotokezaka&Nakar}
and $f_\mathrm{th}$ is thermalization efficiency
for $\gamma$-rays and electrons.
Previous studies \citep{2023_Tarumi+,2024_Sneppen+_c,2026_Sneppen+_a} adopted
$\dot{q} \approx 1\,t_\mathrm{d}^{-1.3}\,\unit{eV.s^{-1}.ion^{-1}}$,
which corresponds to the heating rate 
evaluated with
$\mu_\mathrm{ion} = 100$, $X_\mathrm{He} \ll 1$, and $f_\mathrm{th} = 1$,
appropriate for a typical condition of early-phase kilonova ejecta.
In this study, we adopt $f_\mathrm{th} = 1$ 
and assume that 
the energies of both primary $\beta$-decay
electrons and secondary electrons generated 
via Compton scattering of $\gamma$-rays are
instantaneously and locally deposited
(see \citealt{2024_Wollaeger+, 2026_Andalman+} for the detailed studies of
the transport of $\beta$-decay electrons in the kilonova ejecta).
We also generalize $\mu_\mathrm{ion}$ to extend its applicability 
to a wide range of composition parameters:
\begin{equation}
  \mu_\mathrm{ion} = 
  \left(\frac{X_\mathrm{He}}{A_\mathrm{He}} + \frac{X_\mathrm{Sr}}{A_\mathrm{Sr}} 
  + \frac{1-X_\mathrm{He}-X_\mathrm{Sr}}{A_\mathrm{env}}\right)^{-1},
\end{equation}
where $X_\mathrm{Sr}$ is mass fraction of Sr, 
$A_\mathrm{He} = 4$ and $A_\mathrm{Sr} = 88$ are 
the atomic mass numbers of He and Sr,
and $A_\mathrm{env}$ is a weighted-average mass number of the other environmental species (we adopt $A_\mathrm{env} = 100$ as a representative of $r$-process elements).

When solving the rate equation, 
the number density of free electrons is evaluated as follows:
\begin{equation}
  \begin{aligned}
    n_\mathrm{e} 
    &= \frac{X_\mathrm{He}\rho_\mathrm{ej}}{A_{\rm He}m_\mathrm{u}}f_\mathrm{e}^\mathrm{\, (He)}
    + \frac{X_\mathrm{Sr}\rho_\mathrm{ej}}{A_{\rm Sr} m_\mathrm{u}}f_\mathrm{e}^\mathrm{\, (Sr)} \\
    &\quad + \frac{\left(1-X_\mathrm{He}-X_\mathrm{Sr}\right)\rho_\mathrm{ej}}{A_\mathrm{env}\,
    m_\mathrm{u}}f_\mathrm{e}^\mathrm{\, (env)}.
  \end{aligned}
    \label{eq:ne}
\end{equation}
Here, $f_{\rm e}^{(X)}$ is 
the mean number of free electrons contributed 
per atom of the element $X$.
We assume that the ionization degree of the environmental species 
is the same as that of Sr (i.e., $f_{\rm e}^{\rm (env)} = f_{\rm e}^{\rm (Sr)}$), 
which is a sound approximation to estimate the number of electrons.

\subsection{Non-LTE Calculations for He}
As an atomic model for He, 
we consider a system consisting of 21 states in total: 
19 bound states of \ionname{He}{I} up to 
the principal quantum number $n=4$ (neglecting fine structure), 
and the ground states of \ionname{He}{II} and \ionname{He}{III}.
For all transitions between these states, 
we include excitation/de-excitation, ionization, and recombination 
through both radiative and collisional (electron-impact) processes.
In addition to these thermal processes\footnote{
In this study, radiative processes are induced by 
diluted blackbody radiation.},
we also consider non-thermal ionization induced by high-energy electrons.
As illustrated in \myreffig{fig:atom_structure}, 
the $\qty{1}{\mu m}$ feature arises from the transition 
between highly excited states 
(\atomterm{2}{3}{S}~--~\atomterm{2}{3}{P}, \qty{1.08}{\mu m}), 
which cannot be populated by thermal processes 
in a typical temperature in the kilonova ejecta.
Thus, non-LTE treatment is crucial to consider the contribution of He to the kilonova spectra.

Assuming that atomic processes occur on timescales 
much shorter than the dynamical timescale of the ejecta, 
the level populations of He can be evaluated 
with a steady state approximation at each time.
Given the total number density of He, $n_\mathrm{He}$,
the number densities of the 21 individual states, $\{n_j\}$, 
can be obtained by solving the rate equations 
that describe the balance between the transitions 
among all the included states:
\begin{equation}
  \label{eq:req_formal}
  \sum_{j=1}^{21}\left(\Lambda_{kj}n_j - \Lambda_{jk}n_k\right) = 0 
  \quad (k = 1, 2, \dots, 21),
\end{equation}
where $n_j$ is the number density of the state $j$ and 
$\Lambda_{kj}$ is the transition rate from the state $j$ to the state $k$.
We can decompose this transition rate matrix, $\Lambda$,
to the each atomic process:
\begin{equation}
  \Lambda =  R_\mathrm{bb} + C_\mathrm{bb} 
  + R_\mathrm{bf/fb} + C_\mathrm{bf/fb} + \Gamma_\mathrm{nt},
\end{equation} 
where $R$ and $C$ correspond to radiative and collisional processes,
the subscript ``bb'' and ``bf/fb'' represent bound-bound transition and
bound-free/free-bound transition, and 
$\Gamma_\mathrm{nt}$ is the contribution of non-thermal ionization. 
More details of the included atomic processes are discussed in Appendix \ref{app:sec:atomic_process}.

Atomic data required for the He calculation are obtained from various sources: 
radiative ionization and recombination rates from \citet{2010_Nahar} through the NORAD database \citep{2020_Nahar,2024_Nahar&Hinojosa-Aguirre};
bound-bound radiative transitions and energy levels 
from the NIST Atomic Spectra Database \citep[ASD,][]{NIST_ASD}; 
and electron-impact excitation and ionization data from \citet{2008_Ralchenko+}.
We use the values of $w_i$ for He (\myrefeq{eq:def-nonth_ion_rate})
taken from \citet{2023_Tarumi+}:
$w_\mathrm{\ionname{He}{I}} = \qty{593}{eV}$ and $w_\mathrm{\ionname{He}{II}} = \qty{3076}{eV}$.

\subsection{Non-LTE Calculations for Sr}
Heavy elements such as Sr have many electrons and their atomic structures 
are significantly more complex compared to those of lighter elements such as He.
Due to this complexity, 
accurate and complete atomic data are generally unavailable, 
both theoretically and experimentally.
As a result, it is not feasible to explicitly account for all the transitions 
among individual atomic states as done for He
(but see \citealt{2024_Mulholland+_a, 2025_Dougan+} for recent progress).
In this study, we adopt a simplified non-LTE ionization model
which include the ionization state of Sr up to Sr\,XX
(only the state up to \ionname{Sr}{VII} are found to be 
important in this study).

The balance between each ionization state can be described as
\begin{equation}
  n^{(i+1)}n_\mathrm{e}\alpha_{i+1} = n^{(i)} \left(R_{\mathrm{bf}, i}+\Gamma_i\right),
\end{equation}
where $n^{(i)}$ is the number density of ion $i$\footnote{
Here we denote the number density of ion $i$ as $n^{(i)}$
rather than $n_i$ (as in the case for He) 
as we consider only ionization states (not excited states) for Sr.},
$\alpha_{i+1}$ is the recombination coefficient 
from the ion $i+1$ to the ion $i$,
and $R_{\mathrm{bf}, i}$ is the thermal photoionization rate 
from the ion $i$ to the ion $i+1$.
For the early-phase kilonova ejecta, the thermal ionization
is governed by photoionization induced by diluted blackbody radiation
from the photosphere %
($R_{\mathrm{bf/fb}, i} \gg C_{\mathrm{bf/fb}, i}$).
From the detailed balance between bound-free and free-bound transitions,
we can evaluate the ratio of the number density in each ionization state as follows:
\begin{equation}
  \frac{n^{(i+1)}}{n^{(i)}} = W\left(\frac{n^{(i+1)}}{n^{(i)}}\right)^*
  +\frac{1}{n_\mathrm{e}\alpha_{i+1}}\frac{\dot{q}}{w_i},
 \label{eq:ionization}  
\end{equation}
where 
$\left(\frac{n^{(i+1)}}{n^{(i)}}\right)^*$ is the Saha population ratio.
Here $W$ is a geometric dilution factor \citep{1978_Mihalas}:
\begin{equation}
  W (v) = 0.5 
  \left(1-\sqrt{1-\left(\frac{v_\mathrm{ph}}{v}\right)^2}\right).
\end{equation}
As we mainly consider the plasma just above the photosphere,
we adopt $W=0.5$ in this work.

For the level populations of Sr II, 
we assume the following relations
\citep[e.g.,][]{1985_Abbott&Lucy, 1999_Lucy_b}:
\begin{equation}
    \label{eq:SrII_excited_pop}
    \begin{aligned}
        \frac{n_\mathrm{4D}}{n_\mathrm{\ionname{Sr}{II}}} 
        &= \frac{g_\mathrm{4D}}{U_\mathrm{\ionname{Sr}{II}}} \exp\left(-\frac{E_\mathrm{4D}}{k_\mathrm{B}T}\right) \\
        \frac{n_\mathrm{5P}}{n_\mathrm{\ionname{Sr}{II}}} 
        &= W\frac{g_\mathrm{5P}}{U_\mathrm{\ionname{Sr}{II}}} \exp\left(-\frac{E_\mathrm{5P}}{k_\mathrm{B}T}\right),
    \end{aligned}
\end{equation}
where $n_\mathrm{4D/5P}$ is the number density of each state, $g_\mathrm{4D/5P}$ 
is the statistical weight,
$E_\mathrm{4D/5P}$ is the excitation energy,
and $U_\mathrm{\ionname{Sr}{II}}$ is 
the partition function of \ionname{Sr}{II}.
We consider only five states in total 
($\mathrm{5S}_{1/2}$, $\mathrm{4D}_{3/2}$, $\mathrm{4D}_{5/2}$,
$\mathrm{5P}_{1/2}$, $\mathrm{5P}_{3/2}$). 
In a temperature range adopted in this work
(see \myreftab{tab:fit_params}),
the partition function is mainly determined only 
by the ground state, i.e., 
$U_\mathrm{\ionname{Sr}{II}} \simeq g_\mathrm{5S}$.

The assumption above is a sound approximation 
because \ionname{Sr}{II} 4D and 5P states lie within a few eV 
from the ground state (see \myreffig{fig:atom_structure}).
In a typical plasma conditions in the early phase of kilonova, the 4D states 
(responsible for the \qty{1}{\mu m} feature) are mainly populated 
through the radiative transition from the upper 5P states 
because the radiative transitions from the ground 5S state to the 4D states are forbidden.
In other words, the 5P states act as a bypass \citep{2023_Tarumi+}:
the 5P states are well connected with both the ground 5S state and 4D states 
through the radiative transitions\footnote{
A typical radiative transition rate is about $10^{4-5}\, {\rm s^{-1}}$ 
in optically thin case. The Sobolev optical depth for the transition 
between the 5S and 5P states is at most $\tau_{\rm Sob} \sim 10^3$
with the reasonable Sr abundance, 
which corresponds to the escape probability of $\beta_{\rm esc} \sim 10^{-3}$.
Thus, even in the optically thick case, 
the effective radiative transition rate is about $10^{1-2}\, {\rm s^{-1}}$.
Therefore, the radiative transitions are  faster than collisional excitation 
by non-thermal electrons 
(at most \qty{1}{s^{-1}} even when all the radioactive energy goes to the excitation) 
and by thermal electrons ($\sim \qty{0.1}{s^{-1}}$, 
see \citealt{2024_Mulholland+_a,2026_Arya+}).}.
Thus, both $n_\mathrm{5P}/n_\mathrm{5S}$ and $n_\mathrm{5P}/n_\mathrm{4D}$ follow 
the Boltzmann ratio weighted by the dilution factor $W$.
As a result, $n_\mathrm{4D}/n_\mathrm{5S}$ and 
thus $n_\mathrm{4D}/n_\mathrm{\ionname{Sr}{II}}$ simply follow 
the Boltzmann ratio as in \myrefeq{eq:SrII_excited_pop} 
because the dilution factor $W$ is canceled out.

The atomic data required for Sr calculations, 
i.e., energy levels and ionization potentials, 
are taken from the NIST ASD \citep{NIST_ASD}.
Radiative bound-bound transition rates of \ionname{Sr}{II} are also taken from 
the NIST ASD to evaluate the depth of the absorption feature. 
Recombination rate coefficients, $\{\alpha_{i+1}\}$, 
are evaluated from the value for hydrogen 
with appropriate corrections
(\citealt{1962_Bates+,2021_Nahar}, and see also \citealt{2023_Tarumi+}).
The values of $w_i$ up to Sr\,V
are taken from \citet[Appendix A]{2023_Tarumi+}:
$w_\mathrm{\ionname{Sr}{I}} = \qty{124}{eV}$, $w_\mathrm{\ionname{Sr}{II}} = \qty{272}{eV}$,
$w_\mathrm{\ionname{Sr}{III}} = \qty{444}{eV}$, $w_\mathrm{\ionname{Sr}{IV}} = \qty{608}{eV}$
and $w_\mathrm{Sr\,V} = \qty{822}{eV}$, respectively.
For ionization higher state, we assume $w_i = 30 I_i$ 
($I_i$ is the ionization potential of ion $i$),
which is a sound approximation in the kilonova ejecta
\citep{2021_Hotokezaka+,2026_Brethauer+}.

%%%%%%%%%%%%%%%%%%%%%%%%%%%%%%%%%%%%%%%%
% Section: 
%%%%%%%%%%%%%%%%%%%%%%%%%%%%%%%%%%%%%%%%

\section{Results} \label{sec:results}

\subsection{Level Populations of He}
\label{subsec:pop_frac_He}

\begin{figure}
  \includegraphics[width=\linewidth]{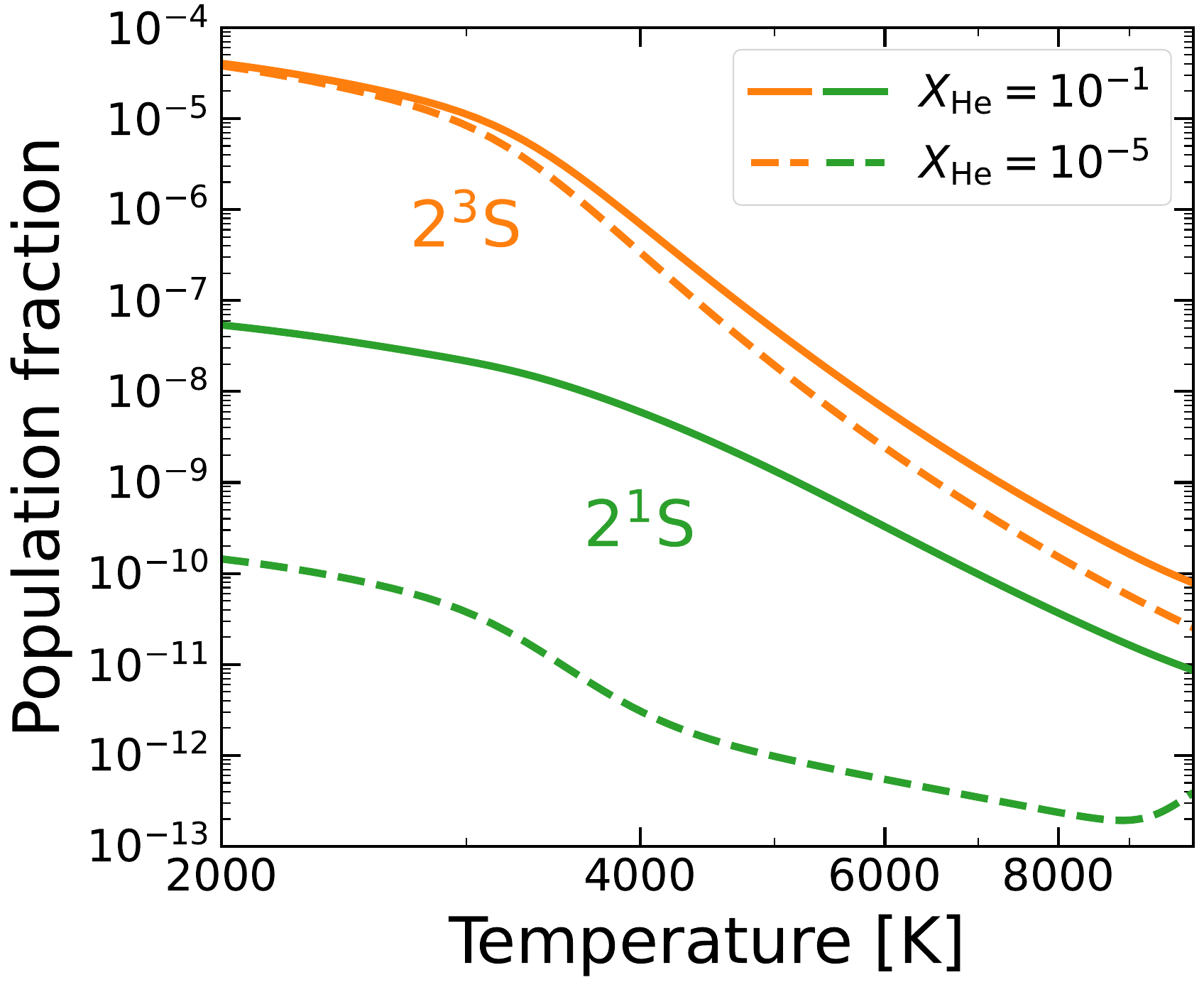}
  \caption{Population fractions of the \ionandterm{He}{I}{2S} states 
  as a function of temperature.
  The mass density is fixed to be $\rho = \qty{1e-14}{g.cm^{-3}}$. 
  The orange and green lines show the population fractions of 2$^3$S and 2$^1$S states, respectively. 
  The solid and dashed lines show the results with the He mass fractions of
  $X_{\rm He}=10^{-1}$ and $10^{-5}$, respectively.
  \label{fig:pop_frac_HeI}}
\end{figure}

\begin{figure}
  \centering
  \includegraphics[width=\linewidth]{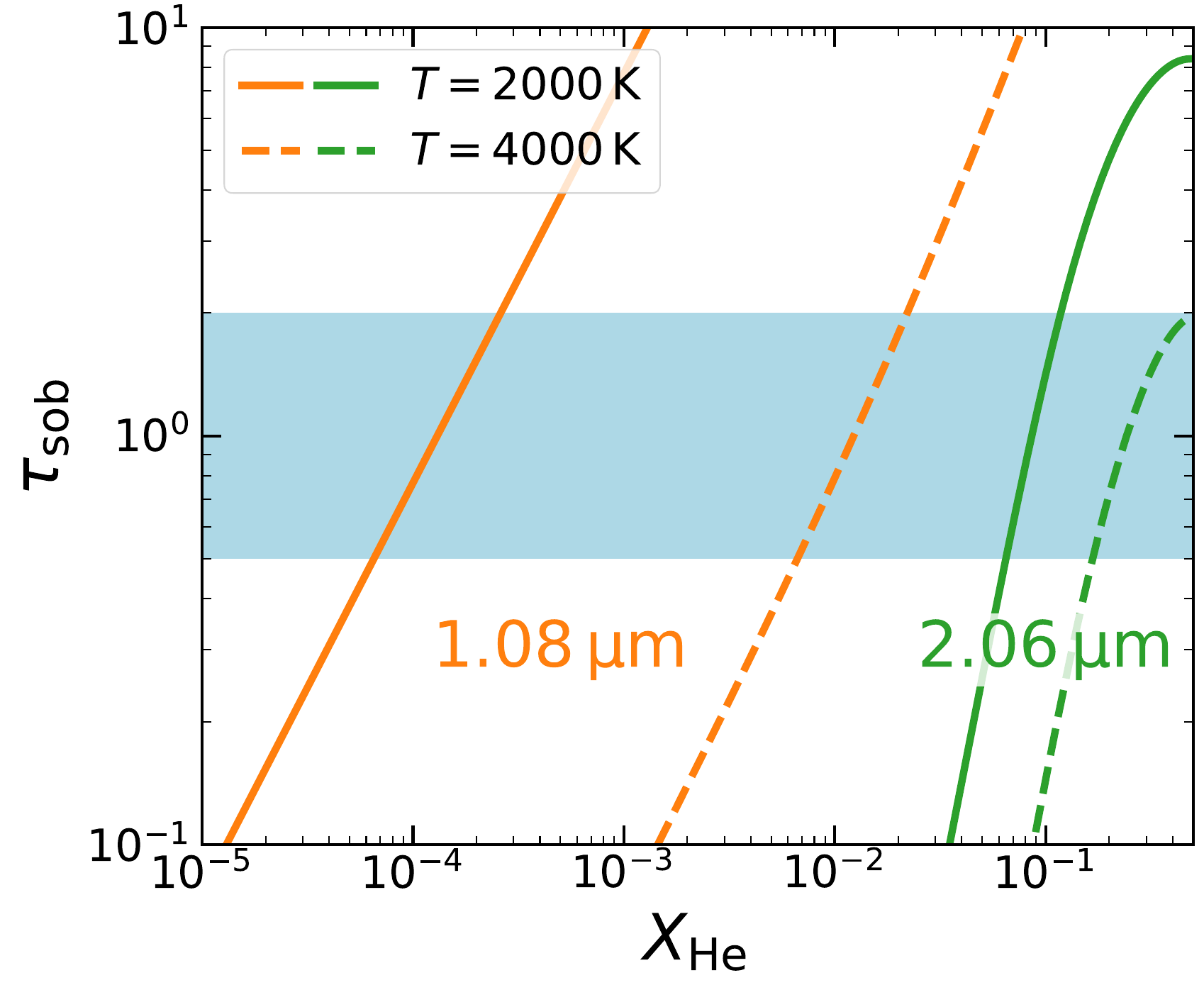}
  \caption{Sobolev optical depths of the \ionname{He}{I} lines
           as a function of mass fraction. 
           The mass density is fixed to be $\rho = \qty{1e-14}{g.cm^{-3}}$ 
           and the Sobolev optical depth is evaluated for $t = 1$ day.
           The orange and green lines show the optical depths for 1.08 $\mu$m and 2.06 $\mu$m lines, respectively. The solid and dashed lines show the cases for $T=2000$ and \qty{4000}{K}, respectively.
           The blue shaded region represents 
           the range of $0.5 \le \tau_\mathrm{sob} \le 2.0$.
           \label{fig:tau_HeI}}
\end{figure}

Here we first study the level populations of He.
To understand the behaviors of the He lines as a function of temperature 
and mass fraction, we only consider He and environmental elements 
in this section.
The mass density of the plasma is fixed to be 
$\rho = \qty{1e-14}{g.cm^{-3}}$.
The number density of the free electrons is evaluated by assuming $X_{\rm Sr} = 0$ and $f_{\rm e}^{\rm env} = 3$ in Equation \ref{eq:ne},
and the heating rate is evaluated according to Equation \ref{eq:def-heat_rate}.

\myreffig{fig:pop_frac_HeI} shows 
the calculated population fractions\footnote{
Hereafter, ``population fraction'' is defined as 
the level population normalized by 
the total number density of the corresponding element.}
of \ionandterm{He}{I}{2S} states as a function of temperature.
The orange and green lines show
the population fraction of the spin-triplet 
(\ionandterm{He}{I}{\atomterm{2}{3}{S}}) and
the spin-singlet 
(\ionandterm{He}{I}{\atomterm{2}{1}{S}}) states, respectively.
Overall, the population of the triplet state is much higher than 
that of the singlet state 
as the triplet state is weakly connected to the ground state
by the selection rule (i.e., forbidden line).
Population fractions of both singlet and triplet states 
decrease with temperature:
this is because photoexcitation and photoionization are
dominant channels to depopulate \ionandterm{He}{I}{2S} states
and depopulation flow increases for a higher temperature.
As discussed in \citet{2024_Sneppen+_c},
the populations of the 2S states decrease 
not only by the direct photoionization 
but also by photoexcitation to the 2P states.

\begin{figure*}
  \centering
  \epsscale{1.1}
  \plottwo{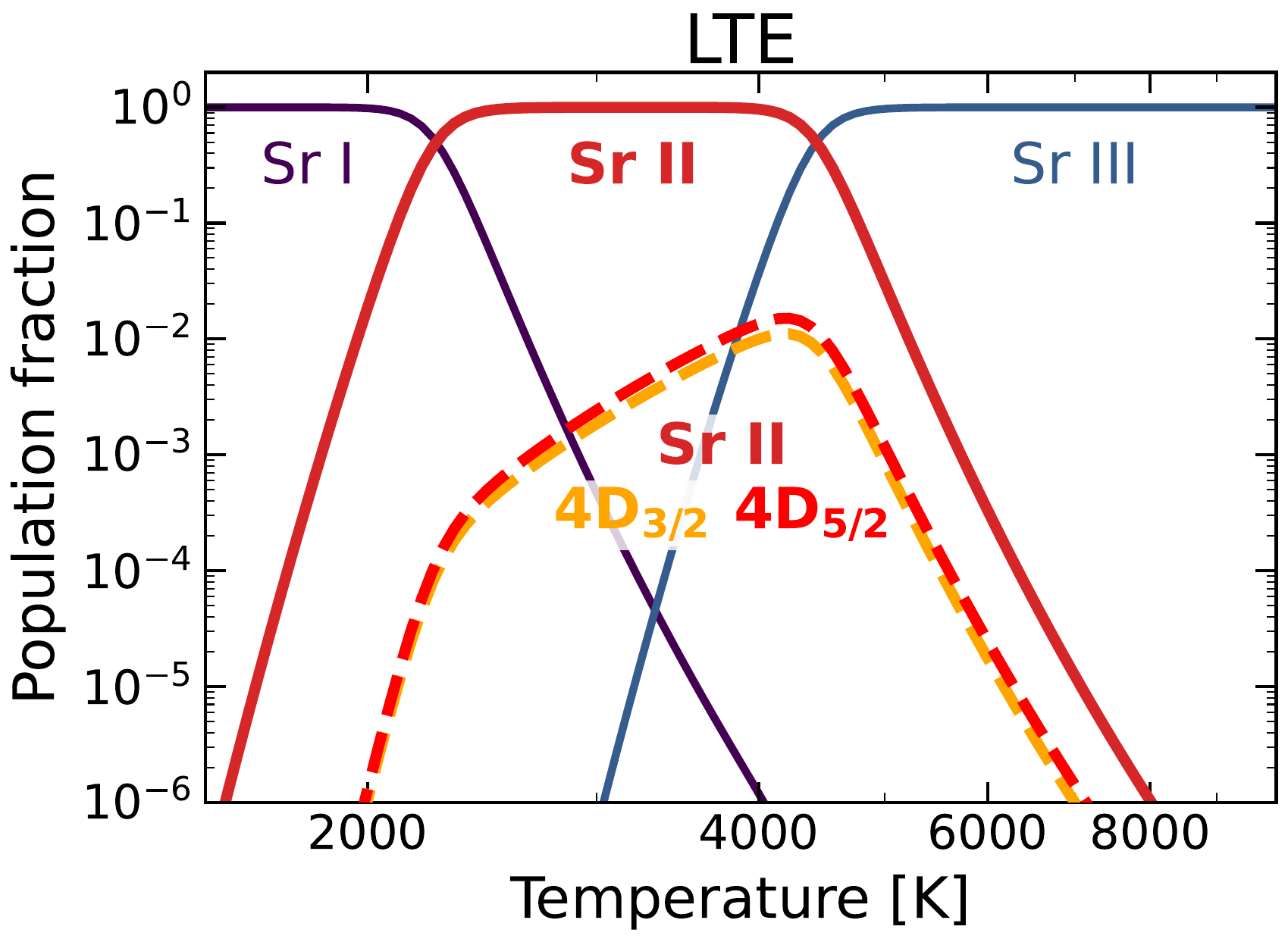}{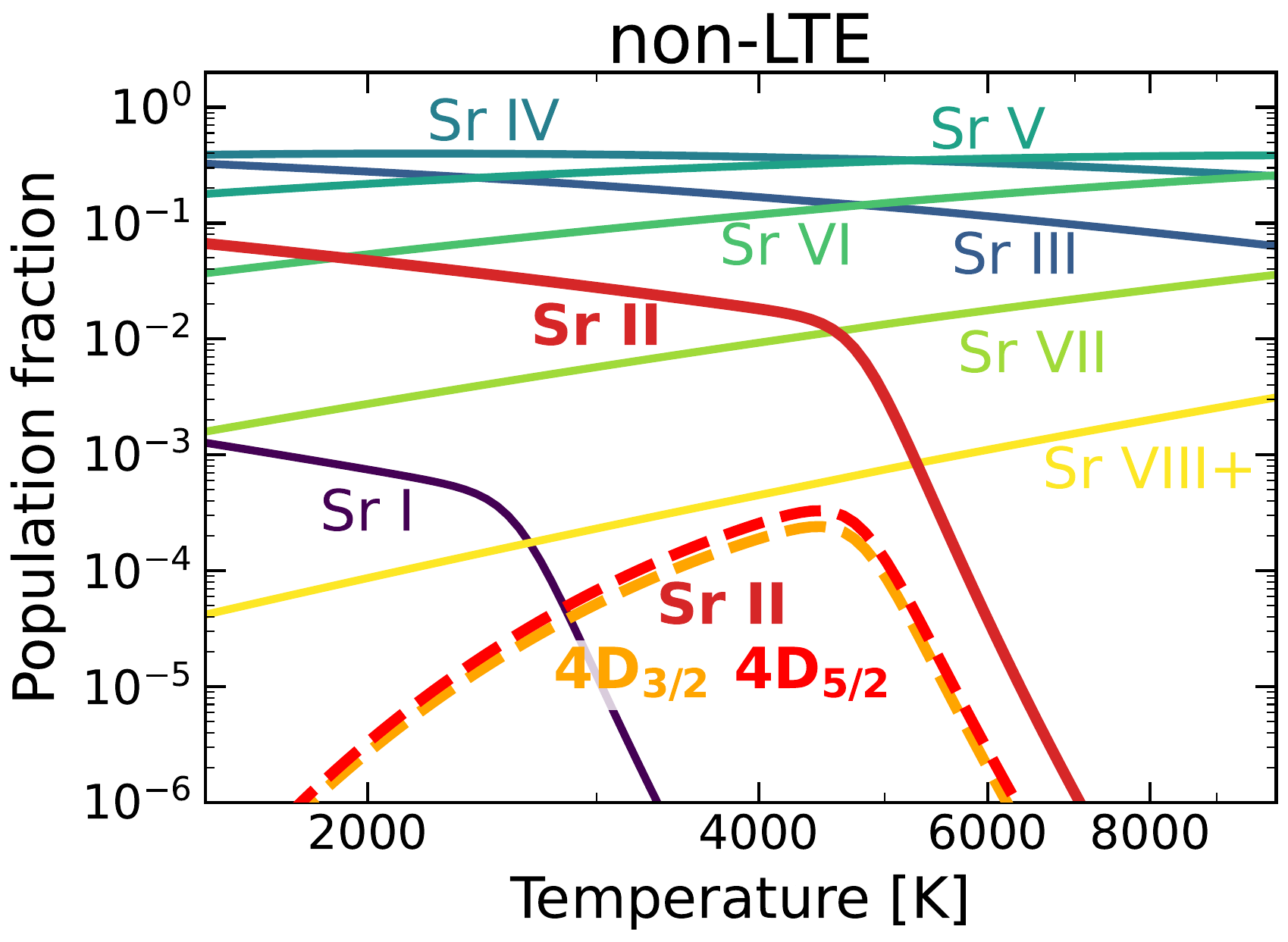}
  \caption{
  Population fractions of each ionization state of Sr
  and \ionandterm{Sr}{II}{4D} states as a function of temperature
  for the case with $n_\mathrm{e} = \qty{1e8}{cm^{-3}}$.
  The left panel shows a LTE case while the right panel shows a non-LTE case 
  with a radioactive heating rate of 
  $\dot{q} = 1.0\,\mathrm{eV\, s^{-1}\, \mathrm{ion}^{-1}}$. 
  The solid lines show the ionization fraction 
  for each ionization state given in the panels
  (Sr~VIII+ represents the total fraction of 
  all the ionization states higher than VII) and 
  the dashed line shows the population fraction of 
  the \ionname{Sr}{II} 4D states among all Sr.
  \label{fig:pop_frac_Sr}}
\end{figure*}

\myreffig{fig:tau_HeI} shows the Sobolev optical depth of NIR lines
corresponding to \ionandterm{He}{I}{2S} states as a function of 
the mass fraction of He.
Orange and green lines represent the Sobolev optical depths of \qty{1.08}{\mu m} 
and \qty{2.06}{\mu m} lines, respectively.
As shown in \myreffig{fig:pop_frac_HeI}, 
the mass fraction of He ($X_\mathrm{He}$) 
giving $\tau \sim 1$ is higher at a higher temperature,
as expected from the population fractions.
Also, there are the large differences between the Sobolev optical depths of 
the \qty{1.08}{\mu m} and \qty{2.06}{\mu m} lines:
when $\tau \sim 1$ for the \qty{2.06}{\mu m} line at $T=\qty{2000}{K}$,
the optical depth for the \qty{1.08}{\mu m} line reach 
to $\tau \sim$ \numrange{e2}{e3}.
Therefore, if the \qty{2.06}{\mu m} absorption feature were detected,
the \qty{1.08}{\mu m} feature would be much stronger.

\subsection{Ionization Degree of Sr}
\label{subsec:pop_frac_Sr}

In this section,
we demonstrate the impacts of non-thermal ionization 
on the ionization degree of Sr.
Here, for simplicity, we treat the electron density as an independent parameter
(i.e., we do not use \myrefeq{eq:ne}).
Moreover, in this non-LTE calculation,
the radioactive heating rate per ion is 
fixed to be $\dot{q} = 1.0\,\mathrm{eV\, s^{-1}\, \mathrm{ion}^{-1}}$,
assuming $t=\qty{1}{d}$ and $\mu_\mathrm{ion} \approx 100$
(see Section \ref{subsec:ionization}).
In this setup, the results are independent on the mass fraction of Sr.

\myreffig{fig:pop_frac_Sr} shows
the population fractions of each ionization state of Sr (solid lines)
and \ionname{Sr}{II} 4D states (among all Sr, dashed lines) 
as a function of temperature 
for the case with $n_\mathrm{e} = \qty{1e8}{cm^{-3}}$.
When LTE is assumed (left panel), the ionization fraction is 
strongly dependent on the temperature
by the exponential dependence on temperature in Saha's equation.
On the other hand, in the non-LTE case (right panel), 
non-thermal ionization weakens this dependency
as the second term in \myrefeq{eq:ionization} 
is nearly temperature independent\footnote{
The recombination rate coefficient is weakly 
dependent on temperature.}.
As a result, ions in different ionization state coexist,
and highly-ionized ions can appear even at a low temperature
\citep{2023_Tarumi+,2026_Brethauer+}.
The population of the Sr II 4D state follows the ionization fraction of Sr II, but its population is suppressed at a lower temperature due to the small Boltzmann factor (dashed lines).

\begin{figure*}
  \centering
  \epsscale{1.1}
  \plottwo{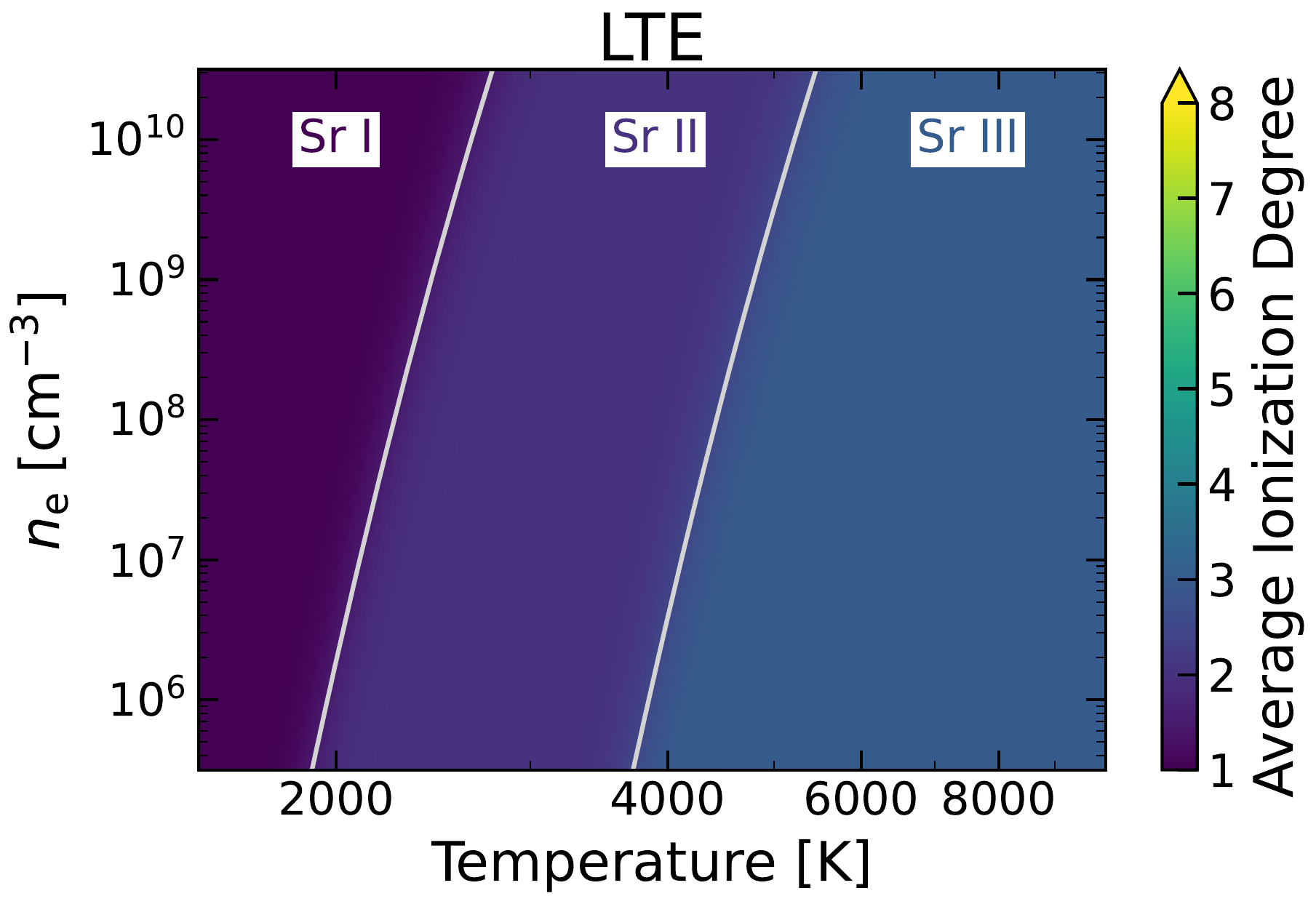}{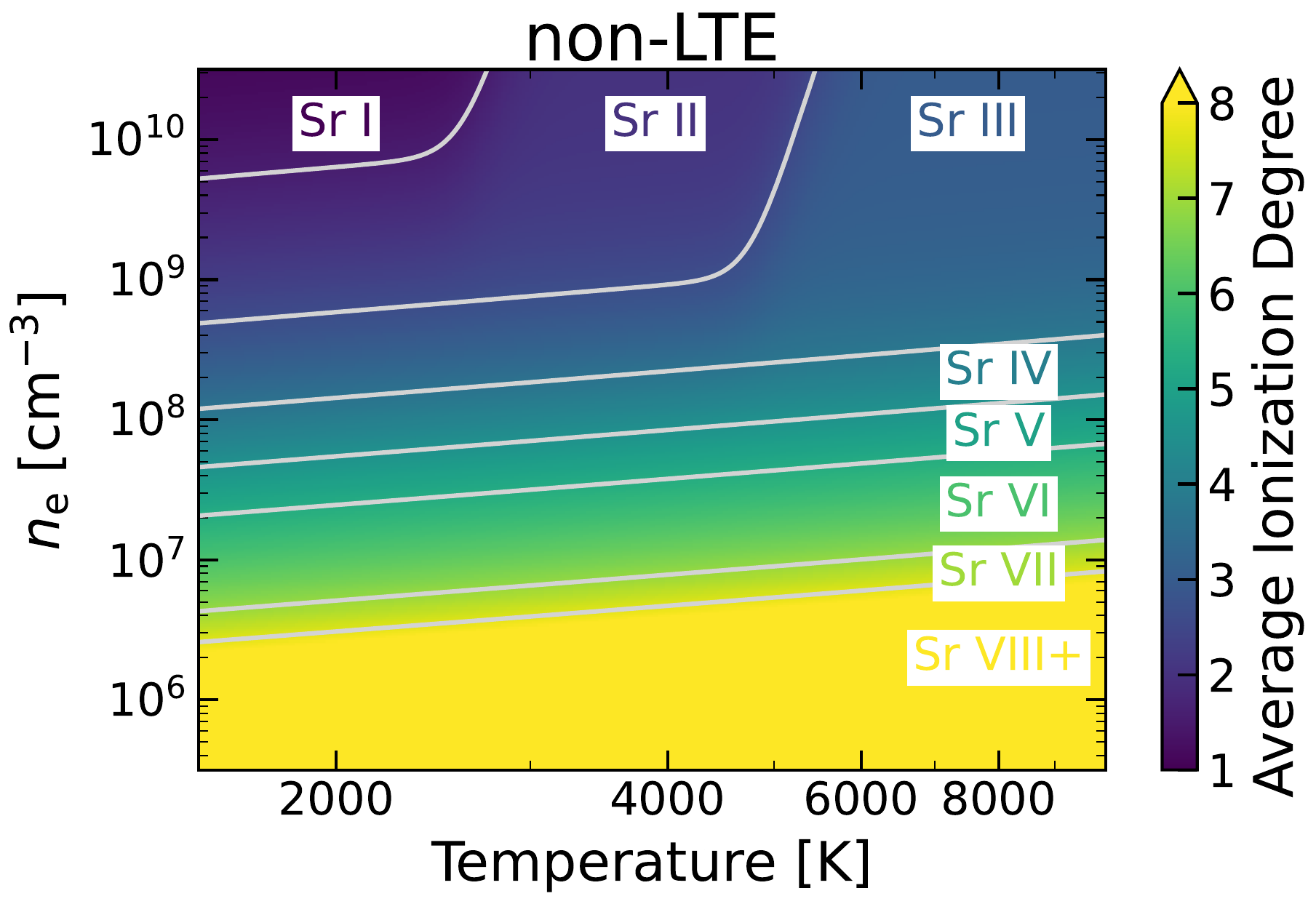}
  \caption{Average ionization degree of Sr
  as a function of electron density and temperature. 
  The left panel shows a LTE case while the right panel shows a non-LTE case 
  with a radioactive heating rate of 
  $\dot{q} = \qty{1.0}{eV.s^{-1}.ion^{-1}}$. 
  The solid lines show the boundaries to give equal ionization fractions 
  for consecutive charge states.
  \label{fig:average_ionization}}
\end{figure*}

Figure \ref{fig:average_ionization} shows 
the average ionization degree of Sr 
for a wide range of temperature and electron density. 
As discussed above, the ionization degree has 
a sharp dependence on the temperature in the LTE case (left panel). 
On the other hand, in the non-LTE case (right panel), 
the ionization degree is more sensitive to the electron density. 
In particular, 
at a low density ($n_{\rm e} \lesssim \qty{e9}{cm^{-3}}$), 
the ionization degree is mainly determined by the non-thermal effects 
(the second term in the right hand side of \myrefeq{eq:ionization}),
which gives a strong dependence to the density. 
At a high density ($n_{\rm e} \gtrsim \qty{e9}{cm^{-3}}$), 
the thermal ionization is more important, and thus, 
the ionization degree is similar to those in the LTE case.

\begin{figure}
    \centering
    \epsscale{1.1}
    \plotone{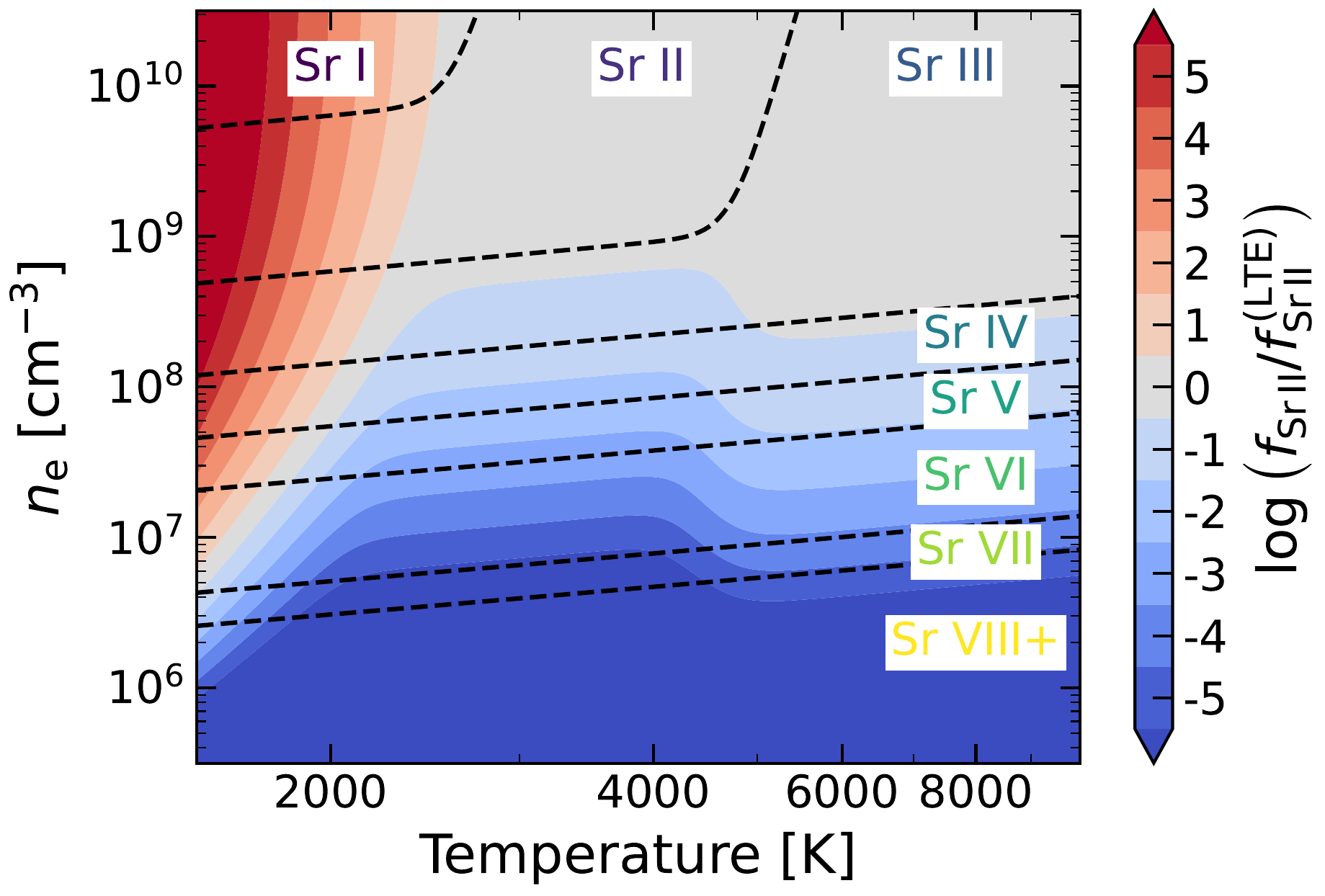}
    \caption{Departure coefficient of \ionname{Sr}{II}
    as a function of electron density and temperature.
    The setup of the calculation is the same as 
    in \myreffig{fig:average_ionization}.
    The dashed lines show the boundaries 
    to give equal ionization fractions for consecutive charge states.
    \label{fig:dep_coeff_Sr}}
\end{figure}

To evaluate the deviation from the LTE population of \ionname{Sr}{II},
we define a departure coefficient for \ionname{Sr}{II} as
$f_\mathrm{\ionname{Sr}{II}}/f_\mathrm{\ionname{Sr}{II}}^\mathrm{(LTE)}$,
where $f_\mathrm{\ionname{Sr}{II}}^\mathrm{(LTE)}$ 
and $f_\mathrm{\ionname{Sr}{II}}$ is 
the ionization fractions of \ionname{Sr}{II}
in the LTE and non-LTE cases, respectively.
The departure coefficient for different electron density and temperature is shown in Figure \ref{fig:dep_coeff_Sr}.
In the ejecta at a few days after the merger
(with the typical electron temperature of \qtyrange{3000}{4000}{K} 
and the typical electron number density of
\qtyrange{e7}{e8}{cm^{-3}}),
the departure coefficient for \ionname{Sr}{II} 
is on the order of \numrange{e-4}{e-2}.
This indicates that abundance of \ionname{Sr}{II} is strongly suppressed 
due to the ionization by high-energy electrons.
Accordingly, the lower state of the $\qty{1}{\mu m}$ feature
is also suppressed (dashed lines in Figure \ref{fig:pop_frac_Sr}). 
As we assume the Boltzmann distribution for excited states of Sr,
the population of these states are also suppressed by following
the departure coefficient of ionization degree.
This suppression implies that more Sr is
required to explain a certain absorption feature
in the non-LTE case.

\begin{figure*}
  \centering
  \epsscale{1.1}
  \plottwo{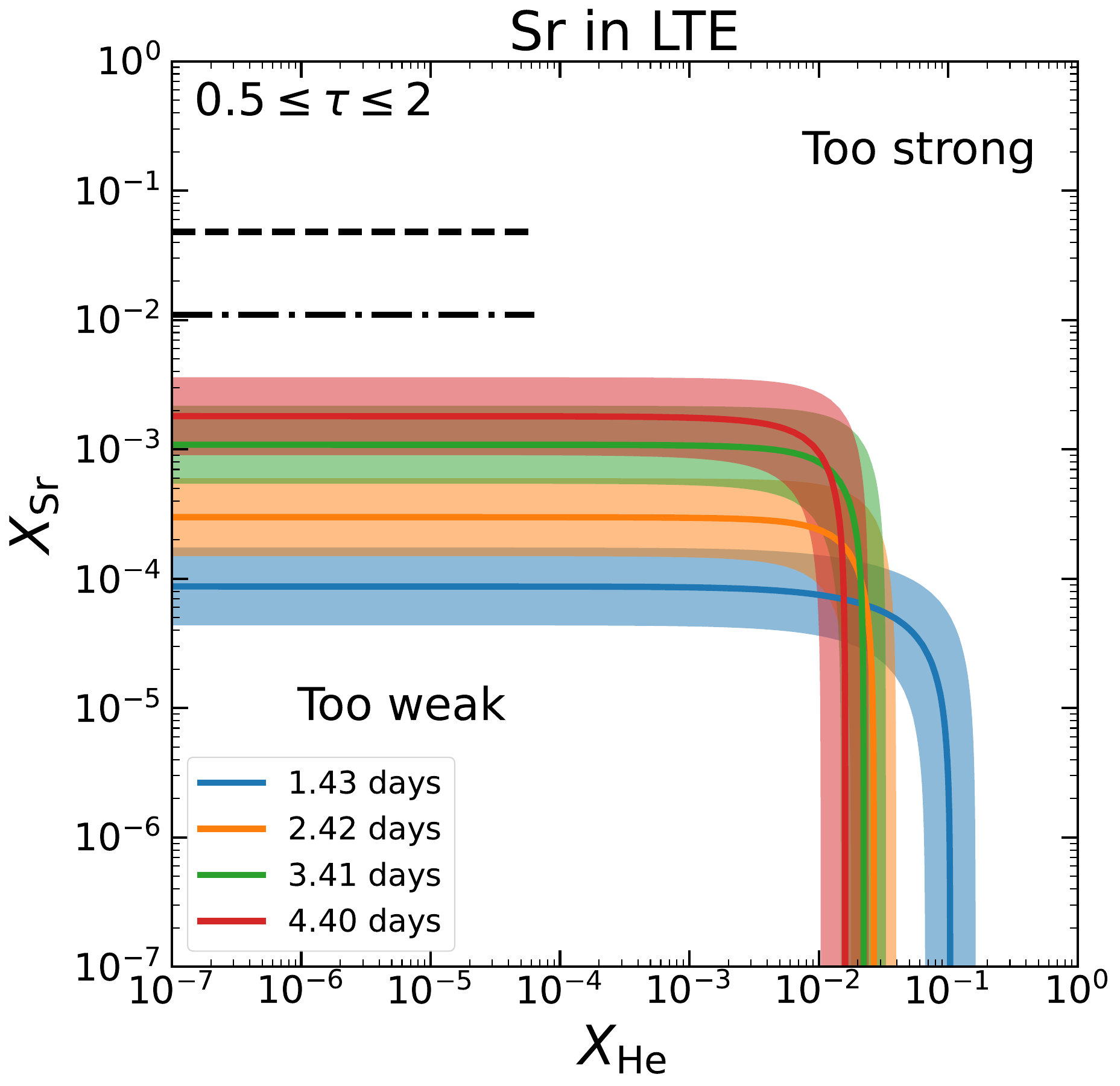}{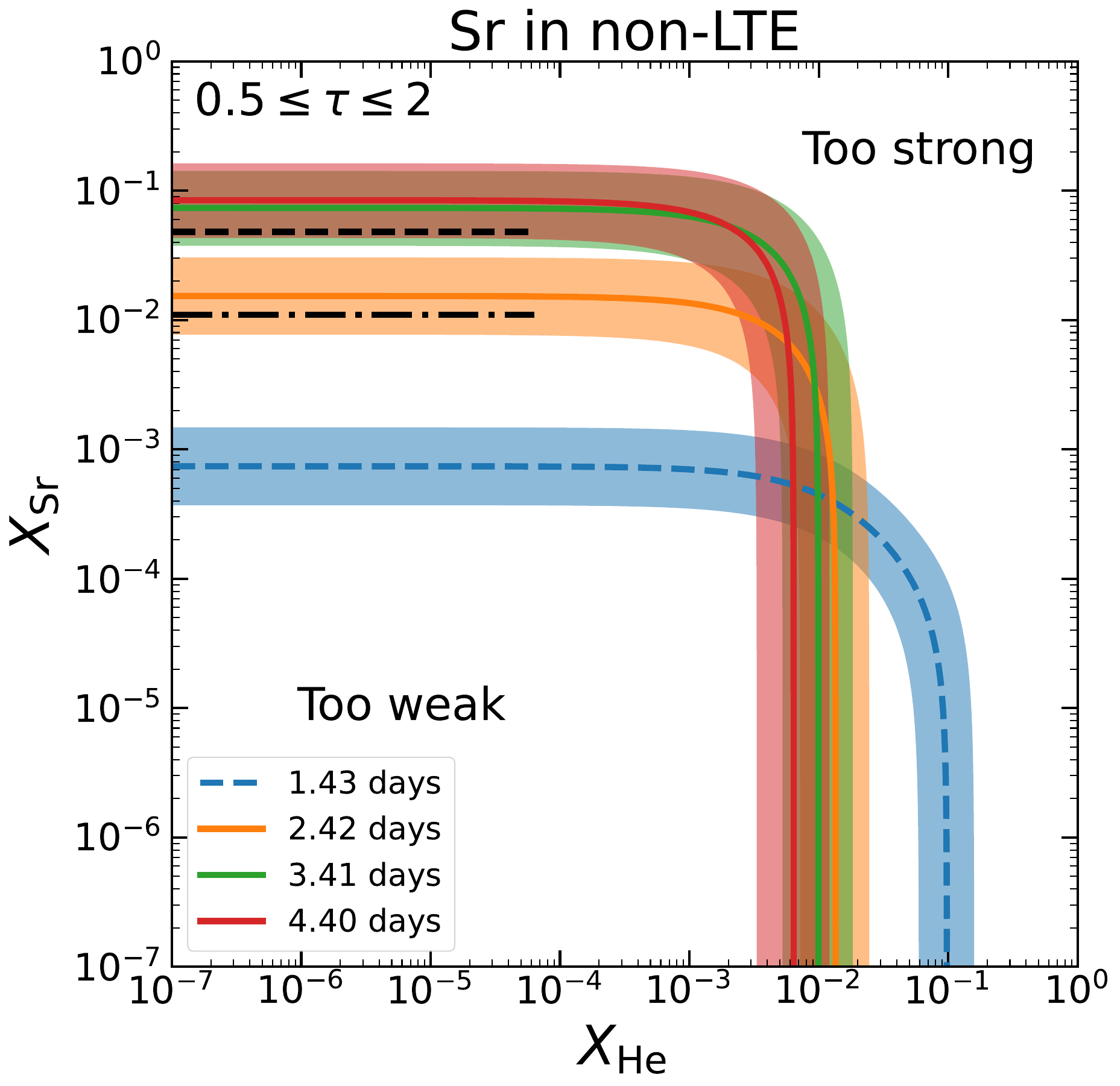}
  \caption{
  Constraints on the mass fractions of He and Sr 
  at the photosphere in each epoch.
  The solid curves correspond to $\tau = 1$ and the shaded region show 
  the region for $0.5 \le \tau \le 2$.
  The constraint for each epoch is shown in different colors
  according to the legend.
  The left panel shows the constraints in the LTE case for Sr, 
  while the right panel shows those in the non-LTE case for Sr.
  The horizontal dashed and dashed-dotted lines show the Sr mass fractions 
  in the solar $r$-process pattern \citep{2020_Prantzos+},
  assuming the minimum mass numbers of $r$-process elements to be 
  $A_{\rm min} = 88$ and 69, respectively.
  For the non-LTE case, our constraints are 
  rather uncertain at the earliest epoch ($t =1.43$ days, dashed lines) 
  due to the uncertainty in the total mass density.
  \label{fig:constraints-X_He-X_Sr}}
\end{figure*}

\subsection{Abundance Constraints in GW170817} 
\label{subsec:abundance_constraints}
To constrain the abundances of He and Sr, 
we apply our models to the spectral time series of AT2017gfo.
We calculate the ionization and excitation 
at $t = 1.43,\, 2.42,\, 3.41,$ and \qty{4.40}{days} after the merger.
%For each epoch, we perform local, non-LTE calculations.
%The non-LTE calculations for each epoch are performed under the one-zone approximation.
For each epoch, by adopting appropriate mass density 
and temperature,
we perform the non-LTE calculations at the photosphere.
To evaluate the mass density at the photosphere for each epoch,
we first assume an overall density profile.
We adopt a power-law density profile of the ejecta as:
\begin{equation}
  \label{eq:density_profile}
  \rho(v, t) = \left\{
    \begin{aligned}
      & \rho_0 (v/v_0)^{-s} (t/t_0)^{-3} & (0.05 \le v/c \le 0.35) \\
      & 0 & \text{(otherwise)}.
    \end{aligned}
  \right.
\end{equation}
Here we adopt $s=3$, and $\rho_0$, $v_0$ and $t_0$ are 
taken to give a total ejecta mass 
of $M_\mathrm{ej} = \qty{0.03}{M_\odot}$,
which is known to provide a reasonable agreement with 
the bolometric luminosity of AT2017gfo
\citep[e.g.,][]{2017_Tanaka+_b,2018_Kawaguchi+}.
Then, according to this profile, we set the density
at the photosphere for each epoch.
For the photospheric temperature,
we use the temperature derived by the spectral fitting at each epoch (\myreftab{tab:fit_params}).
In this model sequence, we determine
both the level populations of He 
and the ionization fractions of Sr simultaneously.
We solve the rate equations iteratively to obtain the self-consistent electron number density (Equation \ref{eq:ne}).

\myreffig{fig:constraints-X_He-X_Sr} shows mass fractions of He and Sr at the photosphere 
to reproduce the observed $\qty{1}{\mu m}$ feature at each epoch.
The solid curves correspond to $\tau = 1$ and the shaded region show the region for 
$0.5 \le \tau \le 2$.
We can divide each constraint curve into three parts 
based on the contribution to the observed \qty{1}{\mu m} feature: 
Sr-dominated region (horizontal part), 
He-dominated region (vertical part) and the region in between.

In the Sr-dominated region, the required mass fraction of Sr 
in the non-LTE case at each epoch is 
up to two orders of magnitudes larger than in the LTE case. 
This is because the ionization fraction of \ionname{Sr}{II} in the non-LTE case
becomes lower as compared to that in the LTE case due to overionization (see \myrefsec{subsec:pop_frac_Sr}).
As a result, in the non-LTE case, the required mass fraction of Sr is similar to the mass fraction in the solar $r$-process pattern (horizontal dashed/dashed-dotted lines in Figure \ref{fig:constraints-X_He-X_Sr}).
\citet{2026_Arya+} also reached a similar conclusion by obtaining 
a similar constraint on the Sr abundance with their non-LTE modeling.

In the He-dominated region, 
the required mass fraction of He decreases with time.
As discussed in \myrefsec{subsec:pop_frac_He}, the relative population of 
the \atomterm{2}{3}{S} state is higher for a lower temperature 
due to the suppressed depopulation through the photoionization and photoexcitation.
As a result, the relative population of this state is higher for a later epoch with a lower temperature (Table \ref{tab:fit_params}).

\begin{figure*}
  \centering
  \epsscale{0.9}
  \plotone{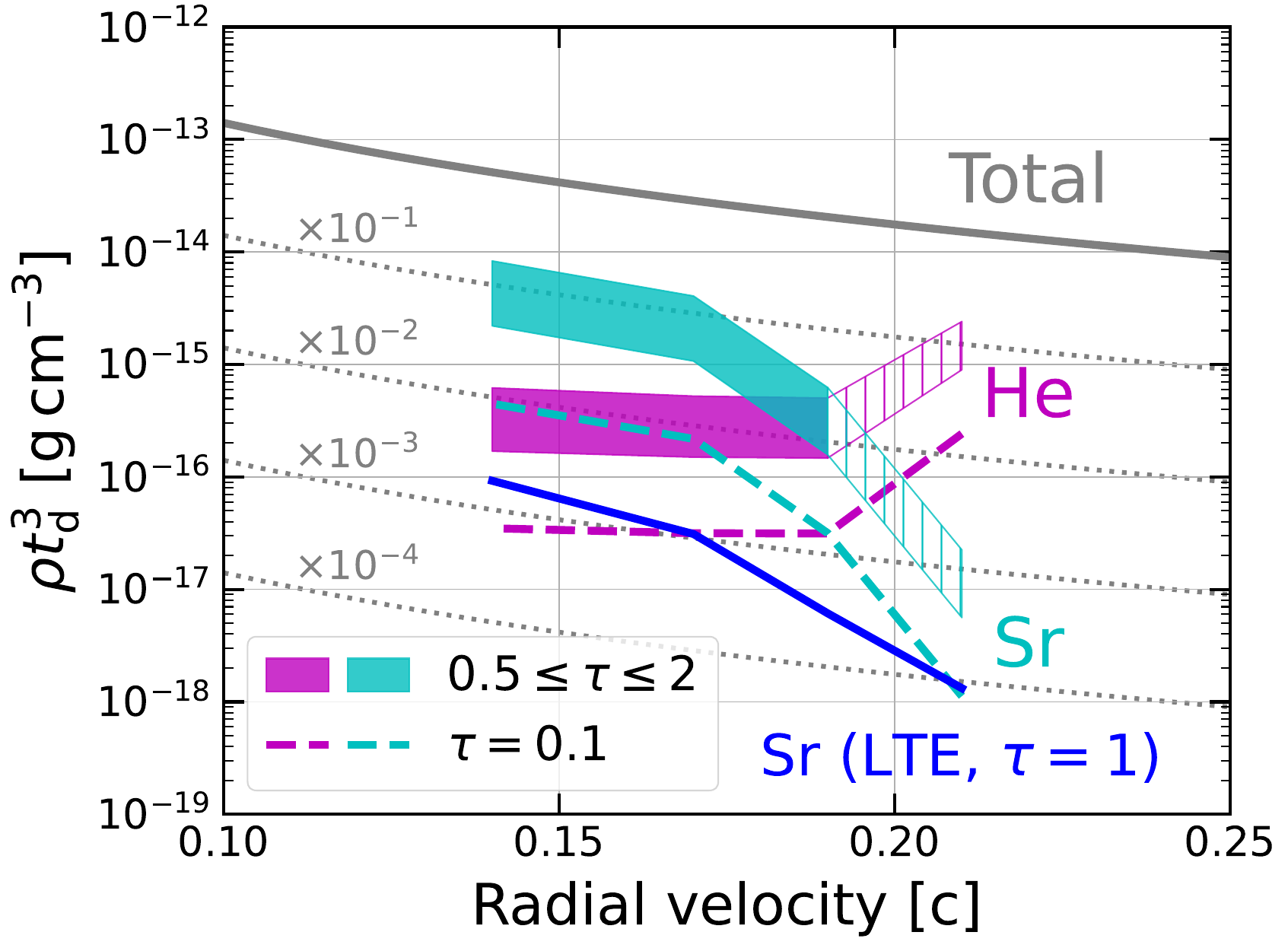}
  \caption{Constraints on the mass-density profiles of 
           He (magenta) and Sr (cyan).
           The density is scaled to that at $t= \qty{1}{day}$.
           Shaded regions correspond to the required densities 
           in the case of $0.5\le \tau \le 2$.
           The constraints at the highest velocity are shown with hatched region as the constraints are rather uncertain.
           The bold dashed lines show the mass-density giving $\tau = 0.1$.
           The blue solid line shows the constraints on Sr in the LTE case.
           The gray solid line shows the adopted total density profile (see \myrefeq{eq:density_profile})
           and the gray dashed lines show the total density scaled by the values in the figure. \label{fig:constraints-rho_profiles}}
\end{figure*}

Based on the abundance constraint from the spectrum at each epoch 
(\myreffig{fig:constraints-X_He-X_Sr}),
we can give constraints on the mass-density profile of He and Sr.
Since the photosphere moves inward in the velocity coordinates with time, 
we can obtain the required mass density of these elements
as a function of velocity, from outside to inside.
\myreffig{fig:constraints-rho_profiles} shows the inferred mass-density profiles
scaled at 1 day after the merger.  
The shaded regions represent the ranges where 
the condition of $0.5 \le \tau \le 2$ is satisfied.
Following the trend in the mass fraction (\myreffig{fig:constraints-X_He-X_Sr}),
the required mass density of Sr drops toward the higher velocity
because of the efficient excitation of the 4D states at an earlier epoch 
(with a higher temperature, 
but see below for the uncertainty due to the total mass density).
On the other hand, the required mass density of He is higher toward higher velocity
because of the strong depopulation of the \atomterm{2}{3}{S} state 
at an early epoch (with a higher temperature).
Table \ref{tab:constraints_density} summarizes the derived constraints 
on the mass density of He and Sr.

Note that the constraints on the mass density of each element 
are dependent on the assumed total mass density $\rho_\mathrm{tot}$
(see Appendix \ref{app:density_dependence} for more details).
For Sr, a lower total mass density leads to more significant non-thermal ionization.
As a result, the required Sr mass density becomes higher
(roughly scales with $\rho_\mathrm{tot}^{-2}$ in the relevant density range).
For He, the dependence on the mass density 
is not as strong as in Sr (roughly scales with $\rho_\mathrm{tot}^{-1}$).
This is because the dominant ionization degree of He is the singly ionized state, which is coupled with the excited states of the neutral He.

In addition, we emphasize that the total mass density 
is more uncertain at higher velocity. 
In this work, total ejecta mass ($M_\mathrm{ej}$), 
inner velocity ($v_0$) and slope of the density profile ($s$) characterize 
the total mass-density profile of the ejecta
(see \myrefeq{eq:density_profile}).
The former two parameters are more strongly constrained,
because they are more directly linked to the bolometric luminosity of a kilonova. 
Given $M_\mathrm{ej}$ and $v_0$ to reproduce
the bolometric luminosity of AT2017gfo, 
the slope $s$ can still vary over a wide range.
This degree of freedom in the slope $s$ results in a larger uncertainty in the mass density at higher velocity.
Given the relatively strong dependence of population fraction to the total mass density, 
our constraints are most uncertain at the earliest epoch (dashed curve in Figure \ref{fig:constraints-X_He-X_Sr}) 
and at the highest velocity 
(hatched regions in Figure \ref{fig:constraints-rho_profiles}).

\begin{deluxetable*}{cccccccccc}[htbp]
\tablecaption{Summary of mass density constraints from our non-LTE calculations \label{tab:constraints_density}}
\tablehead{
    \colhead{\multirow{2}{*}{$v_\mathrm{ph}$ [$c$]} \tablenotemark{a}}
    & \colhead{\multirow{2}{*}{$\rho_\mathrm{tot} t_{\rm d}^3 $ [$10^{-14}\, \mathrm{g\, cm^{-3}}$]} \tablenotemark{b}}
    & \multicolumn{4}{c}{$\rho_\mathrm{He} t_{\rm d}^3$ [$10^{-14}\, \mathrm{g\, cm^{-3}}$] \tablenotemark{c}}
    & \multicolumn{4}{c}{$\rho_\mathrm{Sr} t_{\rm d}^3$ [$10^{-14}\, \mathrm{g\, cm^{-3}}$] \tablenotemark{d}} \\
    &
    & \colhead{$\tau=0.1$} & \colhead{$\tau=0.5$} 
    & \colhead{$\tau=1$} & \colhead{$\tau=2$} 
    & \colhead{$\tau=0.1$} & \colhead{$\tau=0.5$} 
    & \colhead{$\tau=1$} & \colhead{$\tau=2$}
}
% \colnumbers
\startdata
\multicolumn{1}{c}{0.21} & \multicolumn{1}{c}{1.5}
& \num{2e-02} & \num{9e-02} & \num{1e-01} & \num{2e-01} 
& \num{1e-04} & \num{6e-04} & \num{1e-03} & \num{2e-03}\\
\multicolumn{1}{c}{0.19} & \multicolumn{1}{c}{2.0}
& \num{3e-03} & \num{1e-02} & \num{3e-02} & \num{5e-02} 
& \num{3e-03} & \num{2e-02} & \num{3e-02} & \num{6e-02} \\
\multicolumn{1}{c}{0.17} & \multicolumn{1}{c}{2.9}
& \num{3e-03} & \num{1e-02} & \num{3e-02} & \num{5e-02} 
& \num{2e-02} & \num{1e-01} & \num{2e-01} & \num{4e-01} \\
\multicolumn{1}{c}{0.14} & \multicolumn{1}{c}{5.1}
& \num{3e-03} & \num{2e-02} & \num{3e-02} & \num{6e-02} 
& \num{4e-02} & \num{2e-01} & \num{4e-01} & \num{8e-01}
\enddata
\tablecomments{
    \tablenotetext{a}{Velocity at the photosphere}
    \tablenotetext{b}{Total mass density (scaled at $t= \qty{1}{day}$)}
    \tablenotetext{c}{Helium mass density 
    for different $\tau$ (scaled at $t= \qty{1}{day}$)}
    \tablenotetext{d}{Strontium mass density 
    for different $\tau$ (scaled at $t= \qty{1}{day}$)}
}
\end{deluxetable*}

%%%%%%%%%%%%%%%%%%%%%%%%%%%%%%%%%%%%%%%%
% Section:
%%%%%%%%%%%%%%%%%%%%%%%%%%%%%%%%%%%%%%%%

\begin{figure*}
  \centering
  \epsscale{1.1}
  \plottwo{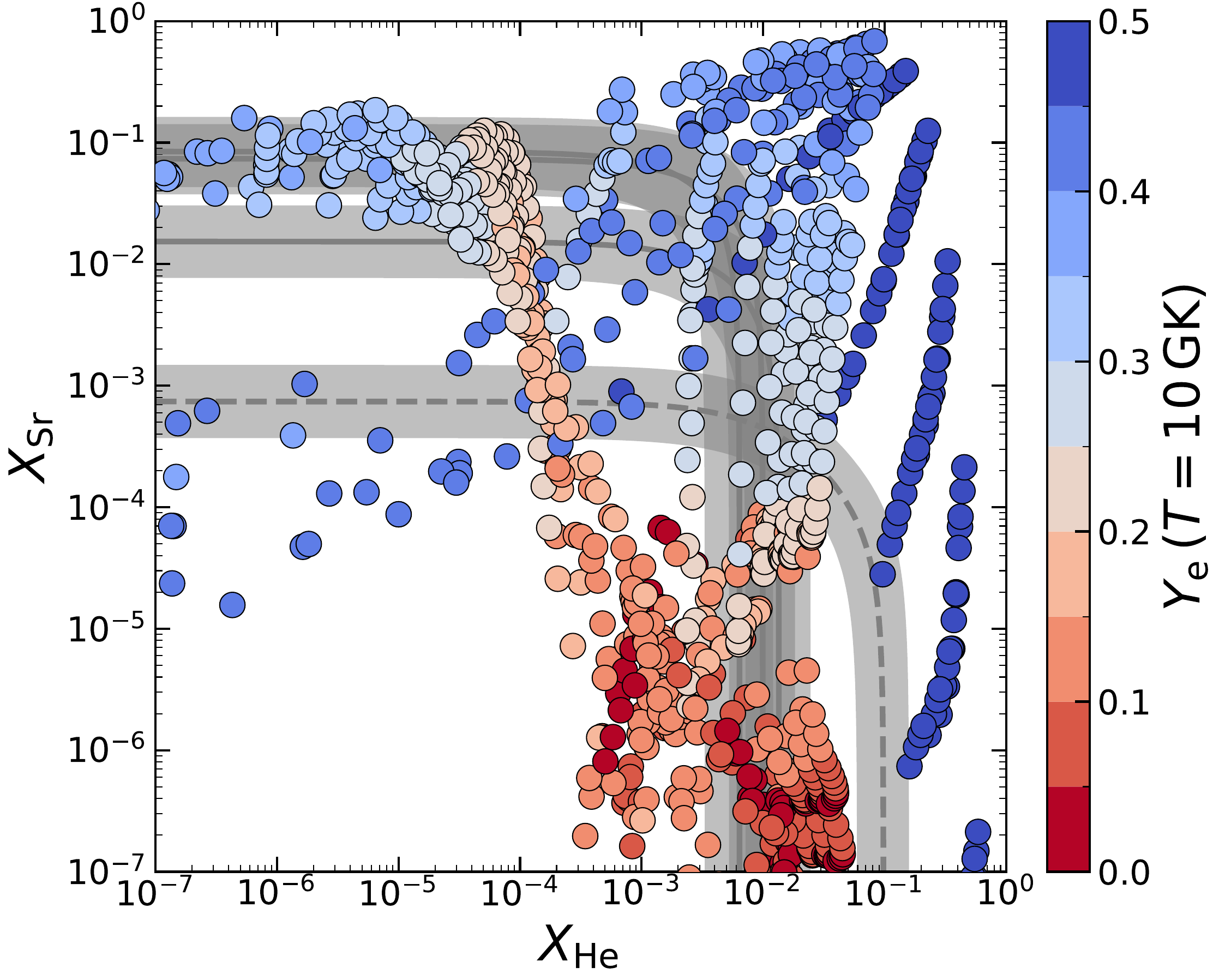}{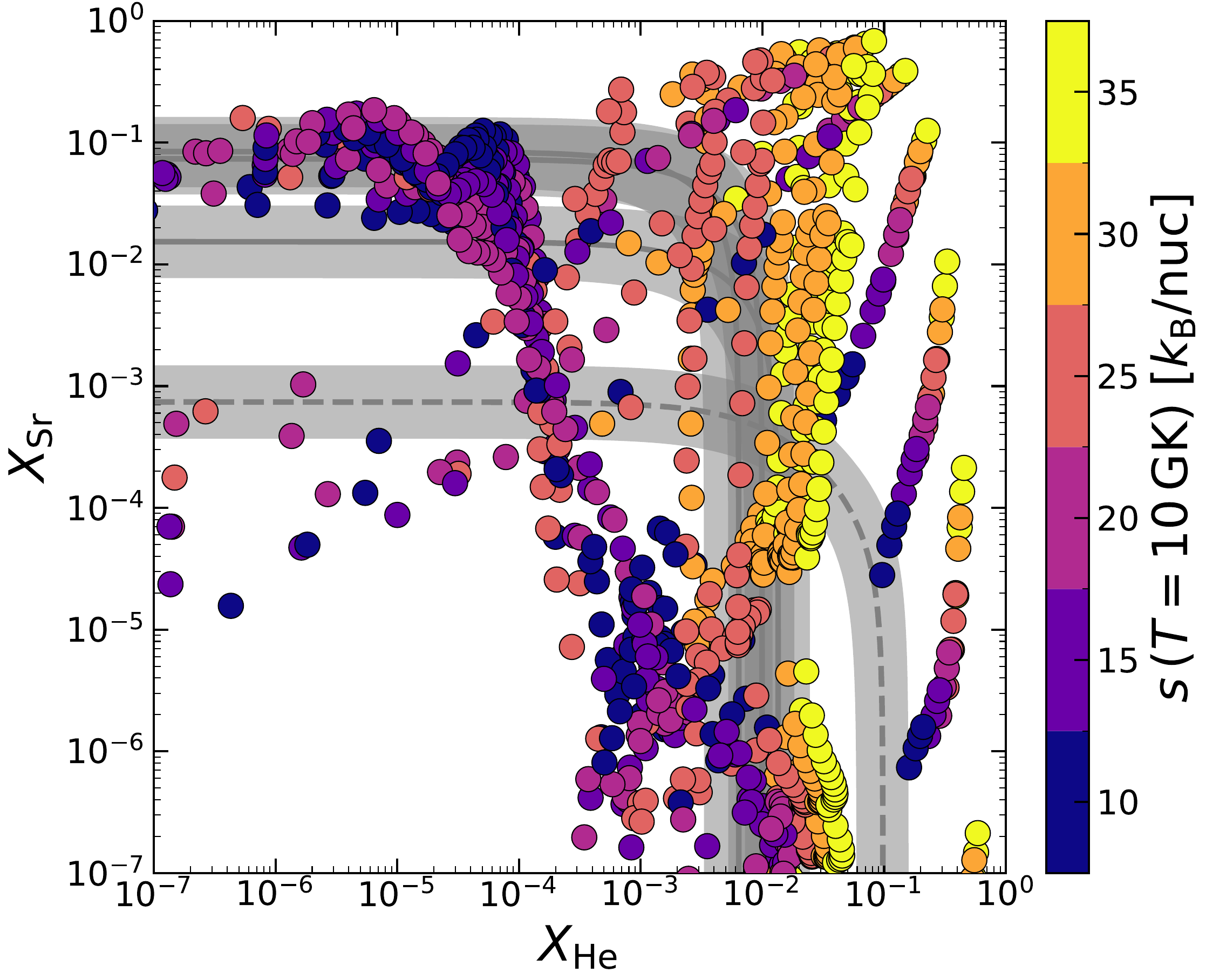}
  \caption{Comparison of our constraints 
          (gray shaded area, see Figure \ref{fig:constraints-X_He-X_Sr})
          with nucleosynthetic calculations \citep{2018_Wanajo}. 
          For nucleosynthesis calculations, mass fractions at 1 day are shown by circles, which are color-coded according to the electron fraction (left) and entropy (right). 
          \label{fig:comparison-X_He-X_Sr}}       
\end{figure*}

\section{Discussion} \label{sec:discussion}

\subsection{Implications to nucleosynthesis conditions} 
\label{subsec:nucleosynthesis}

Based on our constraints shown 
in \myrefsec{subsec:abundance_constraints}, 
we discuss nucleosynthesis conditions in GW170817. 
\myreffig{fig:comparison-X_He-X_Sr} shows comparison of 
our constraints on the mass fractions 
(see \myreffig{fig:constraints-X_He-X_Sr}) with 
parametric nucleosynthetic calculations by \citet{2018_Wanajo}.
In the nucleosynthetic calculations, initial electron fraction ($Y_{\rm e}$, with an interval of 0.01), initial entropy ($s$, with an interval of $5\, k_\mathrm{B}/\mathrm{nucleon}$), and expansion velocity ($v_{\rm exp}$, with an interval of $0.05\, c$) are parameterized in a framework of the free-expansion (FE) model.
The colors of the circles are given according to 
electron fraction (left panel) and entropy (right panel).
To compare with our constraints, we show the nucleosynthesis results
at $t= \qty{1}{day}$ and with the expansion velocity between $0.1\,c$ and $0.25\,c$.

The nucleosynthesis pattern 
has a strong dependence on the electron fraction
(left panel of \myreffig{fig:comparison-X_He-X_Sr}).
Our constraints are broadly consistent with the nucleosynthesis yields
with a relatively low electron fraction ($Y_{\rm e} \lesssim 0.35$).
On the other hand, the yields with $Y_{\rm e} \gtrsim 0.35$ show 
either a very high He abundance 
($X_{\rm He} \gtrsim 10^{-1}$, rightmost part of the figure)
or a high abundance of both He and Sr 
($X_{\rm He/Sr} \gtrsim 10^{-2}$, 
upper right part of the figure),
leading to a too strong $\qty{1}{\mu m}$ feature.
This overproduction is due to $\alpha$-rich freeze-out \citep{1997_Hoffman+,1998_Meyer+}.

The production of He and Sr also depends 
the entropy (right panel of Figure \ref{fig:comparison-X_He-X_Sr}).
In particular, the nucleosynthesis yields with 
$s \gtrsim 35\, k_{\mathrm{B}}/\mathrm{nucleon}$ tends to overproduce 
both He and Sr (right upper part of the figure) 
as the effects of $\alpha$-rich freeze-out is more pronounced 
with a higher entropy.
These comparison implies that 
the \qty{1}{\mu m} feature is best explained 
by the nucleosynthesis with a relatively low electron fraction
($Y_{\rm e} \lesssim 0.35$) and low entropy ($s \lesssim 30 \, k_{\mathrm{B}}/\mathrm{nucleon}$).

Note that our constraints on the electron fraction and entropy 
seem opposite to those indicated by \citet{2021_Domoto+}
using the same nucleosynthetic calculations by \citet{2018_Wanajo}.
With the results of LTE modeling,  \citet{2021_Domoto+} gave constraints on the entropy ($s \gtrsim 25\, k_\mathrm{B}/\mathrm{nucleon}$) based on the 
 absence of the Ca lines. 
They focus on high $Y_{\rm e}$ conditions ($Y_{\rm e} \gtrsim 0.4$), 
which tend to produce both Sr and Ca ($^{48}$Ca) efficiently. 
Under such conditions, 
a high entropy is necessary to suppress the production of Ca.
In this work, on the other hand, 
we use the constraints on the abundances of 
both Sr and He using non-LTE calculations, 
which prefer a low electron fraction ($Y_{\rm e} \lesssim 0.35$)
to reconcile with the observed \qty{1}{\mu m} feature. 
Under this condition, 
the overproduction of $^{48}$Ca is naturally avoided (see Figure 10 of \citealt{2021_Domoto+}).

Among the preferred nucleosynthesis conditions 
(i.e., $Y_{\rm e} \lesssim 0.35$ and $s \lesssim 30 \, k_{\rm B}/\mathrm{nucleon}$),
there are two possible solutions to explain the \qty{1}{\mu m} feature in AT2017gfo.
For an intermediate electron fraction ($Y_{\rm e} \sim 0.15-0.35$),
the \qty{1}{\mu m} feature is likely to be reproduced 
by Sr at all the epochs except for the earliest epoch 
(see the left upper part in \myreffig{fig:comparison-X_He-X_Sr}).
As mentioned in \myrefsec{subsec:abundance_constraints},
the Sr mass fraction in this case nicely agrees with that in the solar $r$-process pattern.
If this is the case, it supports 
BNS mergers as the main sources of the $r$-process elements in the Universe.

On the other hand, 
for a very low electron fraction ($Y_{\rm e} \lesssim 0.15$),
the \qty{1}{\mu m} feature is likely to be reproduced by He
(see the right bottom part in \myreffig{fig:comparison-X_He-X_Sr}).
In this condition, He is mainly produced by 
$\alpha$-decay of trans-Pb nuclei \citep{2022_Perego+}.
Therefore, if the \qty{1}{\mu m} feature is mainly attributed to He,
this feature serves as an indirect signature for the production of elements beyond the third $r$-process peak.

\begin{figure}
  \centering
   \includegraphics[width=0.9\linewidth]{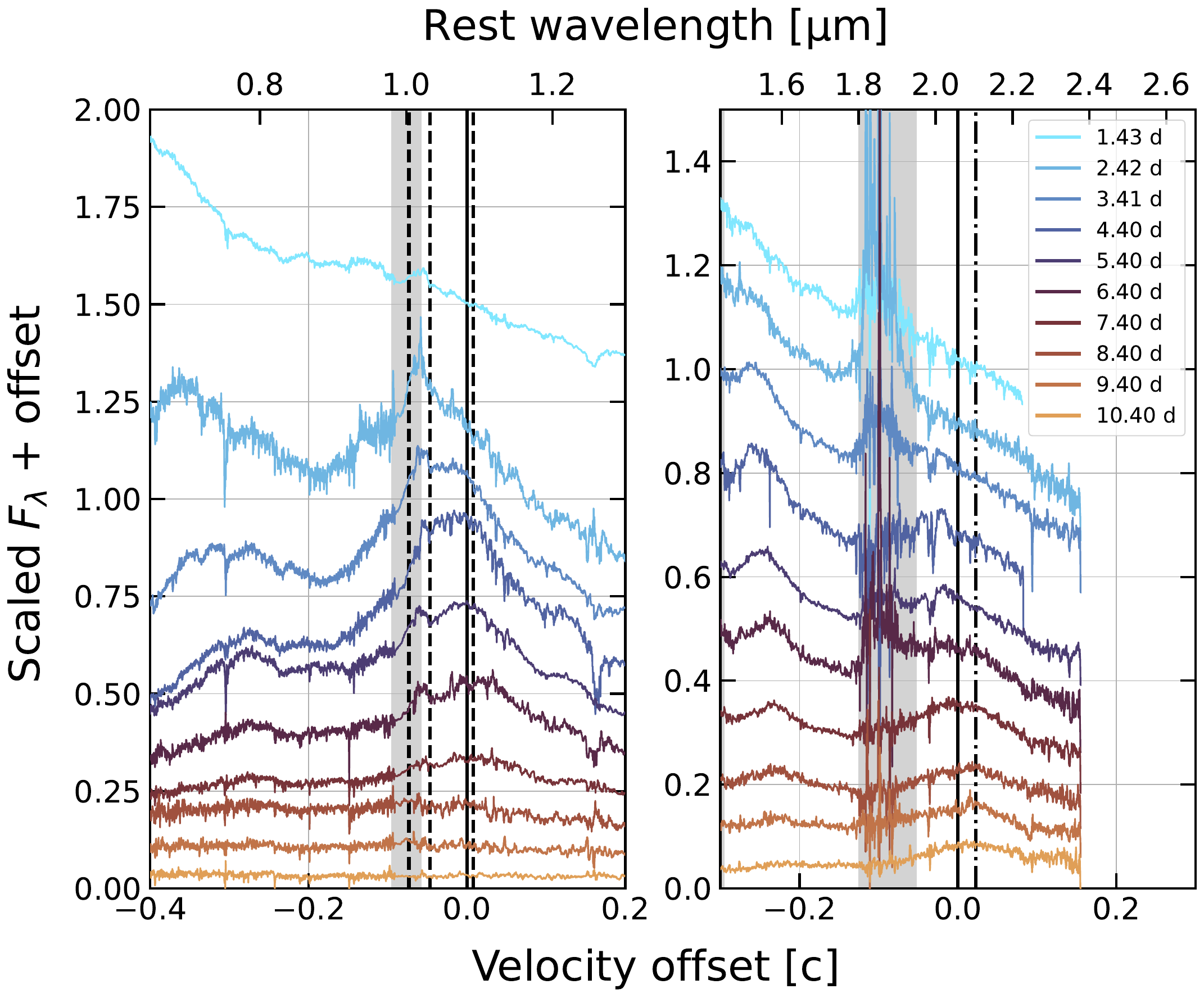}

  \caption{Spectral evolution of AT2017gfo 
  around \qty{1}{\mu m} and \qty{2}{\mu m}.
  The gray shaded regions are the same as in \myreffig{fig:obs_feature}.
  The velocity offsets in each panel are measured from the \ionname{He}{I} lines (\qty{1.08}{\mu m} and \qty{2.06}{\mu m}, the vertical solid lines).
  For comparison, the wavelengths of the \ionname{Sr}{II} lines 
  are shown as the vertical dashed lines in the left panel and
  that of the [\ionname{Te}{III}] line is shown as 
  the vertical dashed-dotted line in the right panel.
  \label{fig:spec_17gfo}}
\end{figure}

Our work alone cannot fully resolve 
the degeneracy of the He and Sr contributions to the \qty{1}{\mu m} feature.
However, the degeneracy can be potentially solved 
by using other lines of either He or Sr.
For Sr, there are strong doublet transitions (4078 and 4216 \AA), 
but the features are heavily blended with other lines in kilonova spectra. 
For He, it is worth exploring the singlet \ionname{He}{I} line 
at 2.06 $\mu$m (2$^1$S--2$^1$P).
In fact, AT2017gfo (and GRB 230307A / AT2023vfi) shows 
an emission feature around 2 $\mu$m at later phases.
Although this feature is mainly attributed to a [Te~III] line (\citealt{2023_Hotokezaka+,2024_Levan+,2025_Gillanders&Smartt}), 
the peak wavelength of this feature also agrees well with that of 
the  \ionname{He}{I} line (see \myreffig{fig:spec_17gfo}), 
as pointed out by \citet{2024_Gillanders+}.

As shown in Section \ref{subsec:pop_frac_He}, 
in a typical condition of kilonova ejecta, 
the \ionname{He}{I} 2.06 $\mu$m line does not produce strong features.
The \ionname{He}{I} 2.06 $\mu$m line may become stronger 
if the density is higher,
but such a condition would produce an even stronger \qty{1.08}{\mu m} line.
It is worth performing more detailed calculations by taking into account realistic ejecta structure to study the He emission feature.

Also, when He is produced via $\alpha$-decays, 
a large fraction of lanthanide is also produced, 
providing a large opacity. 
This seems to contradict with the observed ``blue'' kilonova component, 
which is considered to be 
the emission from lanthanide-free ejecta
\citep[e.g.,][]{2017_Villar+_b, 2017_Kasen+}.
However, the origin of blue kilonova component 
is still under debate 
\citep[e.g.,][]{2020_Kawaguchi+,2023_Just+,2025_Kitamura+}.
In addition, constraints on the lanthanide fraction
are obtained based on the LTE opacity of heavy elements, 
and such constraints may be strongly affected by
the non-LTE effects (see \citealt{2026_Brethauer+}).
Therefore, it is also important to consistently model the light curves as well as spectral features of heavy elements in non-LTE regime.

%%%%%%%%%%%%%%%%%%%%%%%%%%%

\subsection{Implications to mass-ejection mechanism}
\label{subsec:mass_ejection}

In this section, we discuss implications of our abundance constraints
to the mass ejection in GW170817.
For this purpose, 
we use results of long-term viscous hydrodynamics simulations:
DD2-135-135 model as a case leaving a long-lived NS remnant
\citep{2020_Fujibayashi+_b}. The post-merger mass ejection in this model 
is simulated in two-dimensional (2D) axisymmetric space, 
with initial data mapped from a matter profile of 
a three-dimensional (3D) BNS merger simulation in the early post-merger phase. 
The nucleosynthetic yields are calculated using a nuclear reaction network 
along the Lagrangian evolution of passive tracer particles:
the dynamical ejecta are traced in 3D, whereas the post-merger component is traced in 2D. 
The resulting yields are then projected onto a 2D mesh for 
a long-term hydrodynamics simulation for modeling a kilonova \citep{2021_Kawaguchi+}.

This model gives a relatively small dynamical ejecta (\qty{1e-3}{M_{\odot}}) 
and a large post-merger ejecta (\qty{5e-2}{M_{\odot}}).
We adopt this model as the model reproduces the observed light curve of AT2017gfo 
by the large contribution from the post-merger ejecta \citep{2022_Kawaguchi+}.
On the other hand, BNS merger models leaving a short-lived NS tend 
to underproduce the total ejecta due to the small contribution 
of the post-merger ejecta \citep{2023_Kawaguchi+_b}.

\begin{figure}
  \includegraphics[width=\linewidth]{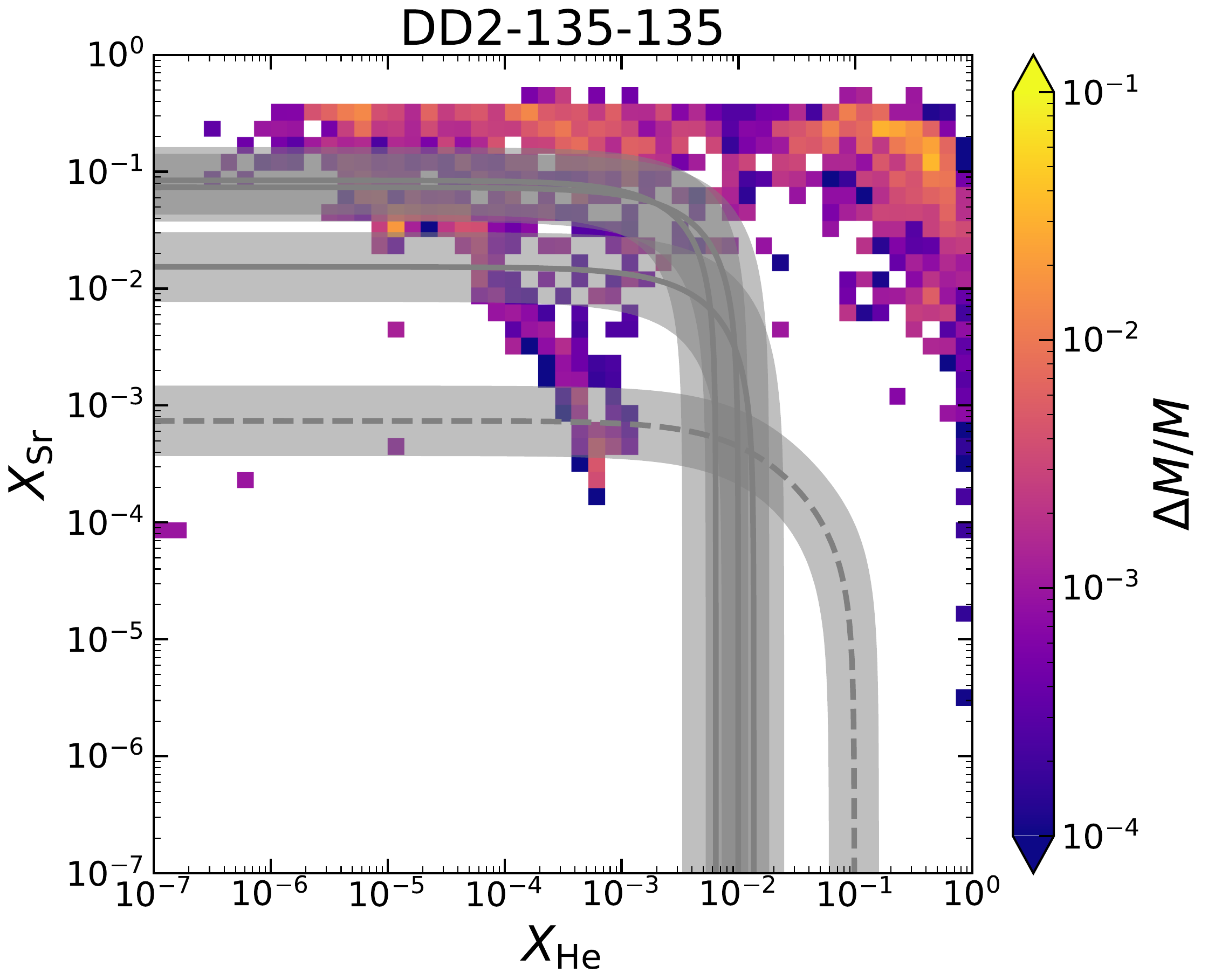}
  \caption{Comparison of our observational constraints with nucleosynthetic results in DD2-135-135 model. The nucleosynthetic results are shown as a color map according to the relative mass in the simulation (expansion velocity between 0.1 $c$ and 0.25 $c$). \label{fig:comparison-X_He-X_Sr_DD2}
  }
\end{figure}

Figure \ref{fig:comparison-X_He-X_Sr_DD2} shows the comparison between the abundance constraints and nucleosynthesis yields in the DD2-135-135 model. 
For the nucleosynthesis yields, we use the abundances in each mesh in long-term hydrodynamics simulations followed until the homologous expansion phase \citep{2022_Kawaguchi+}.
The color map in the panel shows the relative mass 
between the expansion velocity of $v=0.1\, c$ and $0.25\, c$.
As clearly shown in the figure, the model gives too much
He and Sr, which is likely to overproduces 
the \qty{1}{\mu m} feature\footnote{
Although the ejecta in this model is dominated 
by the post-merger ejecta, there is also a small amount of 
dynamical ejecta with a low $Y_{\rm e}$,
which show a high $X_{\rm He}$ and a low $X_{\rm Sr}$
in the original nucleosynthesis calculations.
Since the nucleosynthesis calculations are mapped into the 2D meshes in long-term hydrodynamics simulations, spiral patterns in the dynamical ejecta with low $Y_{\rm e}$ are smeared out. This further suppresses the ejecta component with a high $X_{\rm He}$ and a low $X_{\rm Sr}$ in Figure \ref{fig:comparison-X_He-X_Sr_DD2}. However, as the absorption features in the photospheric spectra is sensitive to the global abundance above the photosphere (rather than small scale patterns), the comparison in Figure \ref{fig:comparison-X_He-X_Sr_DD2} is reasonable.}.
The overproduction is due to the high electron fraction ($Y_{\rm e} \gtrsim 0.35$) and high entropy ($s \gtrsim 35 \, k_{\rm B}/\mathrm{nucleon}$) in the high-velocity post-merger ejecta in this model.

\begin{figure}
  \centering
  \epsscale{1.1}
  \plotone{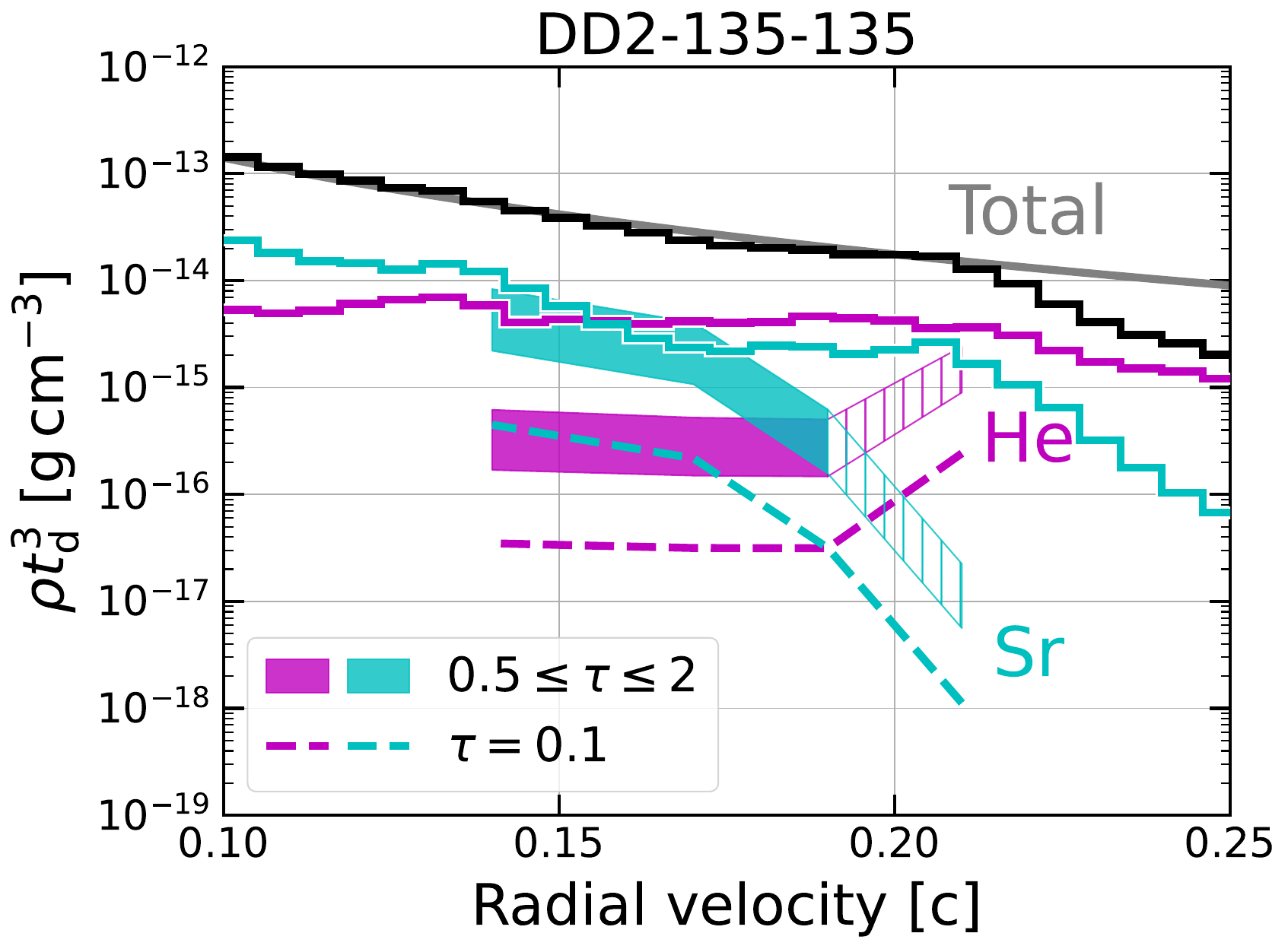}
  \caption{Comparison of mass-density profiles with the results of numerical simulations (DD2-135-135). The color scheme is the same as in Figure \ref{fig:constraints-rho_profiles}: 
  black, magenta, and cyan lines show the mass-density profile of 
  the total, He and Sr in the model.
  \label{fig:comparison-rho_profiles}}
\end{figure}

In \myreffig{fig:comparison-rho_profiles}, 
we compare our constraints 
in the density profile with 
the spherically averaged mass-density profile of 
the DD2-135-135 model 
(see \myrefapp{app:rho_profiles_2D}
for inclination dependence of the mass-density profiles).
In general, the strength of absorption feature is more directly linked to
the mass densities of each element 
($\rho_\mathrm{He/Sr} = \rho X_\mathrm{He/Sr}$)
rather than the mass fractions ($X_\mathrm{He/Sr}$).
As also seen in the mass fraction, the model overproduces 
both He (at $v \sim 0.14-0.19 \, c$) and Sr (at $v \sim 0.19 \, c$).

Although we focus on comparison with a specific model, similar trend is obtained 
in various numerical simulations.
In general, total ejecta mass and the electron fraction tend to be higher 
for the merger with a longer-lived NS remnant 
\citep[e.g.,][]{2023_Just+,2025_Bernuzzi+,2026_Jacobi+}.

In fact, there are several previous works discussing the outflow properties 
based on the abundances inferred from spectral analysis. 
For example, \citet{2022_Perego+} discussed 
a wind active timescale to explain 
the bolometric luminosity of AT2017gfo and the Sr abundance. 
They pointed out that, 
to supply a sufficient ejecta mass explaining the bolometric luminosity, 
a wind active timescale should be about \qty{100}{ms}. 
But in such a case, the expected Sr mass is 
about one order of magnitude larger 
than that inferred from the LTE spectral modeling 
of AT2017gfo \citep{2019_Watson+}.
However, as discussed in \myrefsec{subsec:abundance_constraints}, 
the inferred Sr mass density in the non-LTE case 
is at least one order of magnitude 
larger than that in the LTE case. As a result, 
the constraint on the wind active timescale discussed 
by \citet{2022_Perego+} becomes less stringent.

Similarly, \citet{2026_Sneppen+_a} also discussed
the mass-ejection timescale in terms of the elemental abundance,
focusing on the enhancement of the He abundance in the ejecta.
They pointed out that the post-merger winds from NS remnants 
with their lifetimes longer than $\sim \qty{100}{ms}$
make the ejected material too enriched with He, 
resulting in the too strong $\qty{1}{\mu m}$ feature. 
Based on their numerical simulations, 
they constrained the NS remnant lifetime in GW170817 
is shorter than \qtyrange{20}{30}{ms}.

However, we argue that such short-lived models cannot reproduce 
the observed light curve of AT2017gfo due to 
the insufficient total ejecta mass.
The total ejecta masses in such short-lived models are 
at most around \qty{0.01}{M_{\odot}},
whereas that in GW170817 estimated from the light curve is \qtyrange{0.03}{0.05}{M_{\odot}} \citep[e.g.,][]{2018_Waxman+,2020_Hotokezaka&Nakar}.

In summary, for the light curve, 
a large ejecta mass (\qtyrange{0.03}{0.05}{M_{\odot}}) is necessary, suggesting
the significant post-merger mass ejection.
For the spectral features, our modeling shows that 
the ejecta around $v = 0.15 \, c$ 
show the compositions consistent with a relatively low electron fraction
($Y_\mathrm{e} \lesssim 0.35$) and a relatively low entropy 
($s \lesssim 30\, k_{\rm B}/\mathrm{nucleon}$).
Therefore, we suggest that a significant mass ejection 
in GW170817
occurs under a relatively low electron fraction and low entropy condition, as compared with
the expectation from the currently available long-lived BNS merger models.

This may imply that the post-merger mass ejection takes place in a shorter timescale than the simulations (with viscous hydrodynamics modeling) before $Y_\mathrm{e}$ becomes too high.
Interestingly, recent long-term magneto-hydrodynamics simulations 
of BNS mergers show such a trend \citep{2023_Kiuchi+,2024_Kiuchi+}:
a significant mass ejection happens 
in a relatively short timescale,
which results in the mass ejection with a somewhat lower $Y_\mathrm{e}$ mass ejection.
Detailed comparison between our constraints and 
such realistic simulations
will be interesting to establish 
the physical picture of the mass ejection.

%%%%%%%%%%%%%%%%%%%%%%%%%%%%%%%%%%%%%%%%
% Section:
%%%%%%%%%%%%%%%%%%%%%%%%%%%%%%%%%%%%%%%%

\section{Conclusions} \label{sec:conclusions}

In this work, we develop non-LTE ionization models
to constrain elemental abundances of He and Sr,
which are considered to be the origin of  
the prominent \qty{1}{\mu m} feature
in the early-phase spectra of AT2017gfo.
We show that the level populations of first excited states of \ionname{He}{I}
are strongly dependent on temperature,
as also shown in \citet{2023_Tarumi+} and \citet{2024_Sneppen+_c}.
Non-LTE effect is also crucial for the ionization states of Sr.
The ionization by high-energy electrons results in the overionization of Sr II,
increasing the required mass of Sr to reproduce the observed feature.
By modeling the observed spectra, 
we find that about \qty{1}{\%} of He or \qtyrange{1}{10}{\%} of Sr 
(in mass fraction) is required in the ejecta moving at $v \sim 0.15c$.
This mass fraction of Sr agrees with that in the solar $r$-process pattern, supporting a hypothesis that BNS mergers are the main origin of $r$-process elements.

Based on our constraints, we discuss the nucleosynthetic conditions and 
mass-ejection mechanisms in GW170817.
In terms of the elemental composition, 
our constraints are consistent with
nucleosynthesis yields with a relatively low electron fraction ($Y_\mathrm{e} \lesssim 0.35$) 
and a relatively low entropy ($s \lesssim 30\, k_{\rm B}/\mathrm{nucleon}$).
Interestingly, for $Y_{\rm e} < 0.15$, the \qty{1}{\mu m} feature can be reproduced by He 
from the $\alpha$-decay of trans-Pb nuclei.
If this is confirmed, it gives an indirect signature for the production of elements 
beyond the third $r$-process peak.
To fully test this hypothesis, 
consistent modeling of He emission lines, spectral features of heavy elements, 
and multi-band light curves is necessary.

Comparing our constraints with currently available viscous hydrodynamics simulations,
a long-lived BNS merger model, which gives sufficient post-merger mass ejection to reproduce the bolometric luminosity of AT2017gfo, tend to overproduce both He and Sr.
Therefore, we suggest that a significant mass ejection in GW170817
should occur under a relatively low $Y_\mathrm{e}$ and low entropy condition, as compared with
the finding from the current long-lived BNS models.
This may imply that the post-merger mass ejection takes place in a shorter timescale, as demonstrated by recent magneto-hydrodynamics simulations.

It should be noted that our non-LTE modeling adopts several simplifications.
One factor is the approximated treatment of non-thermal ionization of Sr.
In particular, dielectric recombination is important for heavy elements,
but the recombination rates for heavy elements are still largely unavailable.
Also, recombination photons from other elements may affect the ionization degree.
Another possibly important factor is the time-dependent effects.
Although steady state approximation is adopted in our work,
recombination timescale in kilonova ejecta might become longer than
the dynamical timescale.
Compared with the case of supernova, this effect may be more important
as the density of kilonova ejecta is lower than that of supernova ejecta.
Our present work demonstrates that non-LTE spectral modeling gives unique constraints
on the mass-ejection mechanism of BNS merger.
To obtain more reliable constraints, more realistic non-LTE modeling including these effects is necessary.

\begin{acknowledgments}
We are grateful to the anonymous referee for carefully 
reading our manuscript and providing valuable comments.
We thank Aayush Arya, Maximilian Jacobi, Kenta Kiuchi, Albino Perego, 
Masaru Shibata, and Albert Sneppen for fruitful discussion.
This work is supported by the Grant-in-Aid for Scientific Research from 
JSPS (grant Nos. 21H04997, 23H00127, 23H04894, 23H04891, 
23H05432, 23H01172, 26K00732, 26K21726, and 26KJ0591)
and the JST FOREST Program (grant No. JPMJFR212Y).

\end{acknowledgments}

\section*{Data Availability}
The rate equation solver and the resulting dataset 
used in this work are available at 
\url{https://github.com/koya-chiba/KN-NLTE-He_and_Sr}.

\appendix
\section{Atomic processes} \label{app:sec:atomic_process}

Here we review formulation of atomic processes included in our non-LTE calculations for He.

\subsection{Radiative Bound-Bound Transition}
The excitation and de-excitation rates for the radiative bound-bound transition
between the state $l$ and the state $u$ are expressed as below:
\begin{equation}
  \begin{gathered}
    \label{eq:rad_bb_rates}
    R_{ul} = A_{ul} + B_{ul}\bar{J}, \\
    R_{lu} = B_{lu}\bar{J},
  \end{gathered}
\end{equation}
where $A_{ul}$, $B_{ul}$ and $B_{lu}$ are 
the Einstein coefficients, 
and $\bar{J}$ is the integrated mean intensity.
Under the Sobolev approximation \citep{1960_Sobolev},
$\bar{J}$ can be approximately evaluated as follows
\citep{1970_Castor,1978_Rybicki&Hummer}:
\begin{equation}
  \label{eq:sobolev_mean_intensity}
  \bar{J} = \beta_\mathrm{esc} W B_{\nu} 
  + (1-\beta_\mathrm{esc}) S,
\end{equation}
where $\beta_\mathrm{esc}$ is the escape probability 
of the line photons, $W$ is the geometric dilution factor,
$B_\nu$ is the Planck function, and
$S$ is the line source function.
Here $\beta_\mathrm{esc}$ is expressed using the Sobolev optical depth, $\tau_\mathrm{sob}$ (see \myrefeq{eq:def_tau_sob}):
\begin{equation}
  \label{eq:escape_probability}
  \beta_\mathrm{esc} 
  = \frac{1-e^{-\tau_\mathrm{sob}}}{\tau_\mathrm{sob}}.
\end{equation}
We adopt $W=0.5$ in this work as we only focus on 
the plasma condition 
at the photosphere.
The source function $S$ is given by 
\begin{equation}
  \label{eq:line_source_function}
  S = \frac{2h\nu^3}{c^2}\left(\frac{g_u}{g_l}\frac{n_l}{n_u}-1\right)^{-1}.
\end{equation}
Based on \myrefeq{eq:sobolev_mean_intensity},
we can rewrite \myrefeq{eq:rad_bb_rates} as follows:
\begin{equation}
  \begin{gathered}
    R_{ul}^\mathrm{eff} = \beta_{lu}\left(A_{ul}+B_{ul}J_\mathrm{inc}\right), \\
    R_{lu}^\mathrm{eff} = \beta_{lu}B_{lu}J_\mathrm{inc},
  \end{gathered}
\end{equation}
where $J_\mathrm{inc} = WB_\nu$ is interpreted 
as the incident mean intensity.

\subsection{Collisional Bound-Bound Transition}
Given the cross-section for electron-impact excitation 
from a lower term $l$ to an upper term $u$, $\sigma_{lu}$,
the Maxwellian-averaged $\sigma_{lu} v$ 
is defined as follows: 
\begin{equation}
  q_{lu} =  \int_{E_{ul}}^\infty \sigma_{lu}(E)v(E)f(E)\mathrm{d}E,
\end{equation}
where $v(E)$ is the velocity of a free electron with its energy $E$
and $f(E)$ is the Maxwellian distribution function.
By introducing the effective collision strength $\Upsilon_{lu}(T)$, 
$q_{lu}$ is expressed as 
\begin{equation}
  q_{lu} = \frac{\num{8.629132e-6}}{g_l \sqrt{T}}\Upsilon_{lu}(T)\exp\left(-\frac{E_{ul}}{kT}\right).
\end{equation}
Also, by the principle of detailed balance,
the Maxwellian-averaged $\sigma_{ul} v$ for electron-impact de-excitation  ($q_{ul}$) can be written as:
\begin{equation}
    q_{ul} = \frac{\num{8.629132e-6}}{g_u \sqrt{T}}\Upsilon_{lu}(T).
\end{equation}

By using $q_{ul}$ and $q_{lu}$ defined above, 
the excitation and de-excitation rates 
for collisional bound-bound transition are expressed as follows:
\begin{equation}
    \begin{gathered}
        C_{ul} = q_{ul} n_\mathrm{e}, \\
        C_{lu} = q_{lu} n_\mathrm{e}.
    \end{gathered}
\end{equation}

\subsection{Radiative Bound-Free/Free-Bound Transition}
Radiative bound-free transition (i.e., photoionization) rate can be expressed as 
\begin{equation}
  R_{l\kappa} = \int_{\nu_{l\kappa}}^{\infty} \frac{4\pi}{c}\frac{J_\nu}{h\nu} \sigma_\nu c\, \mathrm{d}\nu = 4\pi \int_{\nu_{l\kappa}}^{\infty} \frac{\sigma_\nu J_\nu}{h\nu} \mathrm{d}\nu,
\end{equation}
where $\sigma_{\nu}$ is the photoionization cross section.

The radiative recombination rate is expressed as 
\begin{equation}
  R_{\kappa l} = \alpha n_\mathrm{e},
\end{equation}
where $\alpha$ is the recombination rate coefficient derived 
from the Milne relation \citep{1979_Rybicki&Lightman}.

\subsection{Collisional Bound-Free/Free-Bound Transition}
Similarly to the case of collisional bound-bound transition, 
the Maxwellian-averaged $\sigma_{l\kappa} v$
can be evaluated as
\begin{equation}
  q_{l\kappa} = \frac{\num{7.20932e-4}}{I\sqrt{T}}
  \Upsilon_{l\kappa}(T) \exp\left(-\frac{I}{kT}\right),
\end{equation}
where $I$ is ionization potential in units of \unit{eV}, $T$ is temperature, 
and $\Upsilon_{l\kappa}(T)$ is the effective collision strength 
for collisional ionization.ß

Three-body recombination coefficient ($q_{\kappa i}$)
should satisfy the principle of detailed balance:
\begin{equation}
  (n_l)^{*} n_\mathrm{e} q_{l\kappa} 
  = (n_\kappa)^{*} n_\mathrm{e}^2 q_{\kappa l},
\end{equation}
where $(n_l)^{*}$ and $(n_\kappa)^{*}$ are
the thermal populations in the bound state and the ionized state, 
respectively. Thus,
\begin{equation}
  \begin{aligned}
    q_{\kappa l}
    &= \left\{\left(\frac{n_\kappa n_\mathrm{e}}{n_l}\right)^{*}\right\}^{-1} q_{l\kappa} \\
    &= \frac{g_l}{g_\kappa} \frac{\num{1.49281e-19}}{I T^2}
    \Upsilon_{l\kappa}(T) \exp\left(-\frac{E_l}{kT}\right), 
    \end{aligned}
\end{equation}
where $E_l$ is the excitation energy of the bound state $l$.
By using $q_{l\kappa}$ and $q_{\kappa l}$ defined above, 
the collisional bound-free/free-bound transition rates 
are expressed as follows:
\begin{equation}
    \begin{gathered}
        C_{l\kappa} = q_{l\kappa} n_\mathrm{e}, \\
        C_{\kappa l} = q_{\kappa l} n_\mathrm{e}^2.
    \end{gathered}
\end{equation}

Note that these processes are not very important in the kilonova ejecta
as the electron temperature in the ejecta 
is lower than the ionization energy of He.

\section{Density dependence of population fractions}
\label{app:density_dependence}

In Section \ref{subsec:abundance_constraints},
we show the results of non-LTE calculations for He and Sr 
by assuming the photospheric density at each epoch.
However, there is an uncertainty in the photospheric density 
due to the unknown density profile of the ejecta.
Here we discuss impacts of the adopted density to the population fractions.

\begin{figure}
    \centering    
    \plotone{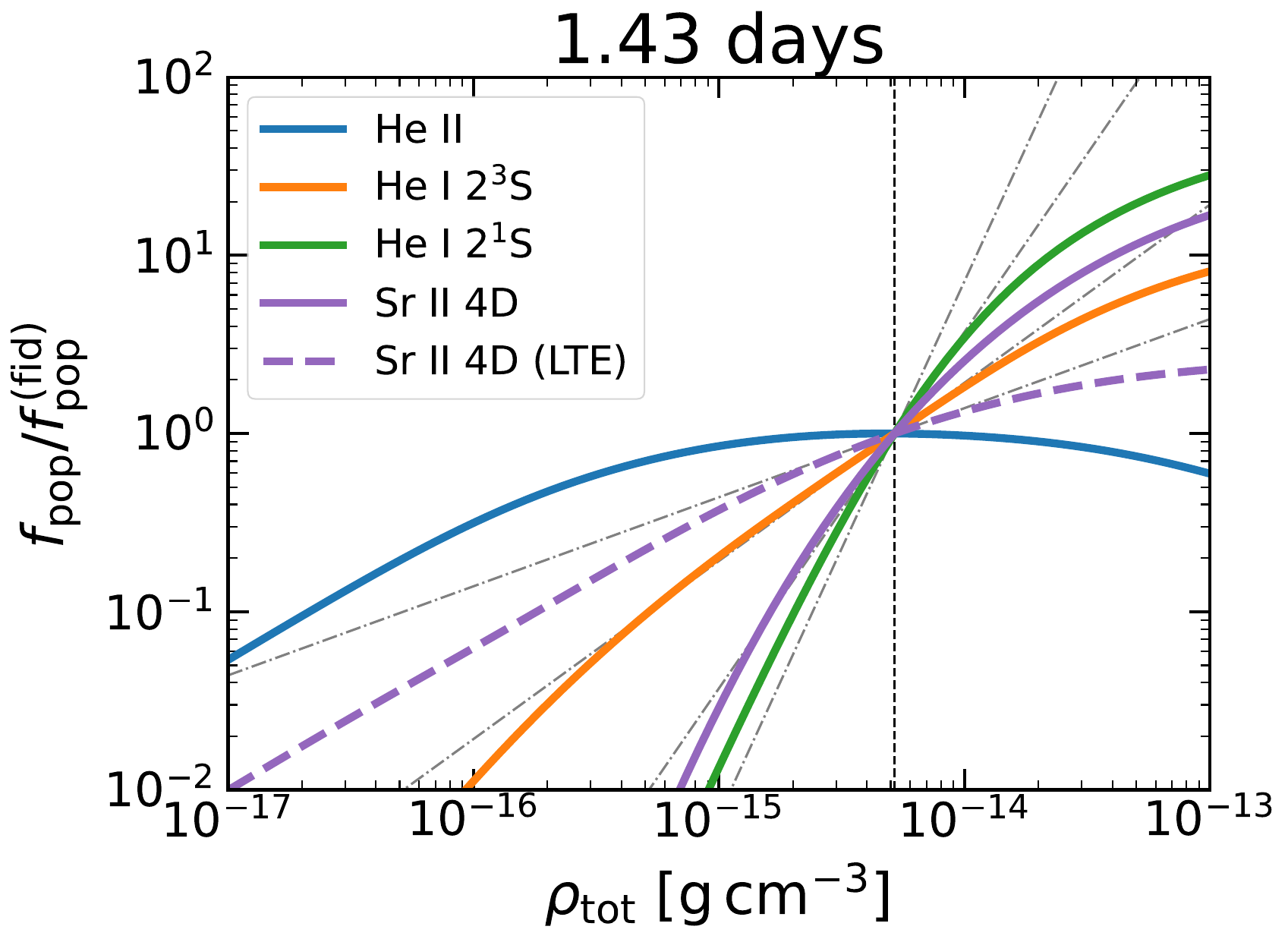}\\
    \medskip    
    \plotone{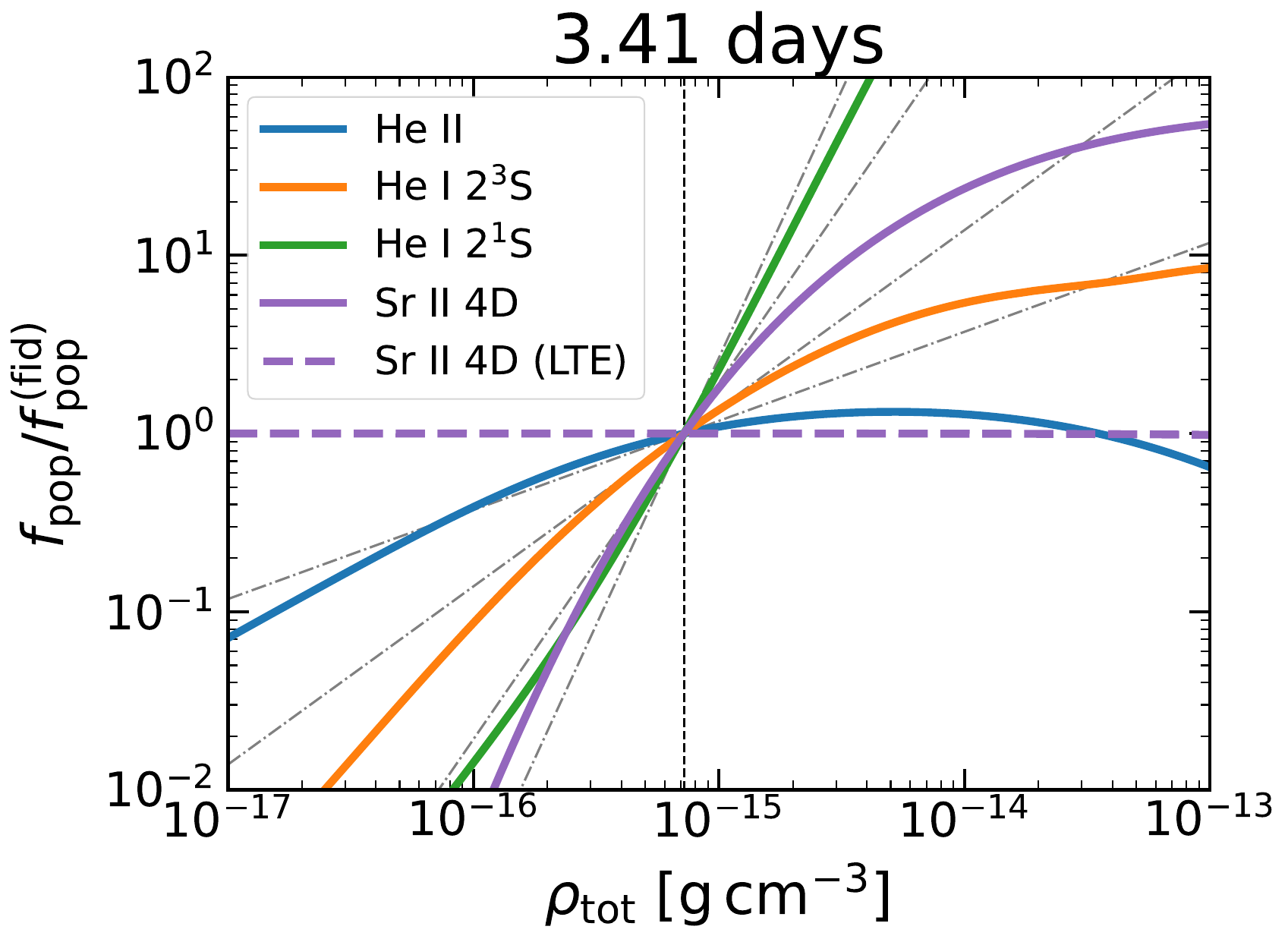}\\
    \caption{The normalized population fractions
    of \ionname{He}{II} (blue), 
    the \ionandterm{He}{I}{\atomterm{2}{1}{S}} 
    and \atomterm{2}{3}{S} states (orange and green), 
    and the \ionandterm{Sr}{II}{4D} state
    (purple, LTE and non-LTE cases)
    as a function of total mass density.
    The top and bottom panels correspond to 
    the conditions of the kilonova ejecta at $t = 1.43$ 
    and 3.41 days, respectively. 
    The fiducial total mass density at each epoch 
    is indicated by a vertical dashed line in the corresponding panel.
    For comparison, several slopes
    ($p = 0.5,\, 1.0,\, 2.0,$ and $3.0$ for $\rho_\mathrm{tot}^p$) 
    are shown as dashed-dotted lines. \label{fig:ion_frac}}
\end{figure}

We calculate population fractions ($f_\mathrm{pop}$)
by varying the total mass density ($\rho_\mathrm{tot}$).
The population fractions calculated with 
the fiducial total mass densities adopted in \myrefsec{subsec:abundance_constraints}
(see also \myreftab{tab:constraints_density}) 
are defined as the ``fiducial'' population fractions 
($f_\mathrm{pop}^\mathrm{(fid)}$).
Figure \ref{fig:ion_frac} shows
the normalized population fractions 
($f_\mathrm{pop}/f_\mathrm{pop}^\mathrm{(fid)}$)
of \ionname{He}{II}, 
the \ionandterm{He}{I}{\atomterm{2}{1}{S}} 
and \atomterm{2}{3}{S} states, 
and the \ionandterm{Sr}{II}{4D} state 
as a function of $\rho_\mathrm{tot}$.
Two panels correspond to the typical conditions of kilonova ejecta 
at $t= 1.43$ and \qty{3.41}{days}, respectively.
Namely, we assume the temperature of $T=4500$ 
and \qty{2800}{K}, respectively.
For the dependence of the He states, 
the calculations are performed with the He mass fractions 
explaining the depth of the \qty{1}{\mu m} feature: 
$X_{\rm He}= 10^{-1}$ and $10^{-2}$, respectively 
(the rest is assumed to be environmental elements, and their ionization degree is evaluated by solving Sr ionization).
For the dependence of the Sr states, the mass fraction of Sr is taken to be $X_{\rm Sr} = 10^{-3}$ and $10^{-1}$, respectively (similarly, the rest is assumed to be environmental elements).
The radioactive heating rate is evaluated according to Equation \ref{eq:def-heat_rate}.

Overall, the population fraction of the He I 2$^3$S state 
(orange line, the lower state of the 1.08 $\mu$m transition) show 
a weaker dependence on the density
as compared with the lower state of the Sr II line (solid purple line). 
At the relevant density range, the dominant ionization state of He is 
always singly ionized state.
The He I 2$^3$S state is well connected to the ground state of 
singly ionized He via the recombination, 
giving the dependence of the population roughly proportional to $n_{\rm e}$ (or $\rho$).
The He 2$^1$S state (the lower state of 2.06 $\mu$m line) shows a stronger dependence on the density
as the density also affects the optical depth of the transition between the ground state and the 2$^1$S state.

For Sr, the dominant ionization degree is higher (Sr IV-VI),
as shown Figure \ref{fig:pop_frac_Sr}.
The fraction of singly ionized Sr has a relatively strong dependence 
on the density ($\propto \rho^{2}$) when they are not dominant states.
The lower state of the \qty{1}{\mu m} feature (the Sr II 4D state) is well 
connected to the ground state of singly ionized Sr.
As a result, the required mass density of Sr for the \qty{1}{\mu m} feature 
has a stronger density dependence than that of He.

\section{Inclination dependence of the mass-density profiles}
\label{app:rho_profiles_2D}
\begin{figure}
\plotone{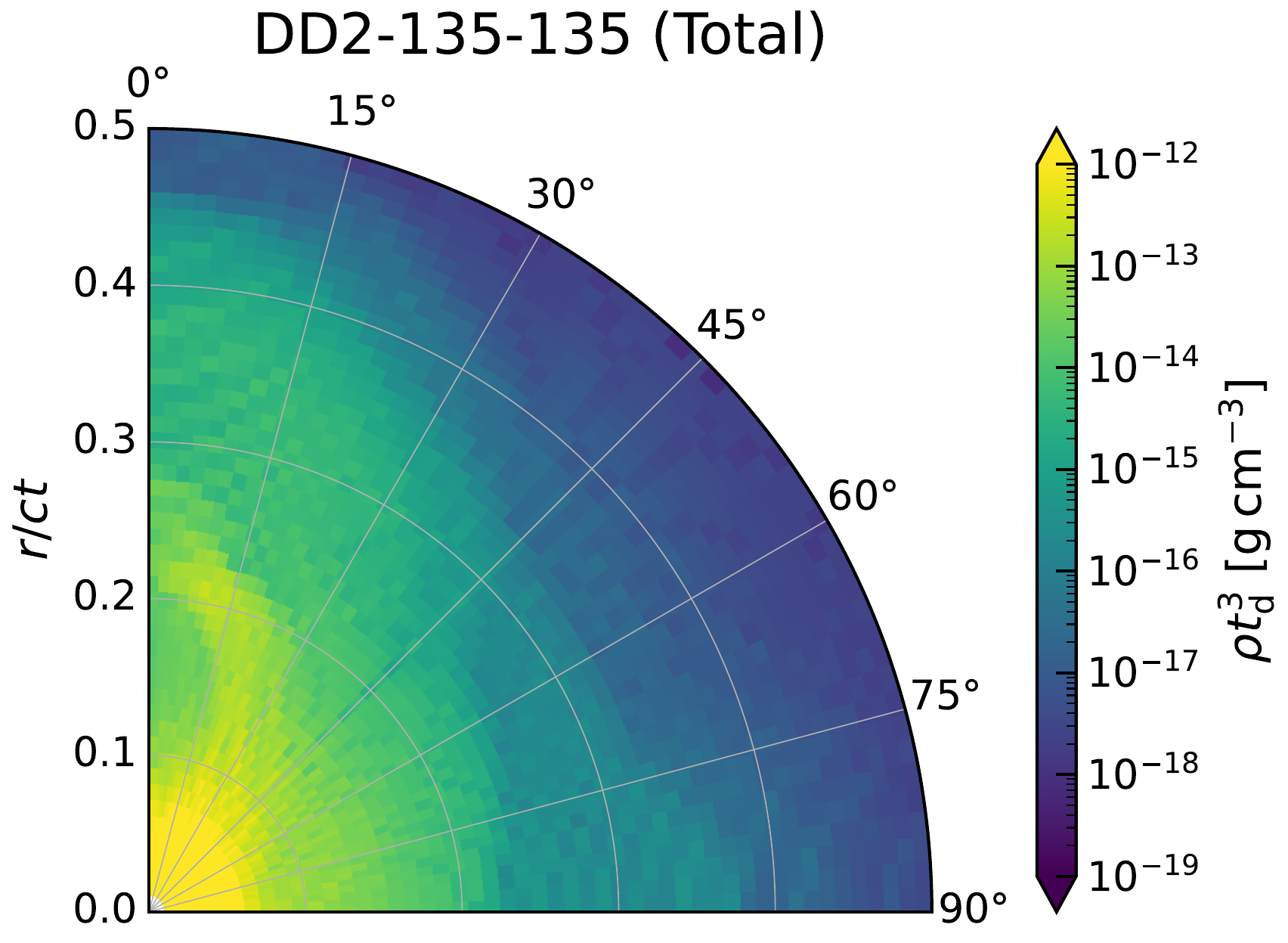}\\
\medskip
\plotone{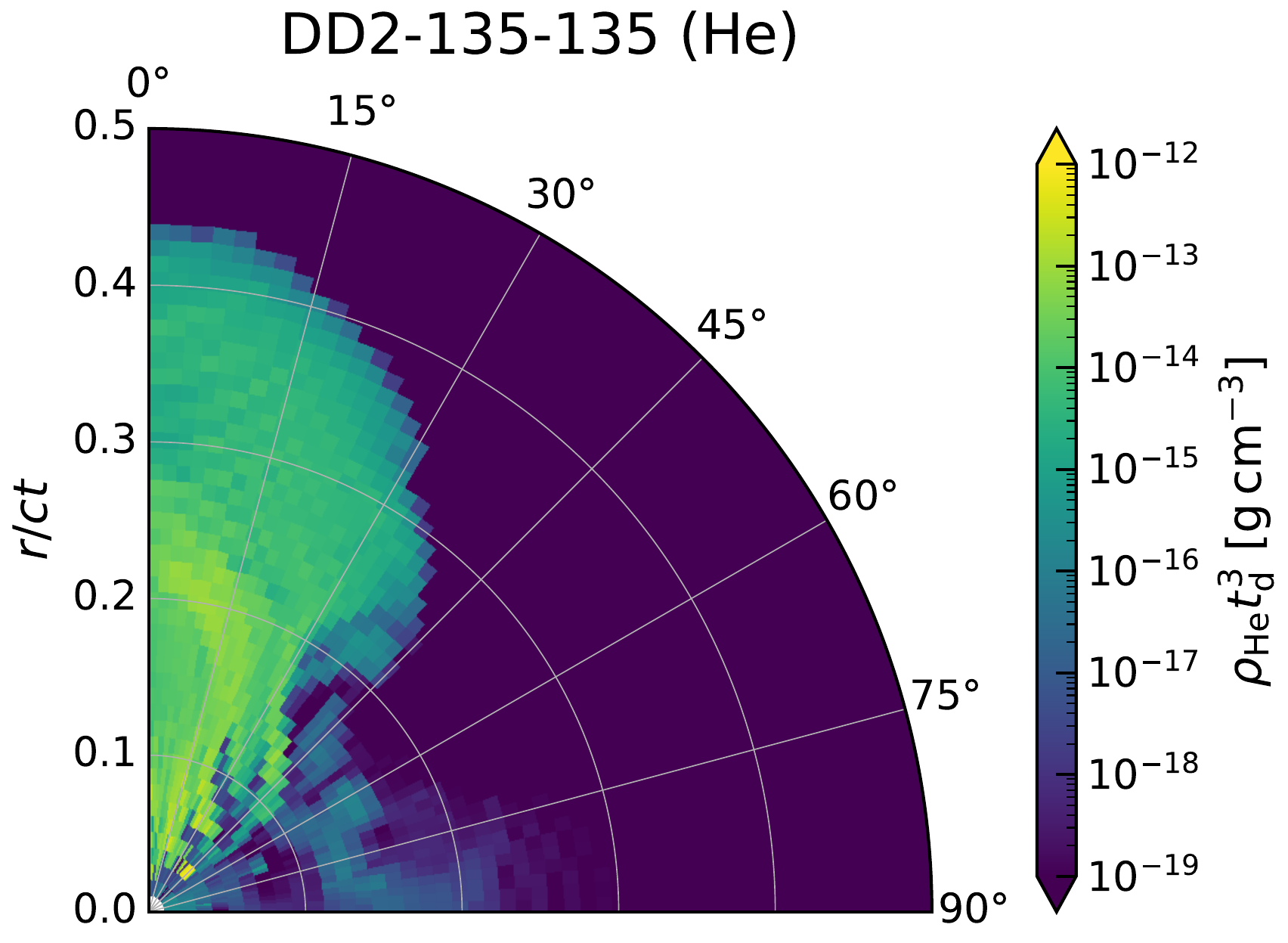}\\
\medskip
\plotone{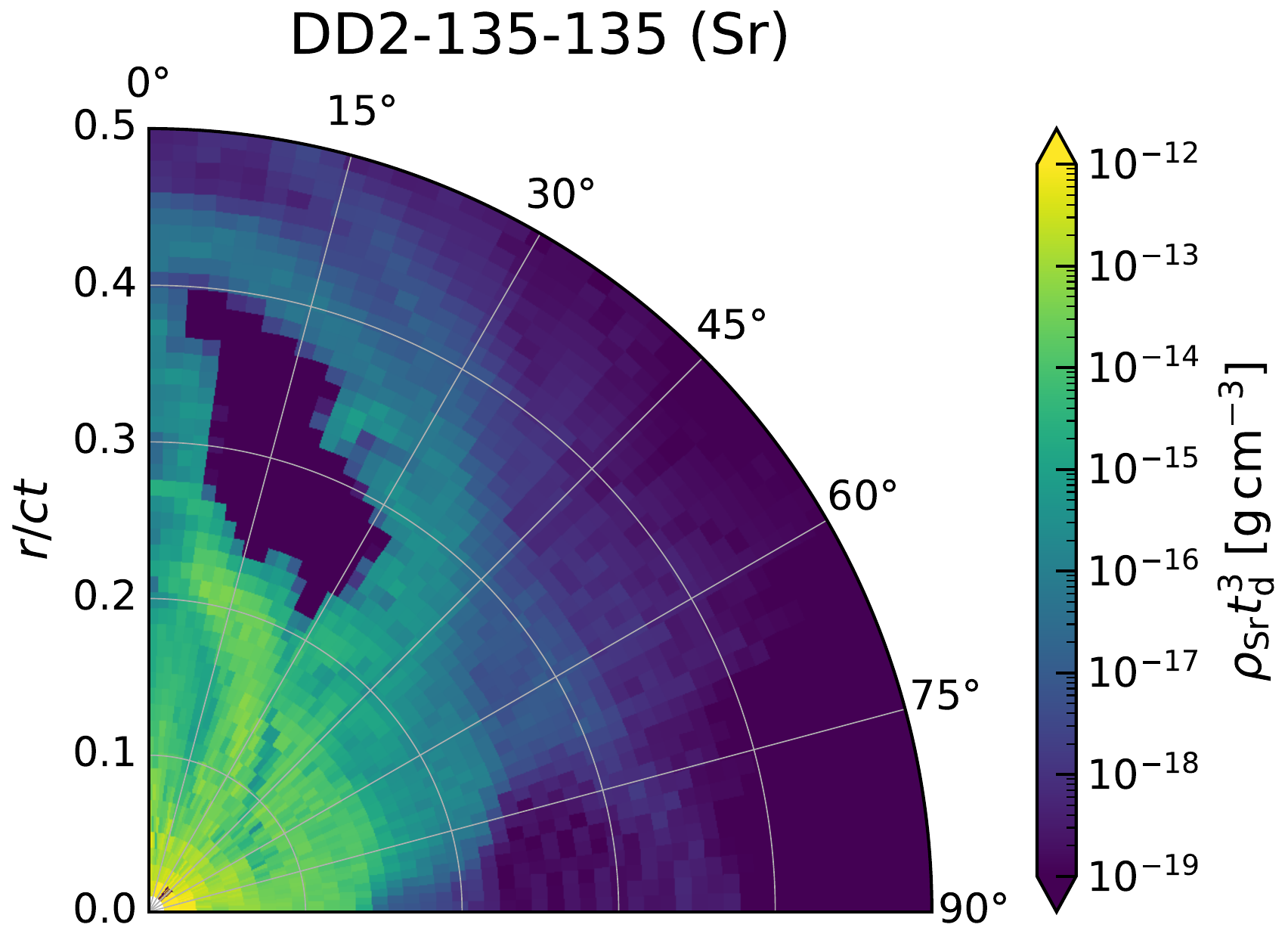}
\caption{2D mass-density profiles of the DD2-135-135 model (top: total mass density, middle: mass density of He, and bottom: mass density of Sr). The density is scaled at $t= \qty{1}{day}$.
\label{fig:rho_profiles_2D}}
\end{figure}

In \myrefsec{subsec:mass_ejection}, 
we compare our constraints with 
the spherically averaged mass-density profiles 
of the DD2-135-135 model. Here we show the inclination dependence of the model.
\myreffig{fig:rho_profiles_2D} shows the 2D mass-density profiles of 
the DD2-135-135 model from \citet{2022_Kawaguchi+}.
The spatial distribution of ejecta (top panel) with the velocity of $v\sim0.15\,c$ 
is highly concentrated in the polar region because this component is 
mainly originated from the post-merger ejecta
expelled to the polar direction.
%and have a higher mass than dynamical ejecta. 
Moreover, due to its relatively high electron fraction ($Y_\mathrm{e}\gtrsim 0.4$),
the mass fraction of He and Sr in the post-merger component is large 
($X_\mathrm{He/Sr} \gtrsim 0.1$, see \myreffig{fig:comparison-X_He-X_Sr_DD2} 
and the discussion in \myrefsec{subsec:mass_ejection}).
As a result, the mass densities of both He and Sr are 
enhanced in the polar region (middle and bottom panels in \myreffig{fig:rho_profiles_2D}).

\begin{figure}
\plotone{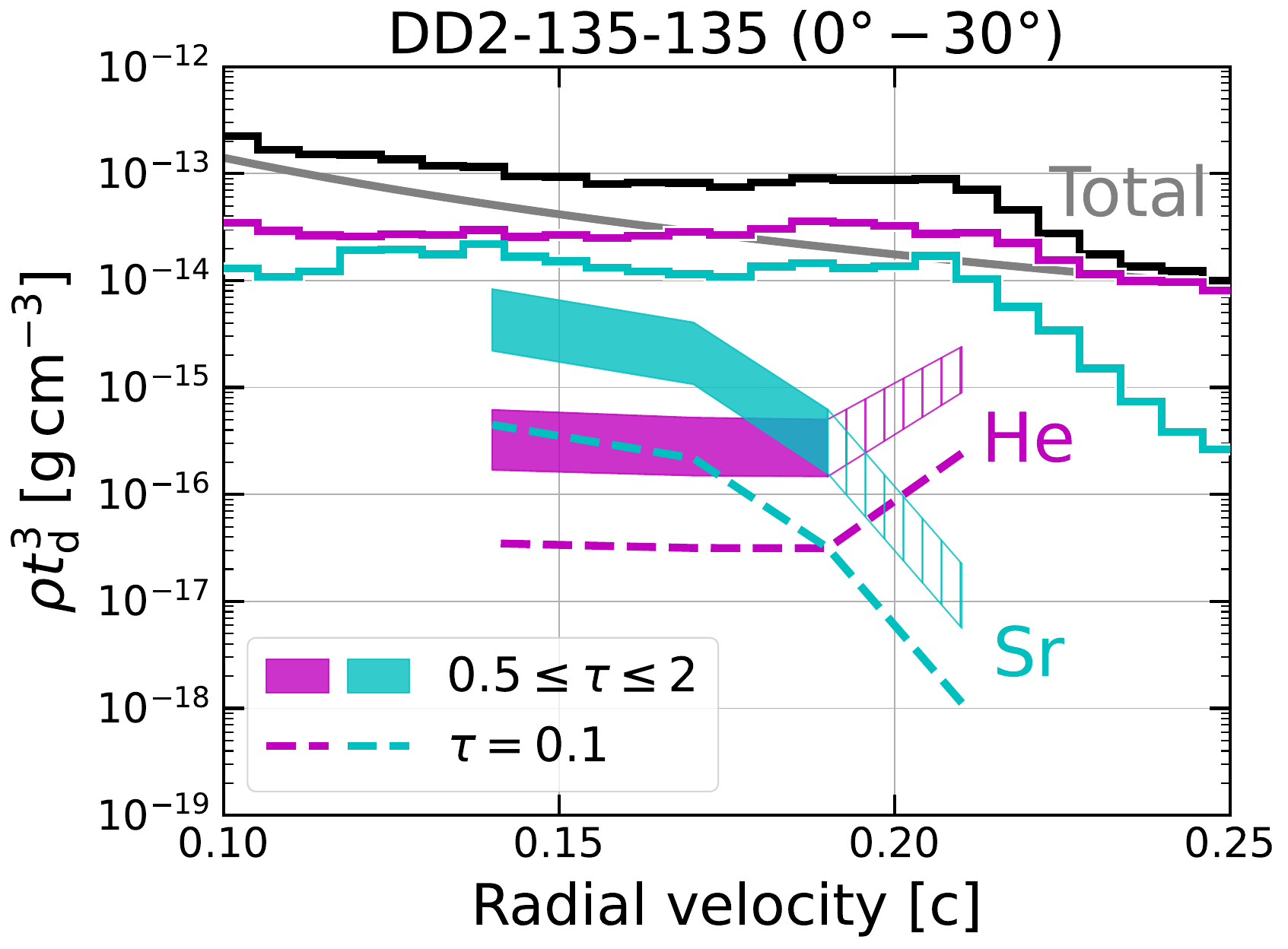}
\caption{Same as Figure \ref{fig:comparison-rho_profiles} 
         but with the model density profiles in the polar region
         (the inclination within $30^\circ$) 
         in the DD2-135-135 model.
         \label{fig:comparison-rho_profiles_polar}}
\end{figure}

\myreffig{fig:comparison-rho_profiles_polar} shows the comparison of our constraints 
with mass-density profiles of the DD2-135-135 model 
in the polar region (the inclination within $30^\circ$).
Due to the spatial enhancement of He and Sr in the polar region,
the polar profiles give higher mass densities of He and Sr 
than the spherically averaged profiles (see \myreffig{fig:comparison-rho_profiles}).
As a result, the discrepancy between our observational constraints and the numerical model
becomes more significant. 
Therefore, our conclusion that the DD2-135-135 model 
overproduces both He and Sr in line-forming region 
remains unchanged even by considering the inclination dependence of the mass-density profiles.

%% For this sample we use BibTeX plus aasjournalv7.bst to generate the

% \bibliography{ref_astr}
\bibliography{ref}

\begin{thebibliography}{}
\expandafter\ifx\csname natexlab\endcsname\relax\def\natexlab#1{#1}\fi
\providecommand{\url}[1]{\href{#1}{#1}}
\providecommand{\dodoi}[1]{doi:~\href{http://doi.org/#1}{\nolinkurl{#1}}}
\providecommand{\doeprint}[1]{\href{http://ascl.net/#1}{\nolinkurl{http://ascl.net/#1}}}
\providecommand{\doarXiv}[1]{\href{https://arxiv.org/abs/#1}{\nolinkurl{https://arxiv.org/abs/#1}}}

% type= article
\bibitem[{B.~P. {Abbott} {et~al.}(2017{\natexlab{a}}){Abbott}, {Abbott}, {Abbott}, {Acernese}, {Ackley}, {Adams}, {Adams}, {Addesso}, {Adhikari}, {Adya}, {Affeldt}, {Afrough}, {Agarwal}, {Agathos}, {Agatsuma}, {Aggarwal}, {Aguiar}, {Aiello}, {Ain}, {Ajith}, {Allen}, {Allen}, {Allocca}, {Altin}, {Amato}, {Ananyeva}, {Anderson}, {Anderson}, {Angelova}, {Antier}, {Appert}, {Arai}, {Araya}, {Areeda}, {Arnaud}, {Arun}, {Ascenzi}, {Ashton}, {Ast}, {Aston}, {Astone}, {Atallah}, {Aufmuth}, {Aulbert}, {AultONeal}, {Austin}, {Avila-Alvarez}, {Babak}, {Bacon}, {Bader}, {Bae}, {Baker}, {Baldaccini}, {Ballardin}, {Ballmer}, {Banagiri}, {Barayoga}, {Barclay}, {Barish}, {Barker}, {Barkett}, {Barone}, {Barr}, {Barsotti}, {Barsuglia}, {Barta}, {Barthelmy}, {Bartlett}, {Bartos}, {Bassiri}, {Basti}, {Batch}, {Bawaj}, {Bayley}, {Bazzan}, {B{\'e}csy}, {Beer}, {Bejger}, {Belahcene}, {Bell}, {Berger}, {Bergmann}, {Bero}, {Berry}, {Bersanetti}, {Bertolini}, {Betzwieser}, {Bhagwat}, {Bhandare}, {Bilenko}, {Billingsley}, {Billman}, {Birch}, {Birney}, {Birnholtz}, {Biscans}, {Biscoveanu}, {Bisht}, {Bitossi}, {Biwer}, {Bizouard}, {Blackburn}, {Blackman}, {Blair}, {Blair}, {Blair}, {Bloemen}, {Bock}, {Bode}, {Boer}, {Bogaert}, {Bohe}, {Bondu}, {Bonilla}, {Bonnand}, {Boom}, {Bork}, {Boschi}, {Bose}, {Bossie}, {Bouffanais}, {Bozzi}, {Bradaschia}, {Brady}, {Branchesi}, {Brau}, {Briant}, {Brillet}, {Brinkmann}, {Brisson}, {Brockill}, {Broida}, {Brooks}, {Brown}, {Brown}, {Brunett}, {Buchanan}, {Buikema}, {Bulik}, {Bulten}, {Buonanno}, {Buskulic}, {Buy}, {Byer}, {Cabero}, {Cadonati}, {Cagnoli}, {Cahillane}, {Calder{\'o}n Bustillo}, {Callister}, {Calloni}, {Camp}, {Canepa}, {Canizares}, {Cannon}, {Cao}, {Cao}, {Capano}, {Capocasa}, {Carbognani}, {Caride}, {Carney}, {Casanueva Diaz}, {Casentini}, {Caudill}, {Cavagli{\`a}}, {Cavalier}, {Cavalieri}, {Cella}, {Cepeda}, {Cerd{\'a}-Dur{\'a}n}, {Cerretani}, {Cesarini}, {Chamberlin}, {Chan}, {Chao}, {Charlton}, {Chase}, {Chassande-Mottin}, {Chatterjee}, {Chatziioannou}, {Cheeseboro}, {Chen}, {Chen}, {Chen}, {Cheng}, {Chia}, {Chincarini}, {Chiummo}, {Chmiel}, {Cho}, {Cho}, {Chow}, {Christensen}, {Chu}, {Chua}, {Chua}, {Chung}, {Chung}, \& {Ciani}}]{2017_Abbott+_b}
{Abbott}, B.~P., {Abbott}, R., {Abbott}, T.~D., {et~al.} 2017{\natexlab{a}}, \bibinfo{title}{{Multi-messenger Observations of a Binary Neutron Star Merger},} \apjl, 848, L12, \dodoi{10.3847/2041-8213/aa91c9}

% type= article
\bibitem[{B.~P. {Abbott} {et~al.}(2017{\natexlab{b}}){Abbott}, {Abbott}, {Abbott}, {Acernese}, {Ackley}, {Adams}, {Adams}, {Addesso}, {Adhikari}, {Adya}, {Affeldt}, {Afrough}, {Agarwal}, {Agathos}, {Agatsuma}, {Aggarwal}, {Aguiar}, {Aiello}, {Ain}, {Ajith}, {Allen}, {Allen}, {Allocca}, {Altin}, {Amato}, {Ananyeva}, {Anderson}, {Anderson}, {Angelova}, {Antier}, {Appert}, {Arai}, {Araya}, {Areeda}, {Arnaud}, {Arun}, {Ascenzi}, {Ashton}, {Ast}, {Aston}, {Astone}, {Atallah}, {Aufmuth}, {Aulbert}, {AultONeal}, {Austin}, {Avila-Alvarez}, {Babak}, {Bacon}, {Bader}, {Bae}, {Bailes}, {Baker}, {Baldaccini}, {Ballardin}, {Ballmer}, {Banagiri}, {Barayoga}, {Barclay}, {Barish}, {Barker}, {Barkett}, {Barone}, {Barr}, {Barsotti}, {Barsuglia}, {Barta}, {Barthelmy}, {Bartlett}, {Bartos}, {Bassiri}, {Basti}, {Batch}, {Bawaj}, {Bayley}, {Bazzan}, {B{\'e}csy}, {Beer}, {Bejger}, {Belahcene}, {Bell}, {Berger}, {Bergmann}, {Bernuzzi}, {Bero}, {Berry}, {Bersanetti}, {Bertolini}, {Betzwieser}, {Bhagwat}, {Bhandare}, {Bilenko}, {Billingsley}, {Billman}, {Birch}, {Birney}, {Birnholtz}, {Biscans}, {Biscoveanu}, {Bisht}, {Bitossi}, {Biwer}, {Bizouard}, {Blackburn}, {Blackman}, {Blair}, {Blair}, {Blair}, {Bloemen}, {Bock}, {Bode}, {Boer}, {Bogaert}, {Bohe}, {Bondu}, {Bonilla}, {Bonnand}, {Boom}, {Bork}, {Boschi}, {Bose}, {Bossie}, {Bouffanais}, {Bozzi}, {Bradaschia}, {Brady}, {Branchesi}, {Brau}, {Briant}, {Brillet}, {Brinkmann}, {Brisson}, {Brockill}, {Broida}, {Brooks}, {Brown}, {Brown}, {Brunett}, {Buchanan}, {Buikema}, {Bulik}, {Bulten}, {Buonanno}, {Buskulic}, {Buy}, {Byer}, {Cabero}, {Cadonati}, {Cagnoli}, {Cahillane}, {Calder{\'o}n Bustillo}, {Callister}, {Calloni}, {Camp}, {Canepa}, {Canizares}, {Cannon}, {Cao}, {Cao}, {Capano}, {Capocasa}, {Carbognani}, {Caride}, {Carney}, {Carullo}, {Casanueva Diaz}, {Casentini}, {Caudill}, {Cavagli{\`a}}, {Cavalier}, {Cavalieri}, {Cella}, {Cepeda}, {Cerd{\'a}-Dur{\'a}n}, {Cerretani}, {Cesarini}, {Chamberlin}, {Chan}, {Chao}, {Charlton}, {Chase}, {Chassande-Mottin}, {Chatterjee}, {Chatziioannou}, {Cheeseboro}, {Chen}, {Chen}, {Chen}, {Cheng}, {Chia}, {Chincarini}, {Chiummo}, {Chmiel}, {Cho}, {Cho}, {Chow}, {Christensen}, {Chu}, {Chua}, \& {Chua}}]{2017_Abbott+_a}
{Abbott}, B.~P., {Abbott}, R., {Abbott}, T.~D., {et~al.} 2017{\natexlab{b}}, \bibinfo{title}{{GW170817: Observation of Gravitational Waves from a Binary Neutron Star Inspiral},} \prl, 119, 161101, \dodoi{10.1103/PhysRevLett.119.161101}

% type= article
\bibitem[{D.~C. {Abbott} \& L.~B. {Lucy}(1985){Abbott} \& {Lucy}}]{1985_Abbott&Lucy}
{Abbott}, D.~C., \& {Lucy}, L.~B. 1985, \bibinfo{title}{{Multiline transfer and the dynamics of stellar winds.},} \apj, 288, 679, \dodoi{10.1086/162834}

% type= article
\bibitem[{Z.~L. {Andalman} {et~al.}(2026){Andalman}, {Fryer}, {Fontes}, {Mumpower}, \& {Wollaeger}}]{2026_Andalman+}
{Andalman}, Z.~L., {Fryer}, C.~L., {Fontes}, C.~J., {Mumpower}, M.~R., \& {Wollaeger}, R.~T. 2026, \bibinfo{title}{{Beta-Particle Transport and Thermalization in Kilonova Ejecta with Detailed Atomic Microphysics},} arXiv e-prints, arXiv:2607.01228, \dodoi{10.48550/arXiv.2607.01228}

% type= article
\bibitem[{A. {Arya} {et~al.}(2026){Arya}, {Damgaard}, {Sneppen}, {Dougan}, {Sim}, {Ballance}, \& {Watson}}]{2026_Arya+}
{Arya}, A., {Damgaard}, R., {Sneppen}, A., {et~al.} 2026, \bibinfo{title}{{Strontium and helium in the kilonova AT2017gfo: Origin of the 1{\ensuremath{\mu}}m feature constrained via NLTE calculations},} arXiv e-prints, arXiv:2604.05812, \dodoi{10.48550/arXiv.2604.05812}

% type= article
\bibitem[{J. {Barnes} {et~al.}(2016){Barnes}, {Kasen}, {Wu}, \& {Mart{\'\i}nez-Pinedo}}]{2016_Barnes+}
{Barnes}, J., {Kasen}, D., {Wu}, M.-R., \& {Mart{\'\i}nez-Pinedo}, G. 2016, \bibinfo{title}{{Radioactivity and Thermalization in the Ejecta of Compact Object Mergers and Their Impact on Kilonova Light Curves},} \apj, 829, 110, \dodoi{10.3847/0004-637X/829/2/110}

% type= article
\bibitem[{D.~R. {Bates} {et~al.}(1962){Bates}, {Kingston}, \& {McWhirter}}]{1962_Bates+}
{Bates}, D.~R., {Kingston}, A.~E., \& {McWhirter}, R.~W.~P. 1962, \bibinfo{title}{{Recombination Between Electrons and Atomic Ions. I. Optically Thin Plasmas},} Proceedings of the Royal Society of London Series A, 267, 297, \dodoi{10.1098/rspa.1962.0101}

% type= article
\bibitem[{S. {Bernuzzi} {et~al.}(2025){Bernuzzi}, {Magistrelli}, {Jacobi}, {Logoteta}, {Perego}, \& {Radice}}]{2025_Bernuzzi+}
{Bernuzzi}, S., {Magistrelli}, F., {Jacobi}, M., {et~al.} 2025, \bibinfo{title}{{Long-lived neutron star remnants from asymmetric binary neutron star mergers: element formation, kilonova signals and gravitational waves},} \mnras, 542, 256, \dodoi{10.1093/mnras/staf1147}

% type= article
\bibitem[{D. {Brethauer} {et~al.}(2026){Brethauer}, {Kasen}, {Margutti}, \& {Chornock}}]{2026_Brethauer+}
{Brethauer}, D., {Kasen}, D., {Margutti}, R., \& {Chornock}, R. 2026, \bibinfo{title}{{Nonthermal Ionization of Kilonova Ejecta: Observable Impacts},} \apj, 996, 64, \dodoi{10.3847/1538-4357/ae1b8d}

% type= article
\bibitem[{M. {Cantiello} {et~al.}(2018){Cantiello}, {Jensen}, {Blakeslee}, {Berger}, {Levan}, {Tanvir}, {Raimondo}, {Brocato}, {Alexander}, {Blanchard}, {Branchesi}, {Cano}, {Chornock}, {Covino}, {Cowperthwaite}, {D'Avanzo}, {Eftekhari}, {Fong}, {Fruchter}, {Grado}, {Hjorth}, {Holz}, {Lyman}, {Mandel}, {Margutti}, {Nicholl}, {Villar}, \& {Williams}}]{2018_Cantiello+}
{Cantiello}, M., {Jensen}, J.~B., {Blakeslee}, J.~P., {et~al.} 2018, \bibinfo{title}{{A Precise Distance to the Host Galaxy of the Binary Neutron Star Merger GW170817 Using Surface Brightness Fluctuations},} \apjl, 854, L31, \dodoi{10.3847/2041-8213/aaad64}

% type= article
\bibitem[{J.~I. {Castor}(1970){Castor}}]{1970_Castor}
{Castor}, J.~I. 1970, \bibinfo{title}{{Spectral line formation in Wolf-Rayet envelopes.},} \mnras, 149, 111, \dodoi{10.1093/mnras/149.2.111}

% type= article
\bibitem[{N. {Domoto} {et~al.}(2022){Domoto}, {Tanaka}, {Kato}, {Kawaguchi}, {Hotokezaka}, \& {Wanajo}}]{2022_Domoto+}
{Domoto}, N., {Tanaka}, M., {Kato}, D., {et~al.} 2022, \bibinfo{title}{{Lanthanide Features in Near-infrared Spectra of Kilonovae},} \apj, 939, 8, \dodoi{10.3847/1538-4357/ac8c36}

% type= article
\bibitem[{N. {Domoto} {et~al.}(2021){Domoto}, {Tanaka}, {Wanajo}, \& {Kawaguchi}}]{2021_Domoto+}
{Domoto}, N., {Tanaka}, M., {Wanajo}, S., \& {Kawaguchi}, K. 2021, \bibinfo{title}{{Signatures of r-process Elements in Kilonova Spectra},} \apj, 913, 26, \dodoi{10.3847/1538-4357/abf358}

% type= article
\bibitem[{D.~J. {Dougan} {et~al.}(2025){Dougan}, {McElroy}, {Ballance}, \& {Ramsbottom}}]{2025_Dougan+}
{Dougan}, D.~J., {McElroy}, N.~E., {Ballance}, C.~P., \& {Ramsbottom}, C.~A. 2025, \bibinfo{title}{{Strontium I, III, IV, and V: electron impact excitation data for Kilonovae and white dwarf diagnostic applications},} \mnras, 541, 367, \dodoi{10.1093/mnras/staf1013}

% type= article
\bibitem[{D. {Eichler} {et~al.}(1989){Eichler}, {Livio}, {Piran}, \& {Schramm}}]{1989_Eichler+}
{Eichler}, D., {Livio}, M., {Piran}, T., \& {Schramm}, D.~N. 1989, \bibinfo{title}{{Nucleosynthesis, neutrino bursts and {\ensuremath{\gamma}}-rays from coalescing neutron stars},} \nat, 340, 126, \dodoi{10.1038/340126a0}

% type= article
\bibitem[{R. {Fern{\'a}ndez} \& B.~D. {Metzger}(2013){Fern{\'a}ndez} \& {Metzger}}]{2013_Fernandez&Metzger}
{Fern{\'a}ndez}, R., \& {Metzger}, B.~D. 2013, \bibinfo{title}{{Delayed outflows from black hole accretion tori following neutron star binary coalescence},} \mnras, 435, 502, \dodoi{10.1093/mnras/stt1312}

% type= article
\bibitem[{S. {Fujibayashi} {et~al.}(2023){Fujibayashi}, {Kiuchi}, {Wanajo}, {Kyutoku}, {Sekiguchi}, \& {Shibata}}]{2023_Fujibayashi+_a}
{Fujibayashi}, S., {Kiuchi}, K., {Wanajo}, S., {et~al.} 2023, \bibinfo{title}{{Comprehensive Study of Mass Ejection and Nucleosynthesis in Binary Neutron Star Mergers Leaving Short-lived Massive Neutron Stars},} \apj, 942, 39, \dodoi{10.3847/1538-4357/ac9ce0}

% type= article
\bibitem[{S. {Fujibayashi} {et~al.}(2020){Fujibayashi}, {Wanajo}, {Kiuchi}, {Kyutoku}, {Sekiguchi}, \& {Shibata}}]{2020_Fujibayashi+_b}
{Fujibayashi}, S., {Wanajo}, S., {Kiuchi}, K., {et~al.} 2020, \bibinfo{title}{{Postmerger Mass Ejection of Low-mass Binary Neutron Stars},} \apj, 901, 122, \dodoi{10.3847/1538-4357/abafc2}

% type= article
\bibitem[{J.~H. {Gillanders} {et~al.}(2024){Gillanders}, {Sim}, {Smartt}, {Goriely}, \& {Bauswein}}]{2024_Gillanders+}
{Gillanders}, J.~H., {Sim}, S.~A., {Smartt}, S.~J., {Goriely}, S., \& {Bauswein}, A. 2024, \bibinfo{title}{{Modelling the spectra of the kilonova AT2017gfo - II. Beyond the photospheric epochs},} \mnras, 529, 2918, \dodoi{10.1093/mnras/stad3688}

% type= article
\bibitem[{J.~H. {Gillanders} \& S.~J. {Smartt}(2025){Gillanders} \& {Smartt}}]{2025_Gillanders&Smartt}
{Gillanders}, J.~H., \& {Smartt}, S.~J. 2025, \bibinfo{title}{{Analysis of the JWST spectra of the kilonova AT 2023vfi accompanying GRB 230307A},} \mnras, 538, 1663, \dodoi{10.1093/mnras/staf287}

% type= article
\bibitem[{J.~H. {Gillanders} {et~al.}(2022){Gillanders}, {Smartt}, {Sim}, {Bauswein}, \& {Goriely}}]{2022_Gillanders+}
{Gillanders}, J.~H., {Smartt}, S.~J., {Sim}, S.~A., {Bauswein}, A., \& {Goriely}, S. 2022, \bibinfo{title}{{Modelling the spectra of the kilonova AT2017gfo - I. The photospheric epochs},} \mnras, 515, 631, \dodoi{10.1093/mnras/stac1258}

% type= article
\bibitem[{S. {Hachinger} {et~al.}(2012){Hachinger}, {Mazzali}, {Taubenberger}, {Hillebrandt}, {Nomoto}, \& {Sauer}}]{2012_Hachinger+}
{Hachinger}, S., {Mazzali}, P.~A., {Taubenberger}, S., {et~al.} 2012, \bibinfo{title}{{How much H and He is 'hidden' in SNe Ib/c? - I. Low-mass objects},} \mnras, 422, 70, \dodoi{10.1111/j.1365-2966.2012.20464.x}

% type= article
\bibitem[{R.~P. {Harkness} {et~al.}(1987){Harkness}, {Wheeler}, {Margon}, {Downes}, {Kirshner}, {Uomoto}, {Barker}, {Cochran}, {Dinerstein}, {Garnett}, \& {Levreault}}]{1987_Harkness+}
{Harkness}, R.~P., {Wheeler}, J.~C., {Margon}, B., {et~al.} 1987, \bibinfo{title}{{The Early Spectral Phase of Type Ib Supernovae: Evidence for Helium},} \apj, 317, 355, \dodoi{10.1086/165283}

% type= article
\bibitem[{R.~D. {Hoffman} {et~al.}(1997){Hoffman}, {Woosley}, \& {Qian}}]{1997_Hoffman+}
{Hoffman}, R.~D., {Woosley}, S.~E., \& {Qian}, Y.~Z. 1997, \bibinfo{title}{{Nucleosynthesis in Neutrino-driven Winds. II. Implications for Heavy Element Synthesis},} \apj, 482, 951, \dodoi{10.1086/304181}

% type= article
\bibitem[{K. {Hotokezaka} \& E. {Nakar}(2020){Hotokezaka} \& {Nakar}}]{2020_Hotokezaka&Nakar}
{Hotokezaka}, K., \& {Nakar}, E. 2020, \bibinfo{title}{{Radioactive Heating Rate of r-process Elements and Macronova Light Curve},} \apj, 891, 152, \dodoi{10.3847/1538-4357/ab6a98}

% type= article
\bibitem[{K. {Hotokezaka} {et~al.}(2021){Hotokezaka}, {Tanaka}, {Kato}, \& {Gaigalas}}]{2021_Hotokezaka+}
{Hotokezaka}, K., {Tanaka}, M., {Kato}, D., \& {Gaigalas}, G. 2021, \bibinfo{title}{{Nebular emission from lanthanide-rich ejecta of neutron star merger},} \mnras, 506, 5863, \dodoi{10.1093/mnras/stab1975}

% type= article
\bibitem[{K. {Hotokezaka} {et~al.}(2023){Hotokezaka}, {Tanaka}, {Kato}, \& {Gaigalas}}]{2023_Hotokezaka+}
{Hotokezaka}, K., {Tanaka}, M., {Kato}, D., \& {Gaigalas}, G. 2023, \bibinfo{title}{{Tellurium emission line in kilonova AT 2017gfo},} \mnras, 526, L155, \dodoi{10.1093/mnrasl/slad128}

% type= article
\bibitem[{D. {Hutsemekers} \& J. {Surdej}(1990){Hutsemekers} \& {Surdej}}]{1990_Hutsemekers&Surdej}
{Hutsemekers}, D., \& {Surdej}, J. 1990, \bibinfo{title}{{Formation of P Cygni Line Profiles in Relativistically Expanding Atmospheres},} \apj, 361, 367, \dodoi{10.1086/169203}

% type= article
\bibitem[{M. {Jacobi} {et~al.}(2026){Jacobi}, {Magistrelli}, {Loffredo}, {Ricigliano}, {Chiesa}, {Bernuzzi}, {Perego}, \& {Arcones}}]{2026_Jacobi+}
{Jacobi}, M., {Magistrelli}, F., {Loffredo}, E., {et~al.} 2026, \bibinfo{title}{{$^{56}$Ni Production in Long-lived Binary Neutron Star Merger Remnants},} \apjl, 999, L16, \dodoi{10.3847/2041-8213/ae4104}

% type= article
\bibitem[{D.~J. {Jeffery}(1993){Jeffery}}]{1993_Jeffery}
{Jeffery}, D.~J. 1993, \bibinfo{title}{{The Relativistic Sobolev Method Applied to Homologously Expanding Atmospheres},} \apj, 415, 734, \dodoi{10.1086/173197}

% type= inproceedings
\bibitem[{D.~J. {Jeffery} \& D. {Branch}(1990){Jeffery} \& {Branch}}]{1990_Jeffery&Branch}
{Jeffery}, D.~J., \& {Branch}, D. 1990, \bibinfo{title}{{Analysis of Supernova Spectra},} in Supernovae, Jerusalem Winter School for Theoretical Physics, ed. J.~C. {Wheeler}, T.~{Piran}, \& S.~{Weinberg}, Vol.~6, 149

% type= article
\bibitem[{A. {Jerkstrand} {et~al.}(2026){Jerkstrand}, {Pognan}, {Banerjee}, {Sterling}, {Grumer}, {Ferguson}, {Butler}, {Gillanders}, {Smartt}, {Kawaguchi}, \& {Vilagos}}]{2026_Jerkstrand+}
{Jerkstrand}, A., {Pognan}, Q., {Banerjee}, S., {et~al.} 2026, \bibinfo{title}{{Infrared spectral signatures of light r-process elements in kilonovae},} \mnras, 548, stag733, \dodoi{10.1093/mnras/stag733}

% type= article
\bibitem[{O. {Just} {et~al.}(2023){Just}, {Vijayan}, {Xiong}, {Goriely}, {Soultanis}, {Bauswein}, {Guilet}, {Janka}, \& {Mart{\'\i}nez-Pinedo}}]{2023_Just+}
{Just}, O., {Vijayan}, V., {Xiong}, Z., {et~al.} 2023, \bibinfo{title}{{End-to-end Kilonova Models of Neutron Star Mergers with Delayed Black Hole Formation},} \apjl, 951, L12, \dodoi{10.3847/2041-8213/acdad2}

% type= article
\bibitem[{D. {Kasen} {et~al.}(2017){Kasen}, {Metzger}, {Barnes}, {Quataert}, \& {Ramirez-Ruiz}}]{2017_Kasen+}
{Kasen}, D., {Metzger}, B., {Barnes}, J., {Quataert}, E., \& {Ramirez-Ruiz}, E. 2017, \bibinfo{title}{{Origin of the heavy elements in binary neutron-star mergers from a gravitational-wave event},} \nat, 551, 80, \dodoi{10.1038/nature24453}

% type= article
\bibitem[{K. {Kawaguchi} {et~al.}(2023){Kawaguchi}, {Fujibayashi}, {Domoto}, {Kiuchi}, {Shibata}, \& {Wanajo}}]{2023_Kawaguchi+_b}
{Kawaguchi}, K., {Fujibayashi}, S., {Domoto}, N., {et~al.} 2023, \bibinfo{title}{{Kilonovae of binary neutron star mergers leading to short-lived remnant neutron star formation},} \mnras, 525, 3384, \dodoi{10.1093/mnras/stad2430}

% type= article
\bibitem[{K. {Kawaguchi} {et~al.}(2022){Kawaguchi}, {Fujibayashi}, {Hotokezaka}, {Shibata}, \& {Wanajo}}]{2022_Kawaguchi+}
{Kawaguchi}, K., {Fujibayashi}, S., {Hotokezaka}, K., {Shibata}, M., \& {Wanajo}, S. 2022, \bibinfo{title}{{Electromagnetic Counterparts of Binary-neutron-star Mergers Leading to a Strongly Magnetized Long-lived Remnant Neutron Star},} \apj, 933, 22, \dodoi{10.3847/1538-4357/ac6ef7}

% type= article
\bibitem[{K. {Kawaguchi} {et~al.}(2021){Kawaguchi}, {Fujibayashi}, {Shibata}, {Tanaka}, \& {Wanajo}}]{2021_Kawaguchi+}
{Kawaguchi}, K., {Fujibayashi}, S., {Shibata}, M., {Tanaka}, M., \& {Wanajo}, S. 2021, \bibinfo{title}{{A Low-mass Binary Neutron Star: Long-term Ejecta Evolution and Kilonovae with Weak Blue Emission},} \apj, 913, 100, \dodoi{10.3847/1538-4357/abf3bc}

% type= article
\bibitem[{K. {Kawaguchi} {et~al.}(2018){Kawaguchi}, {Shibata}, \& {Tanaka}}]{2018_Kawaguchi+}
{Kawaguchi}, K., {Shibata}, M., \& {Tanaka}, M. 2018, \bibinfo{title}{{Radiative Transfer Simulation for the Optical and Near-infrared Electromagnetic Counterparts to GW170817},} \apjl, 865, L21, \dodoi{10.3847/2041-8213/aade02}

% type= article
\bibitem[{K. {Kawaguchi} {et~al.}(2020){Kawaguchi}, {Shibata}, \& {Tanaka}}]{2020_Kawaguchi+}
{Kawaguchi}, K., {Shibata}, M., \& {Tanaka}, M. 2020, \bibinfo{title}{{Diversity of Kilonova Light Curves},} \apj, 889, 171, \dodoi{10.3847/1538-4357/ab61f6}

% type= article
\bibitem[{A. {Kitamura} {et~al.}(2025){Kitamura}, {Kawaguchi}, {Tanaka}, \& {Fujibayashi}}]{2025_Kitamura+}
{Kitamura}, A., {Kawaguchi}, K., {Tanaka}, M., \& {Fujibayashi}, S. 2025, \bibinfo{title}{{Linking Analytic Light-curve Models to Physical Properties of Kilonovae},} \apj, 982, 97, \dodoi{10.3847/1538-4357/adb62c}

% type= article
\bibitem[{K. {Kiuchi} {et~al.}(2023){Kiuchi}, {Fujibayashi}, {Hayashi}, {Kyutoku}, {Sekiguchi}, \& {Shibata}}]{2023_Kiuchi+}
{Kiuchi}, K., {Fujibayashi}, S., {Hayashi}, K., {et~al.} 2023, \bibinfo{title}{{Self-Consistent Picture of the Mass Ejection from a One Second Long Binary Neutron Star Merger Leaving a Short-Lived Remnant in a General-Relativistic Neutrino-Radiation Magnetohydrodynamic Simulation},} \prl, 131, 011401, \dodoi{10.1103/PhysRevLett.131.011401}

% type= article
\bibitem[{K. {Kiuchi} {et~al.}(2024){Kiuchi}, {Reboul-Salze}, {Shibata}, \& {Sekiguchi}}]{2024_Kiuchi+}
{Kiuchi}, K., {Reboul-Salze}, A., {Shibata}, M., \& {Sekiguchi}, Y. 2024, \bibinfo{title}{{A large-scale magnetic field produced by a solar-like dynamo in binary neutron star mergers},} Nature Astronomy, 8, 298, \dodoi{10.1038/s41550-024-02194-y}

% type= article
\bibitem[{C. {Kozma} \& C. {Fransson}(1992){Kozma} \& {Fransson}}]{1992_Kozma&Fransson}
{Kozma}, C., \& {Fransson}, C. 1992, \bibinfo{title}{{Gamma-Ray Deposition and Nonthermal Excitation in Supernovae},} \apj, 390, 602, \dodoi{10.1086/171311}

% type= misc
\bibitem[{A. Kramida {et~al.}(2023)Kramida, {Yu.~Ralchenko}, Reader, \& {and NIST ASD Team}}]{NIST_ASD}
Kramida, A., {Yu.~Ralchenko}, Reader, J., \& {and NIST ASD Team}. 2023, {NIST Atomic Spectra Database (ver. 5.11), [Online]. Available: {\tt{https://physics.nist.gov/asd}} [2024, May 6]. National Institute of Standards and Technology, Gaithersburg, MD.}

% type= article
\bibitem[{S.~R. {Kulkarni}(2005){Kulkarni}}]{2005_Kulkarni}
{Kulkarni}, S.~R. 2005, \bibinfo{title}{{Modeling Supernova-like Explosions Associated with Gamma-ray Bursts with Short Durations},} arXiv e-prints, astro, \dodoi{10.48550/arXiv.astro-ph/0510256}

% type= article
\bibitem[{A.~J. {Levan} {et~al.}(2024){Levan}, {Gompertz}, {Salafia}, {Bulla}, {Burns}, {Hotokezaka}, {Izzo}, {Lamb}, {Malesani}, {Oates}, {Ravasio}, {Rouco Escorial}, {Schneider}, {Sarin}, {Schulze}, {Tanvir}, {Ackley}, {Anderson}, {Brammer}, {Christensen}, {Dhillon}, {Evans}, {Fausnaugh}, {Fong}, {Fruchter}, {Fryer}, {Fynbo}, {Gaspari}, {Heintz}, {Hjorth}, {Kennea}, {Kennedy}, {Laskar}, {Leloudas}, {Mandel}, {Martin-Carrillo}, {Metzger}, {Nicholl}, {Nugent}, {Palmerio}, {Pugliese}, {Rastinejad}, {Rhodes}, {Rossi}, {Saccardi}, {Smartt}, {Stevance}, {Tohuvavohu}, {van der Horst}, {Vergani}, {Watson}, {Barclay}, {Bhirombhakdi}, {Breedt}, {Breeveld}, {Brown}, {Campana}, {Chrimes}, {D'Avanzo}, {D'Elia}, {De Pasquale}, {Dyer}, {Galloway}, {Garbutt}, {Green}, {Hartmann}, {Jakobsson}, {Kerry}, {Kouveliotou}, {Langeroodi}, {Le Floc'h}, {Leung}, {Littlefair}, {Munday}, {O'Brien}, {Parsons}, {Pelisoli}, {Sahman}, {Salvaterra}, {Sbarufatti}, {Steeghs}, {Tagliaferri}, {Th{\"o}ne}, {de Ugarte Postigo}, \& {Kann}}]{2024_Levan+}
{Levan}, A.~J., {Gompertz}, B.~P., {Salafia}, O.~S., {et~al.} 2024, \bibinfo{title}{{Heavy-element production in a compact object merger observed by JWST},} \nat, 626, 737, \dodoi{10.1038/s41586-023-06759-1}

% type= article
\bibitem[{L.-X. {Li} \& B. {Paczy{\'n}ski}(1998){Li} \& {Paczy{\'n}ski}}]{1998_Li&Paczyski}
{Li}, L.-X., \& {Paczy{\'n}ski}, B. 1998, \bibinfo{title}{{Transient Events from Neutron Star Mergers},} \apjl, 507, L59, \dodoi{10.1086/311680}

% type= article
\bibitem[{J. {Lippuner} \& L.~F. {Roberts}(2015){Lippuner} \& {Roberts}}]{2015_Lippuner&Roberts}
{Lippuner}, J., \& {Roberts}, L.~F. 2015, \bibinfo{title}{{r-process Lanthanide Production and Heating Rates in Kilonovae},} \apj, 815, 82, \dodoi{10.1088/0004-637X/815/2/82}

% type= article
\bibitem[{L.~B. {Lucy}(1991){Lucy}}]{1991_Lucy}
{Lucy}, L.~B. 1991, \bibinfo{title}{{Nonthermal Excitation of Helium in Type Ib Supernovae},} \apj, 383, 308, \dodoi{10.1086/170787}

% type= article
\bibitem[{L.~B. {Lucy}(1999){Lucy}}]{1999_Lucy_b}
{Lucy}, L.~B. 1999, \bibinfo{title}{{Improved Monte Carlo techniques for the spectral synthesis of supernovae},} \aap, 345, 211

% type= article
\bibitem[{B.~D. {Metzger} {et~al.}(2010){Metzger}, {Mart{\'\i}nez-Pinedo}, {Darbha}, {Quataert}, {Arcones}, {Kasen}, {Thomas}, {Nugent}, {Panov}, \& {Zinner}}]{2010_Metzger+}
{Metzger}, B.~D., {Mart{\'\i}nez-Pinedo}, G., {Darbha}, S., {et~al.} 2010, \bibinfo{title}{{Electromagnetic counterparts of compact object mergers powered by the radioactive decay of r-process nuclei},} \mnras, 406, 2650, \dodoi{10.1111/j.1365-2966.2010.16864.x}

% type= article
\bibitem[{B.~S. {Meyer} {et~al.}(1998){Meyer}, {Krishnan}, \& {Clayton}}]{1998_Meyer+}
{Meyer}, B.~S., {Krishnan}, T.~D., \& {Clayton}, D.~D. 1998, \bibinfo{title}{{Theory of Quasi-Equilibrium Nucleosynthesis and Applications to Matter Expanding from High Temperature and Density},} \apj, 498, 808, \dodoi{10.1086/305562}

% type= book
\bibitem[{D. {Mihalas}(1978){Mihalas}}]{1978_Mihalas}
{Mihalas}, D. 1978, {Stellar atmospheres}

% type= article
\bibitem[{L.~P. {Mulholland} {et~al.}(2024){Mulholland}, {McElroy}, {McNeill}, {Sim}, {Ballance}, \& {Ramsbottom}}]{2024_Mulholland+_a}
{Mulholland}, L.~P., {McElroy}, N.~E., {McNeill}, F.~L., {et~al.} 2024, \bibinfo{title}{{New radiative and collisional atomic data for Sr II and Y II with application to Kilonova modelling},} \mnras, 532, 2289, \dodoi{10.1093/mnras/stae1615}

% type= article
\bibitem[{L.~P. {Mulholland} {et~al.}(2026){Mulholland}, {Ramsbottom}, {Ballance}, {Sneppen}, \& {Sim}}]{2026_Mulholland+}
{Mulholland}, L.~P., {Ramsbottom}, C.~A., {Ballance}, C.~P., {Sneppen}, A., \& {Sim}, S.~A. 2026, \bibinfo{title}{{Electron impact excitation of Te IV and V and level resolved R-matrix photoionization of Te I─IV with application to modelling of AT2017gfo},} \mnras, 546, stag237, \dodoi{10.1093/mnras/stag237}

% type= article
\bibitem[{S. {Nahar}(2020){Nahar}}]{2020_Nahar}
{Nahar}, S. 2020, \bibinfo{title}{{Database NORAD-Atomic-Data for Atomic Processes in Plasma},} Atoms, 8, 68, \dodoi{10.3390/atoms8040068}

% type= article
\bibitem[{S.~N. {Nahar}(2010){Nahar}}]{2010_Nahar}
{Nahar}, S.~N. 2010, \bibinfo{title}{{Photoionization and electron-ion recombination of He I},} \na, 15, 417, \dodoi{10.1016/j.newast.2009.11.010}

% type= article
\bibitem[{S.~N. {Nahar}(2021){Nahar}}]{2021_Nahar}
{Nahar}, S.~N. 2021, \bibinfo{title}{{Photoionization and Electron-Ion Recombination of n = 1 to Very High n-Values of Hydrogenic Ions},} Atoms, 9, 73, \dodoi{10.3390/atoms9040073}

% type= article
\bibitem[{S.~N. {Nahar} \& G. {Hinojosa-Aguirre}(2024){Nahar} \& {Hinojosa-Aguirre}}]{2024_Nahar&Hinojosa-Aguirre}
{Nahar}, S.~N., \& {Hinojosa-Aguirre}, G. 2024, \bibinfo{title}{{Enhancement of the NORAD-Atomic-Data Database in Plasma},} Atoms, 12, 22, \dodoi{10.3390/atoms12040022}

% type= article
\bibitem[{R. {Narayan} {et~al.}(1992){Narayan}, {Paczynski}, \& {Piran}}]{1992_Narayan+}
{Narayan}, R., {Paczynski}, B., \& {Piran}, T. 1992, \bibinfo{title}{{Gamma-Ray Bursts as the Death Throes of Massive Binary Stars},} \apjl, 395, L83, \dodoi{10.1086/186493}

% type= article
\bibitem[{A. {Perego} {et~al.}(2022){Perego}, {Vescovi}, {Fiore}, {Chiesa}, {Vogl}, {Benetti}, {Bernuzzi}, {Branchesi}, {Cappellaro}, {Cristallo}, {Fl{\"o}rs}, {Kerzendorf}, \& {Radice}}]{2022_Perego+}
{Perego}, A., {Vescovi}, D., {Fiore}, A., {et~al.} 2022, \bibinfo{title}{{Production of Very Light Elements and Strontium in the Early Ejecta of Neutron Star Mergers},} \apj, 925, 22, \dodoi{10.3847/1538-4357/ac3751}

% type= article
\bibitem[{E. {Pian} {et~al.}(2017){Pian}, {D'Avanzo}, {Benetti}, {Branchesi}, {Brocato}, {Campana}, {Cappellaro}, {Covino}, {D'Elia}, {Fynbo}, {Getman}, {Ghirlanda}, {Ghisellini}, {Grado}, {Greco}, {Hjorth}, {Kouveliotou}, {Levan}, {Limatola}, {Malesani}, {Mazzali}, {Melandri}, {M{\o}ller}, {Nicastro}, {Palazzi}, {Piranomonte}, {Rossi}, {Salafia}, {Selsing}, {Stratta}, {Tanaka}, {Tanvir}, {Tomasella}, {Watson}, {Yang}, {Amati}, {Antonelli}, {Ascenzi}, {Bernardini}, {Bo{\"e}r}, {Bufano}, {Bulgarelli}, {Capaccioli}, {Casella}, {Castro-Tirado}, {Chassande-Mottin}, {Ciolfi}, {Copperwheat}, {Dadina}, {De Cesare}, {di Paola}, {Fan}, {Gendre}, {Giuffrida}, {Giunta}, {Hunt}, {Israel}, {Jin}, {Kasliwal}, {Klose}, {Lisi}, {Longo}, {Maiorano}, {Mapelli}, {Masetti}, {Nava}, {Patricelli}, {Perley}, {Pescalli}, {Piran}, {Possenti}, {Pulone}, {Razzano}, {Salvaterra}, {Schipani}, {Spera}, {Stamerra}, {Stella}, {Tagliaferri}, {Testa}, {Troja}, {Turatto}, {Vergani}, \& {Vergani}}]{2017_Pian+}
{Pian}, E., {D'Avanzo}, P., {Benetti}, S., {et~al.} 2017, \bibinfo{title}{{Spectroscopic identification of r-process nucleosynthesis in a double neutron-star merger},} \nat, 551, 67, \dodoi{10.1038/nature24298}

% type= article
\bibitem[{Q. {Pognan} {et~al.}(2023){Pognan}, {Grumer}, {Jerkstrand}, \& {Wanajo}}]{2023_Pognan+}
{Pognan}, Q., {Grumer}, J., {Jerkstrand}, A., \& {Wanajo}, S. 2023, \bibinfo{title}{{NLTE spectra of kilonovae},} \mnras, 526, 5220, \dodoi{10.1093/mnras/stad3106}

% type= article
\bibitem[{Q. {Pognan} {et~al.}(2022{\natexlab{a}}){Pognan}, {Jerkstrand}, \& {Grumer}}]{2022_Pognan+_a}
{Pognan}, Q., {Jerkstrand}, A., \& {Grumer}, J. 2022{\natexlab{a}}, \bibinfo{title}{{On the validity of steady-state for nebular phase kilonovae},} \mnras, 510, 3806, \dodoi{10.1093/mnras/stab3674}

% type= article
\bibitem[{Q. {Pognan} {et~al.}(2022{\natexlab{b}}){Pognan}, {Jerkstrand}, \& {Grumer}}]{2022_Pognan+_b}
{Pognan}, Q., {Jerkstrand}, A., \& {Grumer}, J. 2022{\natexlab{b}}, \bibinfo{title}{{NLTE effects on kilonova expansion opacities},} \mnras, 513, 5174, \dodoi{10.1093/mnras/stac1253}

% type= article
\bibitem[{Q. {Pognan} {et~al.}(2025){Pognan}, {Wu}, {Mart{\'\i}nez-Pinedo}, {da Silva}, {Jerkstrand}, {Grumer}, \& {Fl{\"o}rs}}]{2025_Pognan+}
{Pognan}, Q., {Wu}, M.-R., {Mart{\'\i}nez-Pinedo}, G., {et~al.} 2025, \bibinfo{title}{{Actinide signatures in low electron fraction kilonova ejecta},} \mnras, 536, 2973, \dodoi{10.1093/mnras/stae2778}

% type= article
\bibitem[{N. {Prantzos} {et~al.}(2020){Prantzos}, {Abia}, {Cristallo}, {Limongi}, \& {Chieffi}}]{2020_Prantzos+}
{Prantzos}, N., {Abia}, C., {Cristallo}, S., {Limongi}, M., \& {Chieffi}, A. 2020, \bibinfo{title}{{Chemical evolution with rotating massive star yields II. A new assessment of the solar s- and r-process components},} \mnras, 491, 1832, \dodoi{10.1093/mnras/stz3154}

% type= article
\bibitem[{S. {Rahmouni} {et~al.}(2025){Rahmouni}, {Tanaka}, {Domoto}, {Kato}, {Hotokezaka}, {Aoki}, {Hirano}, {Kotani}, {Kuzuhara}, \& {Tamura}}]{2025_Rahmouni+}
{Rahmouni}, S., {Tanaka}, M., {Domoto}, N., {et~al.} 2025, \bibinfo{title}{{Revisiting Near-infrared Features of Kilonovae: The Importance of Gadolinium},} \apj, 980, 43, \dodoi{10.3847/1538-4357/ada251}

% type= article
\bibitem[{Y. {Ralchenko} {et~al.}(2008){Ralchenko}, {Janev}, {Kato}, {Fursa}, {Bray}, \& {de Heer}}]{2008_Ralchenko+}
{Ralchenko}, Y., {Janev}, R.~K., {Kato}, T., {et~al.} 2008, \bibinfo{title}{{Electron-impact excitation and ionization cross sections for ground state and excited helium atoms},} Atomic Data and Nuclear Data Tables, 94, 603, \dodoi{10.1016/j.adt.2007.11.003}

% type= article
\bibitem[{G.~B. {Rybicki} \& D.~G. {Hummer}(1978){Rybicki} \& {Hummer}}]{1978_Rybicki&Hummer}
{Rybicki}, G.~B., \& {Hummer}, D.~G. 1978, \bibinfo{title}{{A generalization of the Sobolev method for flows with nonlocal radiative coupling.},} \apj, 219, 654, \dodoi{10.1086/155826}

% type= book
\bibitem[{G.~B. {Rybicki} \& A.~P. {Lightman}(1979){Rybicki} \& {Lightman}}]{1979_Rybicki&Lightman}
{Rybicki}, G.~B., \& {Lightman}, A.~P. 1979, {Radiative processes in astrophysics}

% type= article
\bibitem[{G. {Sadeh}(2025){Sadeh}}]{2025_Sadeh}
{Sadeh}, G. 2025, \bibinfo{title}{{Inferring Kilonova Ejecta Photospheric Properties from Early Blackbody Spectra},} \apj, 988, 130, \dodoi{10.3847/1538-4357/adecee}

% type= article
\bibitem[{L.~J. {Shingles} {et~al.}(2023){Shingles}, {Collins}, {Vijayan}, {Fl{\"o}rs}, {Just}, {Leck}, {Xiong}, {Bauswein}, {Mart{\'\i}nez-Pinedo}, \& {Sim}}]{2023_Shingles+}
{Shingles}, L.~J., {Collins}, C.~E., {Vijayan}, V., {et~al.} 2023, \bibinfo{title}{{Self-consistent 3D Radiative Transfer for Kilonovae: Directional Spectra from Merger Simulations},} \apjl, 954, L41, \dodoi{10.3847/2041-8213/acf29a}

% type= article
\bibitem[{S.~J. {Smartt} {et~al.}(2017){Smartt}, {Chen}, {Jerkstrand}, {Coughlin}, {Kankare}, {Sim}, {Fraser}, {Inserra}, {Maguire}, {Chambers}, {Huber}, {Kr{\"u}hler}, {Leloudas}, {Magee}, {Shingles}, {Smith}, {Young}, {Tonry}, {Kotak}, {Gal-Yam}, {Lyman}, {Homan}, {Agliozzo}, {Anderson}, {Angus}, {Ashall}, {Barbarino}, {Bauer}, {Berton}, {Botticella}, {Bulla}, {Bulger}, {Cannizzaro}, {Cano}, {Cartier}, {Cikota}, {Clark}, {De Cia}, {Della Valle}, {Denneau}, {Dennefeld}, {Dessart}, {Dimitriadis}, {Elias-Rosa}, {Firth}, {Flewelling}, {Fl{\"o}rs}, {Franckowiak}, {Frohmaier}, {Galbany}, {Gonz{\'a}lez-Gait{\'a}n}, {Greiner}, {Gromadzki}, {Guelbenzu}, {Guti{\'e}rrez}, {Hamanowicz}, {Hanlon}, {Harmanen}, {Heintz}, {Heinze}, {Hernandez}, {Hodgkin}, {Hook}, {Izzo}, {James}, {Jonker}, {Kerzendorf}, {Klose}, {Kostrzewa-Rutkowska}, {Kowalski}, {Kromer}, {Kuncarayakti}, {Lawrence}, {Lowe}, {Magnier}, {Manulis}, {Martin-Carrillo}, {Mattila}, {McBrien}, {M{\"u}ller}, {Nordin}, {O'Neill}, {Onori}, {Palmerio}, {Pastorello}, {Patat}, {Pignata}, {Podsiadlowski}, {Pumo}, {Prentice}, {Rau}, {Razza}, {Rest}, {Reynolds}, {Roy}, {Ruiter}, {Rybicki}, {Salmon}, {Schady}, {Schultz}, {Schweyer}, {Seitenzahl}, {Smith}, {Sollerman}, {Stalder}, {Stubbs}, {Sullivan}, {Szegedi}, {Taddia}, {Taubenberger}, {Terreran}, {van Soelen}, {Vos}, {Wainscoat}, {Walton}, {Waters}, {Weiland}, {Willman}, {Wiseman}, {Wright}, {Wyrzykowski}, \& {Yaron}}]{2017_Smartt+}
{Smartt}, S.~J., {Chen}, T.~W., {Jerkstrand}, A., {et~al.} 2017, \bibinfo{title}{{A kilonova as the electromagnetic counterpart to a gravitational-wave source},} \nat, 551, 75, \dodoi{10.1038/nature24303}

% type= article
\bibitem[{A. {Sneppen}(2023){Sneppen}}]{2023_Sneppen}
{Sneppen}, A. 2023, \bibinfo{title}{{On the Blackbody Spectrum of Kilonovae},} \apj, 955, 44, \dodoi{10.3847/1538-4357/acf200}

% type= article
\bibitem[{A. {Sneppen} {et~al.}(2024){Sneppen}, {Damgaard}, {Watson}, {Collins}, {Shingles}, \& {Sim}}]{2024_Sneppen+_c}
{Sneppen}, A., {Damgaard}, R., {Watson}, D., {et~al.} 2024, \bibinfo{title}{{Helium features are inconsistent with the spectral evolution of the kilonova AT2017gfo},} \aap, 692, A134, \dodoi{10.1051/0004-6361/202451450}

% type= article
\bibitem[{A. {Sneppen} \& D. {Watson}(2023){Sneppen} \& {Watson}}]{2023_Sneppen&Watson}
{Sneppen}, A., \& {Watson}, D. 2023, \bibinfo{title}{{Discovery of a 760 nm P Cygni line in AT2017gfo: Identification of yttrium in the kilonova photosphere},} \aap, 675, A194, \dodoi{10.1051/0004-6361/202346421}

% type= article
\bibitem[{A. {Sneppen} {et~al.}(2023){Sneppen}, {Watson}, {Poznanski}, {Just}, {Bauswein}, \& {Wojtak}}]{2023_Sneppen+_b}
{Sneppen}, A., {Watson}, D., {Poznanski}, D., {et~al.} 2023, \bibinfo{title}{{Measuring the Hubble constant with kilonovae using the expanding photosphere method},} \aap, 678, A14, \dodoi{10.1051/0004-6361/202346306}

% type= article
\bibitem[{A. {Sneppen} {et~al.}(2026){Sneppen}, {Just}, {Bauswein}, {Damgaard}, {Watson}, {Shingles}, {Collins}, {Sim}, {Xiong}, {Mart{\'\i}nez-Pinedo}, {Soultanis}, \& {Vijayan}}]{2026_Sneppen+_a}
{Sneppen}, A., {Just}, O., {Bauswein}, A., {et~al.} 2026, \bibinfo{title}{{Helium as an indicator of the neutron-star merger remnant lifetime and its potential for equation of state constraints},} \prd, 113, 063038, \dodoi{10.1103/5j43-l6xr}

% type= book
\bibitem[{V.~V. {Sobolev}(1960){Sobolev}}]{1960_Sobolev}
{Sobolev}, V.~V. 1960, {Moving Envelopes of Stars}, \dodoi{10.4159/harvard.9780674864658}

% type= article
\bibitem[{L.~V. {Spencer} \& U. {Fano}(1954){Spencer} \& {Fano}}]{1954_Spencer&Fano}
{Spencer}, L.~V., \& {Fano}, U. 1954, \bibinfo{title}{{Energy Spectrum Resulting from Electron Slowing Down},} Physical Review, 93, 1172, \dodoi{10.1103/PhysRev.93.1172}

% type= article
\bibitem[{E. {Symbalisty} \& D.~N. {Schramm}(1982){Symbalisty} \& {Schramm}}]{1982_Symbalisty&Schramm}
{Symbalisty}, E., \& {Schramm}, D.~N. 1982, \bibinfo{title}{{Neutron Star Collisions and the r-Process},} \aplett, 22, 143

% type= article
\bibitem[{M. {Tanaka} {et~al.}(2017){Tanaka}, {Utsumi}, {Mazzali}, {Tominaga}, {Yoshida}, {Sekiguchi}, {Morokuma}, {Motohara}, {Ohta}, {Kawabata}, {Abe}, {Aoki}, {Asakura}, {Baar}, {Barway}, {Bond}, {Doi}, {Fujiyoshi}, {Furusawa}, {Honda}, {Itoh}, {Kawabata}, {Kawai}, {Kim}, {Lee}, {Miyazaki}, {Morihana}, {Nagashima}, {Nagayama}, {Nakaoka}, {Nakata}, {Ohsawa}, {Ohshima}, {Okita}, {Saito}, {Sumi}, {Tajitsu}, {Takahashi}, {Takayama}, {Tamura}, {Tanaka}, {Terai}, {Tristram}, {Yasuda}, \& {Zenko}}]{2017_Tanaka+_b}
{Tanaka}, M., {Utsumi}, Y., {Mazzali}, P.~A., {et~al.} 2017, \bibinfo{title}{{Kilonova from post-merger ejecta as an optical and near-Infrared counterpart of GW170817},} \pasj, 69, 102, \dodoi{10.1093/pasj/psx121}

% type= article
\bibitem[{Y. {Tarumi} {et~al.}(2023){Tarumi}, {Hotokezaka}, {Domoto}, \& {Tanaka}}]{2023_Tarumi+}
{Tarumi}, Y., {Hotokezaka}, K., {Domoto}, N., \& {Tanaka}, M. 2023, \bibinfo{title}{{Non-LTE analysis for Helium and Strontium lines in the kilonova AT2017gfo},} arXiv e-prints, arXiv:2302.13061, \dodoi{10.48550/arXiv.2302.13061}

% type= article
\bibitem[{L.~N. {Van den Berg} \& K. {Hotokezaka}(2026){Van den Berg} \& {Hotokezaka}}]{2026_VandenBerg&Hotokezaka}
{Van den Berg}, L.~N., \& {Hotokezaka}, K. 2026, \bibinfo{title}{{Spectra and Ionization Efficiencies of Charged Decay Particles in Kilonova Ejecta},} arXiv e-prints, arXiv:2605.26613, \dodoi{10.48550/arXiv.2605.26613}

% type= article
\bibitem[{N. {Vieira} {et~al.}(2023){Vieira}, {Ruan}, {Haggard}, {Ford}, {Drout}, {Fern{\'a}ndez}, \& {Badnell}}]{2023_Vieira+}
{Vieira}, N., {Ruan}, J.~J., {Haggard}, D., {et~al.} 2023, \bibinfo{title}{{Spectroscopic r-Process Abundance Retrieval for Kilonovae. I. The Inferred Abundance Pattern of Early Emission from GW170817},} \apj, 944, 123, \dodoi{10.3847/1538-4357/acae72}

% type= article
\bibitem[{N. {Vieira} {et~al.}(2024){Vieira}, {Ruan}, {Haggard}, {Ford}, {Drout}, \& {Fern{\'a}ndez}}]{2024_Vieira+}
{Vieira}, N., {Ruan}, J.~J., {Haggard}, D., {et~al.} 2024, \bibinfo{title}{{Spectroscopic r-process Abundance Retrieval for Kilonovae. II. Lanthanides in the Inferred Abundance Patterns of Multicomponent Ejecta from the GW170817 Kilonova},} \apj, 962, 33, \dodoi{10.3847/1538-4357/ad1193}

% type= article
\bibitem[{V.~A. {Villar} {et~al.}(2017){Villar}, {Guillochon}, {Berger}, {Metzger}, {Cowperthwaite}, {Nicholl}, {Alexander}, {Blanchard}, {Chornock}, {Eftekhari}, {Fong}, {Margutti}, \& {Williams}}]{2017_Villar+_b}
{Villar}, V.~A., {Guillochon}, J., {Berger}, E., {et~al.} 2017, \bibinfo{title}{{The Combined Ultraviolet, Optical, and Near-infrared Light Curves of the Kilonova Associated with the Binary Neutron Star Merger GW170817: Unified Data Set, Analytic Models, and Physical Implications},} \apjl, 851, L21, \dodoi{10.3847/2041-8213/aa9c84}

% type= article
\bibitem[{S. {Wanajo}(2018){Wanajo}}]{2018_Wanajo}
{Wanajo}, S. 2018, \bibinfo{title}{{Physical Conditions for the r-process. I. Radioactive Energy Sources of Kilonovae},} \apj, 868, 65, \dodoi{10.3847/1538-4357/aae0f2}

% type= article
\bibitem[{S. {Wanajo} {et~al.}(2014){Wanajo}, {Sekiguchi}, {Nishimura}, {Kiuchi}, {Kyutoku}, \& {Shibata}}]{2014_Wanajo+}
{Wanajo}, S., {Sekiguchi}, Y., {Nishimura}, N., {et~al.} 2014, \bibinfo{title}{{Production of All the r-process Nuclides in the Dynamical Ejecta of Neutron Star Mergers},} \apjl, 789, L39, \dodoi{10.1088/2041-8205/789/2/L39}

% type= article
\bibitem[{D. {Watson} {et~al.}(2019){Watson}, {Hansen}, {Selsing}, {Koch}, {Malesani}, {Andersen}, {Fynbo}, {Arcones}, {Bauswein}, {Covino}, {Grado}, {Heintz}, {Hunt}, {Kouveliotou}, {Leloudas}, {Levan}, {Mazzali}, \& {Pian}}]{2019_Watson+}
{Watson}, D., {Hansen}, C.~J., {Selsing}, J., {et~al.} 2019, \bibinfo{title}{{Identification of strontium in the merger of two neutron stars},} \nat, 574, 497, \dodoi{10.1038/s41586-019-1676-3}

% type= article
\bibitem[{E. {Waxman} {et~al.}(2018){Waxman}, {Ofek}, {Kushnir}, \& {Gal-Yam}}]{2018_Waxman+}
{Waxman}, E., {Ofek}, E.~O., {Kushnir}, D., \& {Gal-Yam}, A. 2018, \bibinfo{title}{{Constraints on the ejecta of the GW170817 neutron star merger from its electromagnetic emission},} \mnras, 481, 3423, \dodoi{10.1093/mnras/sty2441}

% type= article
\bibitem[{R.~T. {Wollaeger} {et~al.}(2024){Wollaeger}, {Fryer}, {Chiodi}, {Brady}, {Korobkin}, {Harnish}, {Fontes}, {Haack}, {Chapurin}, {Koshkarov}, {Delzanno}, \& {Livescu}}]{2024_Wollaeger+}
{Wollaeger}, R.~T., {Fryer}, C.~L., {Chiodi}, R.~M., {et~al.} 2024, \bibinfo{title}{{On a Spectral Method for {\ensuremath{\beta}}-particle Bound Excitation Collisions in Kilonovae},} \apj, 966, 177, \dodoi{10.3847/1538-4357/ad37f9}

\end{thebibliography}
\bibliographystyle{aasjournalv7}



\end{document}